%% file: sukhdeep_cmb_paper.tex
\documentclass[a4paper,fleqn,usenatbib]{mnras}

\usepackage{natbib} 
\usepackage{amsmath,amssymb}
\usepackage{epsfig,graphicx}
\usepackage{epstopdf}
\usepackage{subcaption}
\usepackage{float}
\usepackage{lastpage}

\DeclareGraphicsExtensions{.pdf,.png,.jpg}
\input{def.tex}

\graphicspath{{figures/}}

\usepackage{color}
\definecolor{pink}{rgb}{0.9,0.,0.9}

\newcommand{\sukhdeep}[1]{{\textcolor{pink}{#1}}}
\newcommand{\referee}[1]{{\textcolor{black}{#1}}}
\newcommand{\refereeTwo}[1]{{\textcolor{black}{#1}}}

\newcommand{\lcdm}{\ensuremath{\Lambda \text{CDM}}}
\newcommand{\ugg}{\ensuremath{\Upsilon_{gg}}}
\newcommand{\ugm}{\ensuremath{\Upsilon_{gm}}}
\newcommand{\cgm}{\ensuremath{r_\text{cc}}}
\newcommand{\sigmaNFW}{\ensuremath{\Sigma_{gm}^\text{approx}}}
\newcommand{\nside}{{n$_\text{side}$}}
\newcommand{\hmsun}{{$h^{-1}M_{\odot}$}}

\title[SDSS-Planck lensing cross-correlations]{Cross-correlating Planck CMB lensing with SDSS: Lensing-lensing and galaxy-lensing cross-correlations}

\author[Singh \& Mandelbaum]{ Sukhdeep Singh$^1$\thanks{\tt
    sukhdeep@cmu.edu}, Rachel Mandelbaum$^{1}$, Joel R. Brownstein$^2$ \\ $^1$McWilliams
  Center for Cosmology, Department of Physics, Carnegie Mellon
  University, Pittsburgh, PA 15213, USA\\ $^2$Department of Physics and Astronomy, University of Utah, 115 S 1400 E, Salt Lake City, UT 84112, USA} \date{\today}
\date{Accepted XXX. Received YYY; in original form ZZZ}
\pubyear{2016}
\begin{document}
\label{firstpage}
\pagerange{\pageref{firstpage}--\pageref{LastPage}}
\maketitle

\begin{abstract}
	We present results from cross-correlating Planck CMB lensing maps with the Sloan Digital
    Sky Survey (SDSS) galaxy lensing shape
	catalog and BOSS galaxy catalogs. For galaxy position vs.\ CMB lensing cross-correlations, we measure the
    convergence signal around the galaxies in
	configuration space, using the BOSS LOWZ ($z\sim0.30$) and CMASS ($z\sim0.57$) samples. With fixed Planck 2015
	cosmology, doing a
	joint fit with the galaxy clustering measurement, for the LOWZ (CMASS) sample we find a galaxy
    bias $b_g=1.75\pm0.04$ ($1.95\pm 0.02$)
    and galaxy-matter cross-correlation
    coefficient $\cgm=1.0\pm0.2$ ($0.8\pm 0.1$) using $20<r_p<70\mpch$,
    consistent with results from galaxy-galaxy lensing. Using the same scales and including the
    galaxy-galaxy lensing measurements, we constrain $\Omega_m=0.284\pm0.024$ and relative calibration bias between the
    CMB lensing and galaxy lensing to be $b_\gamma=0.82^{+0.15}_{-0.14}$. The combination of galaxy lensing and CMB
    lensing
    also allows us to measure the cosmological distance ratios (with $z_l\sim0.3$, $z_s\sim0.5$)
    $\mathcal R=\frac{D_s D_{l,*}}{D_* D_{l,s}}=2.68\pm0.29$, consistent with predictions from the
    Planck 2015 cosmology ($\mathcal R=2.35$).
	We detect the galaxy position-CMB convergence cross-correlation
   	at small scales, $r_p<1\mpch$, and find consistency with lensing by NFW halos of mass $M_h\sim10^{13}$\hmsun.
	Finally, we measure the CMB lensing-galaxy shear cross-correlation,
	finding an amplitude of
	$A=0.76\pm0.23$ ($z_\text{eff}=0.35$, $\theta<2^\circ$) with respect to Planck 2015 \lcdm\
    predictions ($1\sigma$-level consistency).
	We do not find evidence for relative systematics between the CMB and SDSS galaxy lensing.
\end{abstract}


\section{Introduction}\label{sec:intro}
	As photons travel from the source towards the observer, their paths are distorted by the gravitational
	potential of the intervening matter. This phenomenon, known as gravitational lensing, has become an important
	tool in cosmology to study the growth of structure in the dark matter distribution as well as cosmic
	acceleration \citep{Weinberg2013}. In the weak regime, the gravitational lensing introduces small but coherent
	distortions in the shapes of background galaxies, which can be measured through correlations of galaxy images.
	Weak gravitational lensing also remaps the CMB anisotropies, blurring the acoustic peaks and correlating different 
	modes,
	which can then be exploited to generate mass maps from the CMB observations with good resolution and
	signal-to-noise ratio \referee{\citep{Zaldarriaga1999,Hu2002,Lewis2006}}.

	CMB lensing is most sensitive to structure at high redshifts, $z=1-5$, and thus provides a unique probe to study
	the structure in the dark matter distribution at these high redshifts.
	The lensing of CMB has been robustly detected by several CMB experiments including the Atacama Cosmology Telescope
	\citep[ACT;][]{Das2011,Das2014}, the South Pole Telescope \citep[SPT;][]{Engelen2012}, the
	Planck telescope \citep{Planck2013lensing,Planck2015lensing}, \referee{The POLARBEAR experiment \citep{Ade2014} and 
	the BICEP2/Keck array \citep{Keck_array2016}}, with measurements being \referee{consistent} 
	with the \lcdm\ predictions.
	Cross-correlating CMB lensing maps with galaxy surveys provides opportunities to probe the large-scale
	structure (LSS), calibrate different probes of large-scale structure (particularly galaxy lensing) and carry out
	consistency tests. Several studies cross-correlating CMB lensing with galaxy position catalogs as well as
	galaxy lensing catalogs have already been performed.

	Given that CMB lensing and galaxy lensing have very different lensing kernels, cross-correlating
    the two can help
	constrain the amplitude of matter fluctuations at low redshift (whereas CMB lensing
    auto-correlations are not very sensitive to this). The past year has seen a number of detections
    of this effect in several surveys \citep{Hand2015,Kirk2016,Liu2015,Harnois2016}, typically in
    Fourier space but in the last work, in configuration space as well.  Assuming a fixed cosmological model, these
    cross-correlations also provide a test for relative
	calibration biases between the two lensing maps \citep{Vallinotto2012,Das2013}.  The results of
    recent work has been largely consistent with \lcdm\ predictions, with at most slight tension
    ($\sim 2\sigma$) that has at times \citep[e.g.,][]{Liu2016} been interpreted as residual
    systematics in the galaxy lensing.

	Cross-correlating CMB lensing with galaxy positions also provides a probe of structure growth and of
    the matter distribution around
	galaxies. In addition, when compared with galaxy-galaxy lensing, such correlations can also
    provide a handle on systematics such as biases in photometric
	redshift distributions \citep{Putter2014}. Many cross-correlations of galaxy positions with
    CMB lensing exist in the
	literature
\citep{Smith2007,Hirata2008,Bleem2012,Sherwin2012,Planck2013lensing,Giannantonio2014,Bianchini2015,Pullen2015,Giannantonio2016}.
As for the lensing-lensing cross-correlation, the results of measurement of the galaxy position vs.\ CMB
lensing cross-correlation are typically consistent with \lcdm\ predictions, at times with low-level
($2\sigma$) tension.
For example,  \cite{Pullen2015} used the cross correlations between BOSS CMASS galaxies and CMB lensing in combination with
     galaxy
     clustering and velocity measurements to constrain the theory of gravity on large scales. They found $
     \sim2\sigma$ deviations
     from \lcdm\ predictions, with the discrepancy being primarily driven by the lower amplitude of
     galaxy position vs.\ CMB lensing cross-correlations at $r_p\gtrsim 80\mpch$.
     These cross-correlation measurements provide a valuable consistency check when compared with results
  	using galaxy lensing.

\cite{Miyatake2016} measured the ratio of CMB lensing and galaxy lensing signal around the BOSS
CMASS galaxies \citep{Alam2015}, using Planck 2015 CMB lensing maps \citep{Planck2015lensing} and
galaxy lensing sources from CFHTLens \citep{Erben2013}. Such a ratio depends on the geometric
factors involving the distance between the
     observer, the lens galaxies, the CMB last scattering surface and the source galaxies, providing a measurement of cosmic distance
     ratios \citep{Hu2007a} and hence a strong consistency check on the cosmological model.

	 Most of the CMB lensing vs.\ galaxy position cross-correlation measurements have been
	performed at large scales. Current and next-generation CMB surveys will have sufficiently high
    resolution and low  noise levels
	to measure the lensing signals even on the scales of dark matter halos, and provide mass
    constraints \citep{Hu2007a}.  This has the potential to be particularly powerful at the high
    redshifts that are beyond the reach of galaxy lensing surveys. The first such measurement
	has already been performed by \cite{Madhavacheril2015} by cross-correlating CMB lensing maps from ACTPol with CMASS
	sample galaxies from the SDSS-III Baryon Oscillation Spectroscopic Survey \citep[BOSS;][]{Alam2015}. 
	\referee{Similar measurements using more massive SZ selected clusters have also been performed by 
	\cite{Baxter2015} and \cite{Planck2015Clusters}.}

	In this work, we perform the galaxy-CMB lensing cross-correlations using SDSS-III BOSS galaxies
	\citep{Alam2015}. We compute the projected matter density $\Sigma$ in real comoving space by stacking the
	convergence obtained from Planck 2015 lensing maps around the positions of BOSS galaxies. Using the low redshift
	sample, LOWZ, we also present a
	direct comparison between the results from galaxy lensing and CMB lensing, and test for relative calibration
	biases between the two lensing signals. Using the combination of CMB lensing and galaxy lensing, we also present a
	measurement of the cosmic distance ratio at an effective lens redshift of $0.26$.
	In addition, we also cross-correlate the two lensing measurements in
	configuration space. The SDSS lensing source sample is at relatively low
	redshift compared to other galaxy lensing surveys, where the rapid decrease of the CMB lensing kernel reduces the
	amplitude of
	the cross-correlation signal. However, the large sky area of the SDSS ($\gtrsim8000$ square
	degrees) compared to other existing lensing surveys compensates for the relatively lower expected signal, 
	particularly in
	the case where noise in the CMB lensing maps dominates the statistical error budget.

	This paper is organized as follows: In Sec.~\ref{sec:formalism}, we discuss the theoretical
	background and the estimators used in our measurements. In Sec.~\ref{sec:data}, we describe the
 datasets used in this work.  Our results are in Sec.~\ref{sec:results}, and we conclude in
 Sec.~\ref{sec:conclusions}. Throughout, we use the Planck 2015 cosmology \citep{Planck2015cosmo}, with $
	\Omega_m=0.309$, $n_s=0.967$, $A_s=2.142\times10^{-9}$, $\sigma_8=0.82$. To compute theoretical
    predictions in this paper, we use the linear$+$halofit
   \citep{Smith2003,Takahashi2012}
   \referee{matter} power spectrum generated using the CAMB software \citep{Lewis2002}.

\section{Formalism and Methodology}\label{sec:formalism}
	In this section, we present the theoretical models and estimators used to carry out and interpret the
	measurements.
	\subsection{Weak Lensing Introduction}
		Here we provide a very brief review of weak lensing,  and refer the reader to
        \cite{Bartelmann2001} for details. Gravitational lensing measurements are sensitive to the lensing potential,
        defined as
		\begin{equation}
			\Phi_L=\int \mathrm{d}\chi_l\frac{f_k(\chi_s-\chi_l)}{f_k(\chi_s)f_k(\chi_l)}\Psi(f_k
			(\chi_l)\vec\theta,\chi_l)
		\end{equation}
		where the Weyl Potential $\Psi=\psi+\phi$, $\phi$ and $\psi$ are the Newtonian and
        curvature potentials, $\vec{\theta}$ is the angular coordinate on the sky, $\chi_s,\chi_l$ are comoving radial
        distances to source and lens respectively and $f_k$ are the generalized (not assuming
        flatness) transverse comoving distances \referee{(in the case of a flat universe, $f_k(\chi)=\chi$)}.
		Within \lcdm, $\phi=\psi$ and $\nabla^2\phi=4\pi G\rho_m$.
		When source size is smaller than the angular scales over which lens
		properties change, the relation between source and image positions can be linearized and the Jacobian of the
		image to source transformation is \citep{Bartelmann2001}
    \begin{align}
      A_{ij}&=\frac{\partial(\theta_o^i-\delta\theta_o^i)}{\partial\theta_o^j}\\
      A_{ij}&=\delta_{ij}-\frac{\partial^2\Phi_l}{\partial\theta_o^i\partial\theta_o^j}
    \end{align}
	where the subscript $o$ indicates that the derivative is carried out in observer (or image)
    coordinates.  The matrix $A$
	is\referee{
    \[
    	A=
	      \begin{bmatrix}
    	    1-\kappa-\gamma_1 & -\gamma_2\\
        	-\gamma_2 & 1-\kappa+\gamma_1
	      \end{bmatrix}.
    \]
		In the weak gravitational lensing regime, the primary
		observables are the convergence $\kappa$ for CMB lensing and the shear 
		$\gamma=\gamma_1+\mathrm{i}\gamma_2=|\gamma|e^{2i\theta}$
		for galaxy lensing. The measured $\gamma$ can be rotated into the lens-source frame, $\gamma=\gamma_t+i\gamma_\times$, 
		where 
		$\gamma_t$ is the shear along the line joining the lens and source galaxy while
		$\gamma_\times$ is the shear with respect to the $45^\circ$ lines.}
		These observables relate to the underlying gravitational potential as
		\begin{align}
	      \kappa&=\frac{1}{2}\nabla_\perp^2\Phi_L\\ \label{eq:kappa}
    	  \gamma_t&=\frac{1}{2}(\nabla_{x,x}^2-\nabla_{y,y}^2)\Phi_L\\ \label{eq:gamma_t}
	      \gamma_\times&=\frac{1}{2}(\nabla_{x,y}^2)\Phi_L
    	\end{align}
		where the derivatives are with respect to the plane-of-sky coordinates ($x$ is along the
		line joining the lens and source positions, and $y$ is orthogonal to $x$).
		Note that the lensing observables are sensitive only to the matter density contrast
		($\bar{\rho}_m\delta=\rho_m-\bar{\rho}_m$) and not the mean matter density.
 \referee{ The equivalent expressions for convergence and shear in the Fourier space are}
		\begin{align}
	      \tilde{\kappa}&=\frac{1}{2}k_\perp^2\tilde\Phi_L\\ \label{eq:kappa_fourier}
    	  \tilde{\gamma}_t&=\equiv\frac{1}{2}(k_{\perp,x}^2-k_{\perp,y}^2)\tilde\Phi_L\\ \label{eq:gamma_t_fourier}
	      \tilde{\gamma}_\times&=\frac{1}{2}(k_{\perp,x}k_{\perp,y})\tilde\Phi_L,
    	\end{align}

	\subsection{Lensing-lensing cross-correlation}\label{subsec:ll-theory}
		When cross-correlating CMB lensing with galaxy shear, we will be measuring the tangential
        shear $\gamma_t$ in the galaxies
		around each pixel of the CMB map, and weight that shear with the CMB convergence value $\kappa$ within that
		pixel.
		Under the Limber approximation, using expressions for $\kappa$ and $\gamma_t$ from Eqs.~\eqref{eq:kappa} and
		\eqref{eq:gamma_t}, we can write the cross-correlation function for shear and convergence as
		\begin{align}
			\langle\kappa\gamma\rangle(\theta)&=\int \mathrm{d}z_{{\kappa}} p(z_{{\kappa}})\int \mathrm{d}z_{{
			\gamma}}
			 p(z_{{\gamma}})\nonumber\\
			 &\int
			 \mathrm{d}z_l \frac{H(z_l)}{c}W_L(\chi_{{\kappa}},
			 \chi_l)W_L(\chi_{{\gamma}},\chi_l)\nonumber\\
			&\int \frac{\mathrm{d}k}{2\pi} k P_{\delta\delta}(k)
			J_2\left[k f_k(\chi_l)\theta\right],
			\label{eq:kappa_gamma}
		\end{align}
		where $p({z_{\kappa}})$ and $p({z_{\gamma}})$ are the redshift distribution of the source samples used to
		measure the convergence and shear respectively.
          In the case that $\kappa$ is measured from CMB lensing, $p({z_{\kappa}})=\delta_D(z_{\kappa}-1100)$.
		$\gamma_t$ is the galaxy tangential shear defined with respect to the line joining the galaxy with the pixel
		center of $\kappa$ map.
		The lensing weight $W_L$ is defined as
		\begin{align}
			&W_L(\chi_s,\chi_l)=\frac{3}{2}\frac{H_0^2}{c^2}{\Omega_{m,0} (1+z_l)}\frac{c}{H(z_l)}\frac{f_k
			(\chi_l)
			f_k(\chi_s-\chi_l)}{f_k(\chi_s)}\label{eq:lens_weight}
		\end{align}
		when $\chi_s>\chi_l$ and $W_l=0$ for $\chi_s<\chi_l$.

\begin{figure}
         \centering
         \includegraphics[width=\columnwidth]{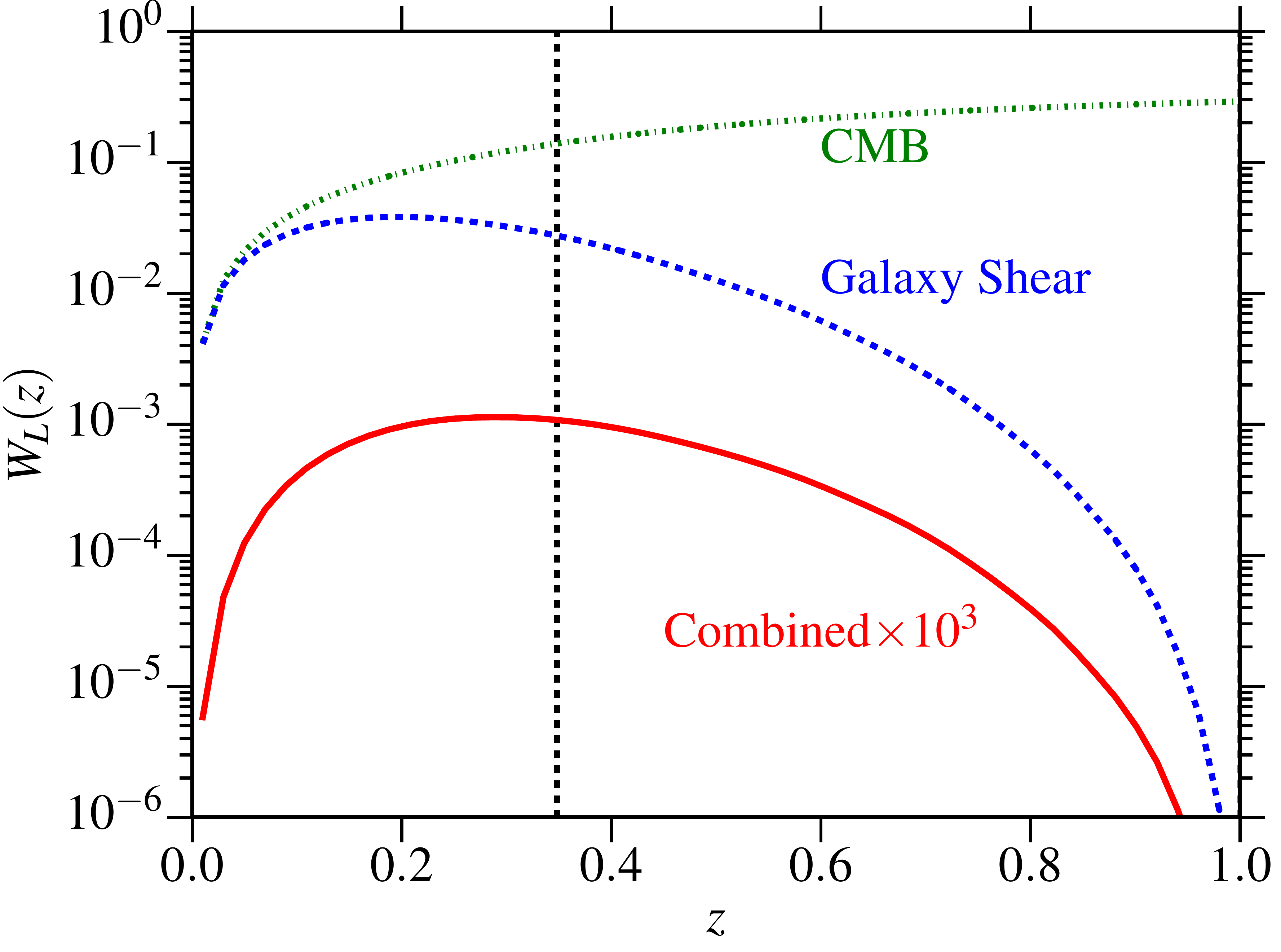}
         \caption{ We show the weight functions that enter the lensing-lensing correlations as defined in
         Eqs.~\eqref{eq:lens_weight} and~\eqref{eq:wt_combined}. The vertical black line marks the effective redshift for the
         lensing-lensing cross-correlation signal, measured using combined weights. For galaxy
         lensing we used the SDSS source sample redshift distribution as defined in Sec.~\ref{ssec:data_source_sample}.
         }
         \label{fig:Lens_weight_z}
      \end{figure}

      The CMB lensing weight (or kernel) decreases sharply at low redshift, while the galaxy lensing weights
       depends on the
      redshift distribution of source galaxies. In Fig.~\ref{fig:Lens_weight_z} we show the redshift-dependent weights 
      of both
      CMB lensing and galaxy lensing
	 (calculated using the SDSS source sample redshift distribution as described in
Sec.~\ref{ssec:data_source_sample})
      as a function of redshift, as well as the combined weight given by
      \begin{equation}\label{eq:wt_combined}
         W_{L}(z)^\text{Combined}=\frac{H(z)}{c}D(z)^2 W_{L}(z)^\text{CMB}W_{L}(z)^\text{galaxy shear}.
      \end{equation}
      $W_L(z)$ for CMB and galaxy shear are defined in Eq.~\eqref{eq:lens_weight}; $D(z)$, the growth
      function normalized to 1 at $z=0$, accounts for the growth of matter perturbations with
       redshifts. In this work,
      we will use source galaxies with photometric redshift values satisfying $z_p>0.1$ (see section ~\ref{ssec:data_source_sample}). The galaxy lensing weight peaks around $z\sim0.2$ and the
      combined weight peaks around $z\sim0.3$, though it is skewed towards higher redshift with
      $\langle z\rangle=0.35$.

	\subsection{Lensing signal around galaxies}\label{ssec:theory_galaxy_lensing}
		When studying the lensing signal around galaxies, we are interested in measuring the projected average surface mass density $\Sigma$
		around
		the lens galaxy sample. We start by writing the convergence and shear as
		\begin{align}
			&\kappa(r_p)=\frac{\Sigma(r_p)}{\Sigma_{c}}\\
			&\gamma_t(r_p)=\frac{\bar{\Sigma}(<r_p)-\Sigma(r_p)}{\Sigma_{c}}
		\end{align}
		$\bar{\Sigma}(<r_p)$ is the mean projected surface mass density within the radius
        $r_p$ (transverse comoving), and the critical surface density is
		defined in comoving coordinates as
		\begin{equation}\label{eqn:sigma_crit}
    		\Sigma_c=\frac{c^2}{4\pi G}\frac{f_k(\chi_s)}{(1+z_l) f_k(\chi_l) f_k(\chi_s-\chi_l)}.
	    \end{equation}
		The	$(1+z_l)$ factor is required to convert the $c^2/G$
		factor to comoving space since it has dimensions of $\left[\frac{\text{Mass}}{\text{length}}\right]$.
		We can write $\Sigma$ in terms of the projected surface mass density as
		\begin{equation}
    		\Sigma(r_p)=\bar\rho_m\int \mathrm{d}\Pi \,\xi_{gm}(r_p,\Pi)=\bar\rho_m w_{gm}(r_p),
	    \end{equation}
		where $\Pi$ is the line of sight distance from the galaxy in redshift space, and $\xi_{gm}$ is the three
		dimensional
		galaxy-matter cross-correlation function.
		 Here we use $\xi_{gm}$ rather than $1+\xi_{gm}$ since the shear is only sensitive to the matter density
		 contrast, i.e., $\Sigma^\text{lensing}=\Sigma(r_p)-\bar{\Sigma}$, where $\bar{\Sigma}$ is the background
		 projected surface mass density. Throughout this paper when using $\Sigma$, we mean $\Sigma^\text{lensing}$.
           The projected matter-galaxy correlation function can be derived from matter power spectrum as
		\begin{align}
			w_{gm}(r_p)=b_g \cgm\int \mathrm{d}z W(z)\int \frac{\mathrm{d}^2k_\perp}{(2\pi)^2}&P_{\delta
			\delta}(\vec{k},z)e^{i(\vec{r}_p.\vec{k}_\perp)}
	    \end{align}
		 where $b_g$ is the linear galaxy bias and $\cgm$ is the galaxy-matter cross-correlation
         coefficient, assumed here to be independent of redshift. The linear bias assumption is only valid at linear
         scales $r_p\gtrsim10\mpch$, and at smaller scales there are substantial contributions from non-linear bias
         \citep{Baldauf2010,Mandelbaum2013}. In this work we primarily use scales where the linear bias model is valid,
         and analysis with a scale-dependent bias will be
         presented in future work.
		 To first order, lensing measurements are not affected by redshift space distortions
		 and hence we do not include any corrections for that. The weight function $W(z)$ depends on the redshift
		 distribution of source galaxies  and on any redshift-dependent weights used
		 when estimating the signal (see Sec.~\ref{ssec:estimators}). We explicitly calculate these weights and
		 include them in the theory calculations, which integrates over the whole redshift range
         while using the weights (but does assume redshift-independent $b$ and \cgm).  For a
         nearly volume-limited sample like LOWZ, this is a quite good assumption, but in general if the
         bias and \cgm\ do evolve with redshift, we will measure an effective value averaged over
         redshift.
	\subsubsection{1-halo term: Galaxy-galaxy lensing}\label{ssec:delta_sigma_1_halo}
        To model the small-scale signal ($r_p\lesssim 1\mpch$), we use the NFW
		profile \citep{Navarro1996}, for which
        the 3D density is
        \begin{equation}
              \rho(r)=\frac{\rho_s}{(r/r_s)(1+r/r_s)^2}
        \end{equation}
		The NFW profile can be integrated over the line-of-sight to get the projected mass density, $\Sigma$
		and then $\Delta\Sigma$
         \begin{align}
              &\Sigma(r_p)=2\int_0^{r_{\text{vir}}} \rho(r=\sqrt{r_p^2+\chi^2})\mathrm{d}\chi.\\
              &\Delta\Sigma(r_p)=\bar\Sigma(<r_p)-\Sigma(r_p)
          \end{align}
          For NFW profiles, we define the concentration, $c_{200b}=r_{200b}/r_s$, and mass, $M_{200b}$, using a
          spherical overdensity of $200$ times the mean density:
            		\begin{equation}
				M_{200b}=\frac{4\pi}{3}r_{200b}^3 (200\bar{\rho}_m)
			\end{equation}
			We use the {\sc colossus}\footnote{\url{http://www.benediktdiemer.com/code/}} software
            \citep{Diemer2015} to compute the NFW profiles with  the mass-concentration relation
            from \cite{Bhattacharya2013}.  
			To avoid the contamination from the host halo (for those galaxies that are satellites
            within some larger host halo) and halo-halo terms, we fit the NFW profile only in the
			range $0.05<r_p<0.3\mpch$ in case of galaxy-galaxy lensing.

		\subsubsection{1-halo term: Galaxy-CMB lensing}\label{ssec:sigma_1halo}
			In the case of CMB lensing, as described in Sec.~\ref{ssec:data_planck}, the Planck CMB lensing maps
			only go up to $l_\text{max}=2048$, which corresponds to a hard edge in $\ell$ that
            effectively smoothes out the configuration-space maps at $\sim6'$
            scales.  To model the
			signal, we then have to smooth out the NFW profile by convolving it with a sinc kernel
			($\propto \frac{J_1(l_\text{max}\theta)}{l_\text{max}\theta}$) and with a tophat kernel
            with size set by the resolution of the healpix map being used.
			However, $6'$ is $\gtrsim$ the virial radius at all
			redshifts and expected halo masses of our samples, and thus
			the measured profiles on all scales will have contributions from satellite and halo-halo
            terms. Since the simple NFW
			profile does not contain these contributions, mass estimates using only the NFW profiles will be biased high. To
			get more correct mass estimates, one should use a halo model. However,
			since the signal-to-noise ratio of the CMB lensing at these scales  is fairly low, and there are additional
			uncertainties in the measurement from the Planck beam and possible leakage of foregrounds and astrophysical
            systematics such the thermal SZ effect \citep{Engelen2014},
			we do not attempt more complicated modeling in this work. Instead, for a qualitative comparison, we will
			present convolved profiles by
			defining $\Sigma_{gg}$ and \sigmaNFW\ as
			\begin{equation}\label{eqn:sigma_gg}
				\Sigma_{gg}(r_p)=\frac{\bar{\rho}_m}{b_g}\wgg(r_p)
			\end{equation}
			\begin{equation}\label{eq:sigma_gg_nfw}
				\sigmaNFW(r_p)=\Sigma_\text{NFW}(r_p)+e^{-(0.5/r_p)^2}\Sigma_{gg}(r_p)
			\end{equation}
			where \wgg\ is measured directly from the data as described in
            Sec.~\ref{ssec:clustering_theory} and~\ref{sssec:estimator_clustering}. At very small
            scales, the \wgg\ measurement is likely to be biased
			because of imperfect corrections for fiber collisions (see Sec.~\ref{ssec:data_boss}). Hence we down-weight
			\wgg\ at these scales
			and use the NFW profile, which should
			provide an adequate estimates for the small-scale profile.  We fit this profile to the data
            with the  NFW halo mass as a free
			parameter. However, these mass constraints could be biased because of several
            approximations being made in this model, so they
            too should be taken as a rough guide.

			Finally, we will also present a comparison with $\Sigma_{gm}$ measured from the
            `Med-Res' $N$-body simulations
			that were first presented in \cite{Reid2014}, using the $z=0.25$ and $z=0.6$ snapshots
            for LOWZ and CMASS galaxies, respectively. The sample of halos we use is
			generated using an HOD model from \cite{Zheng2005} fit to the clustering of galaxies assuming fixed
			abundance of parent halos of the galaxies, with priors on abundance being derived from range of redshift-dependent abundance of galaxies \citep[see][for more details]{Reid2014}.
         To get $\Sigma_{gm}$, we cross-correlate the
			halos with matter particles to obtain $w_{gm}$, which is then multiplied with $\bar{\rho}_m$ to get
			$\Sigma_{gm}$.

         We note that the smoothing kernel is only applied to the theoretical predictions
         when fitting a model to the small-scale signals below the
         resolution of the Planck convergence maps. When fitting large scales ($r_p\gtrsim5\mpch$),
         we have confirmed that the effects of smoothing  are negligible, and hence we do not apply
         the smoothing kernel to the theoretical predictions in those cases.

	\subsection{Galaxy Clustering}\label{ssec:clustering_theory}
		We also measure the two-point correlation function of galaxies to constrain their
        large-scale (linear) bias. In the linear bias regime, the two-point
		correlation function of galaxies is given by
		\begin{align}
			\xi_{gg}(r_p,\Pi)=&b_g^2\int \mathrm{d}z \,W(z)\int \frac{\mathrm{d}^2k_\perp\mathrm{d}k_z}{(2\pi)^3}\\
			&P_{\delta\delta}(\vec{k},z)(1+\beta\mu_k^2)^2 e^{i(\vec{r}_p.\vec{k}_\perp+\Pi k_z)}.
	    \end{align}
		The Kaiser factor,
		$(1+\beta\mu_k^2)$, accounts for the redshift space
		distortions in the linear regime \citep{Kaiser1987}, where $\beta=f(z)/b_g$, $f(z)$ is the linear growth rate
		factor at redshift $z$
		and $\mu_k=k_z/k$. The weight function $W(z)$ is given by
        \citet{Mandelbaum2011} as
			\begin{equation}
				W(z)=\frac{p(z)^2}{\chi^2 (z)\mathrm{d}\chi/\mathrm{d}z} \left[\int \frac{p(z)^2}{\chi^2
				(z)\mathrm{d}\chi/\mathrm{d}z} \mathrm{d}z\right]^{-1}.
			\end{equation}
		Here $p(z)$  is the redshift probability distribution for the galaxy sample.

		Finally we integrate $\xi_{gg}$ over the line-of-sight separation to get the projected correlation
		function
		\begin{equation}
			w_{gg}(r_p)=\int\limits_{-\Pi_\text{max}}^{\Pi_\text{max}}\mathrm{d}\Pi\, \xi_{gg}(r_p,\Pi)
		\end{equation}
		We use $\Pi_\text{max}=100\mpch$ to reduce the effects of redshift space
        distortions \citep{Bosch2013}.

	\subsection{Estimators}\label{ssec:estimators}
		In this section we present the estimators used for measuring various signals. For all  measurements, we use
		100 approximately equal-area ($\sim9^\circ$ on a side) jackknife regions to obtain the jackknife mean and
		errors for each bin, as described in \cite{Singh2015} in more detail. When fitting
        measurements to theoretical predictions, we
		use a weighted least squares method to fit each
		jackknife region using just the diagonal elements of jackknife covariance matrix,
        and then quote the jackknife mean and errors on the best-fitting parameters. When
		doing MCMC fits (Sec.~\ref{ssec:results_lensing_calibration}), we use the jackknife covariance matrix.

		\subsubsection{Galaxy-Galaxy Lensing}\label{ssec:estimator_galaxy_galaxy_lensing}
			For galaxy-galaxy lensing we measure $\Delta \Sigma$ as
    		\begin{equation}
            	\widehat{\Delta \Sigma}(r_p)=\frac{\sum_{ls}w_{ls}\gamma_t^{(ls)}\Sigma_c^{(ls)}}{\sum_{rs}w_{rs}}-\frac{\sum_{rs}w_{rs}\gamma_t^{(rs)}\Sigma_c^{(rs)}}{\sum_{rs}w_{rs}}
        	    \label{eqn:delta_sigma_estimator}
            \end{equation}
            The summation is over all lens-sources (ls) pairs, where 
            the weight $w_{ls}$ for each lens-source pair is defined as
            \citep[see e.g.][for more detail]{Singh2015}
            \begin{equation}\label{eq:delta_sigma_wt}
               w_{ls}=\frac{\Sigma_c^{-2}}{\sigma_\gamma^2+\sigma_{SN}^2}.
            \end{equation}
			$\sigma_{SN}$ is the shape noise and $\sigma_{\gamma}$ is the measurement noise for the
            source galaxy.
            The $\Sigma_c^{-2}$ enters the weight because we defined the $\Delta\Sigma$ in
            Eq.~\eqref{eqn:delta_sigma_estimator} as the maximum-likelihood estimator \citep{Sheldon2004}.
            Note that the denominator in the first term in Eq.~\eqref{eqn:delta_sigma_estimator} has
            a weight $w_{rs}$, measured by using
            random lenses rather than real lenses. This accounts for the dilution of the shear by
            unsheared
            ``source'' galaxies that are actually associated with the lens but are
            put behind the lens due to photometric redshift scatter. The correction factor for this
            effect, $\sum w_{ls}/\sum w_{rs}$, is usually called the boost
            factor \citep{Sheldon2004,Mandelbaum2005}. Finally, the second term in
            Eq.~\eqref{eqn:delta_sigma_estimator} is the subtraction of the $\Delta\Sigma$ measured around the
            random lenses, to remove the effect of spurious shear at large
            scales \citep{Mandelbaum2005}.

            To measure only the tangential shear around galaxies, we use the estimator
            \begin{equation}
            	\widehat{g \gamma_t}(r_p)=\frac{\sum_{ls}w^{\gamma}_{ls}\gamma_t^{(ls)}}{\sum_{rs}w^{\gamma}_{rs}}-\frac{\sum_{rs}w^{\gamma}_{rs}\gamma_t^{(rs)}}{\sum_{rs}w^{\gamma}_{rs}}
        	    \label{eqn:g_gamma_estimator}
            \end{equation}
            where the weight $w^{\gamma}_{ls}$ for galaxy-galaxy lensing (but not cosmography; see Sec.~\ref{ssec:theory_cosmography}) is defined as
            \begin{equation}\label{eq:g_gamma_wt}
               w^\gamma_{ls}=\frac{1}{\sigma_\gamma^2+\sigma_{SN}^2}.
            \end{equation}

        \subsubsection{Cross-correlation between galaxy positions and CMB convergence}
        	CMB lensing measurements provide convergence $\kappa$ measurements on the sky, with the sky being divided
			into equal area pixels using healpix\footnote{\url{http://healpix.sf.net/}\newline
              \url{https://github.com/healpy/healpy}} \citep{Gorski2005}.
			Using these measurements, we can obtain the projected surface mass density
			around lens galaxies as
    		\begin{equation}
            	\widehat{\Sigma}(r_p)=\frac{\sum_{lp}w_{lp}\kappa_{p}\Sigma_{c,{*}}}{\sum_{lp}w_{lp}}
								-\frac{\sum_{Rp}w_{Rp}\kappa_{p}\Sigma_{c,{*}}}{\sum_{Rp}w_{Rp}}
        	    \label{eqn:sigma_cmb}
            \end{equation}
            where the summation is carried over all lens galaxy-CMB pixel pairs ($lp$) separated by comoving projected
            distance
            $r_p\in[r_{p,\text{min}},r_{p,\text{max}}]$ at the lens redshift, where
            $r_{p,\text{min}}$ and $r_{p,\text{max}}$
            define the bin edges. $\Sigma_{c,{*}}$ is the geometric factor defined
            in Eq.~\eqref{eqn:sigma_crit} with CMB as the source. The weight factor is defined as
            \begin{equation}
               w_{lp}=\Sigma_{c,*}^{-2}.
            \end{equation}
            We do not include pixel noise in the weights since each pixel has the same statistical noise.
            Finally we subtract out the signal measured around random galaxies ($Rp$ pairs), to remove the spurious
            signal from noise (more discussion in section~\ref{ssec:results_sigma}).

            To measure the convergence signal around galaxies, we use the estimator
    		\begin{equation}
            	\widehat{g\kappa}(r_p)=\frac{\sum_{lp}w^\kappa_{lp}\kappa_{p}}{\sum_{lp}w^\kappa_{lp}}
								-\frac{\sum_{Rp}w^\kappa_{Rp}\kappa_{p}}{\sum_{Rp}w^\kappa_{Rp}}
        	    \label{eqn:g_kappa_estimator}
            \end{equation}
            The rationale for and effect of subtracting the mean convergence around random points is discussed
            in Sec.~\ref{ssec:results_sigma}.
            Under the assumption that each pixel has the same statistical noise, we adopt uniform weights: $w^\kappa_{lp}=1$.

		\subsubsection{Cross-correlation between galaxy shear and CMB convergence}
			For lensing-lensing cross-correlations, we measure the tangential shear around the
            pixels of the CMB map and
			multiply it with the CMB convergence measured in that pixel.
    		\begin{equation}
            	\widehat{w}_{\kappa\gamma_t}(\theta)=\frac{\sum_{sp}^\theta w_{sp}\gamma_t^{(sp)}\kappa_p}{\sum_{sp}^
				\theta
				w_{sp}}
        	    \label{eqn:wgk}
            \end{equation}
            The summation is over all pixel ($p$) and source galaxy ($s$) pairs with separation
            $\theta\in[\theta_\text{min},\theta_\text{max}]$, where $\theta_\text{min}$ and $\theta_\text{max}$ define
            the bin
            edges.
            The weights are inverse
            variance weights for source galaxies, accounting for shape noise and measurement noise.
            \begin{equation}
               w_{sp}=\frac{1}{\sigma_\gamma^2+\sigma_{SN}^2}
            \end{equation}

		\subsubsection{Galaxy Clustering}\label{sssec:estimator_clustering}
			We use the Landy-Szalay \citep{landy1993} estimator to compute the two-point correlation function:
            \begin{equation}
               \widehat{\xi}_{gg}(r_p,\Pi)=\frac{DD-2DR+RR}{RR}
            \end{equation}
			$DD$ is the count of galaxy-galaxy pairs, $DR$ is count of galaxy-random pairs and $RR$ are the random-
			random pairs.
            The projected correlation function is obtained by integrating over the line-of-sight
            separation ($\Pi$) bins
            \begin{equation}
               \widehat{w}_{gg}(r_p)=\sum_{-\Pi_\text{max}}^{\Pi_\text{max}}\Delta\Pi\,\xi_{gg}(r_p,\Pi)
            \end{equation}
            We use $\Pi_\text{max}=100\mpch$, with 20 line-of-sight bins of size
            $\Delta\Pi=10\mpch$. The choice of bin size does not significantly impact our results since the redshift
            extent of our sample is $\gg10\mpch$. The choice of $\Pi_\text{max}$ also does not affect our measurements;
             measurements with $\Pi_\text{max}=50\mpch$ are not signifcantly different from
             $\Pi_\text{max}=100\mpch$. We do use the correct $\Pi_\text{max}$ in our theory
             predictions to account for redshift-space distortion effects.

	\subsection{Removing small-scale information}\label{ssec:ups_estimator}
		 When measuring galaxy-galaxy lensing for which the observable is $\Delta\Sigma$, information from the matter
         distribution on small scales affects the signal
		 measured at large scales as well. If we have a valid model for  correlation functions on
         all scales, this is not a
		 problem. However, perturbation theory-based methods cannot model signals within the virial
         radius of virialized halos, and even at
		 somewhat larger scales ($r_p\lesssim10\mpch$), lowest order perturbation theory is not
         valid.
           To some extent this problem can be alleviated by using non-linear prescriptions that describe quasi-linear
           scales with some success. \cite{Baldauf2010} suggested an approach for
		 removing the small-scale information by defining a new estimator
		 \begin{equation}
		 	\Upsilon_{gm}(r_p,r_0)=\Delta\Sigma(r_p)-\left(\frac{r_0}{r_p}\right)^2\Delta\Sigma(r_0)
		 \end{equation}
		 $\Sigma(r_p)$ and \wgg\ ($\Sigma_{gg}=\wgg$) can also be converted to $\Upsilon$ using the relations between
		 $\Sigma$ and $\Delta\Sigma$ both in the data and theory. \cite{Baldauf2010} showed that
         using $\Upsilon$ not only removes
         small-scale information that is difficult to robustly model, it also reduces the impact
           of cosmic variance and redshift-space distortions in the projected correlation
           functions. The trade-off made when using $\Upsilon$ is that we are removing signal when
           computing $\Upsilon$, so the
         signal-to-noise ratio ($S/N$) in the measurements decreases,
           especially at scales near $r_0$.
  		\referee{ Thus we want to choose lower $r_0$ to use more signal and have higher $S/N$, and higher $r_0$ to be 
		able to remove non-linear galaxy bias more effectively. \cite{Baldauf2010} suggests using $r_0\gtrsim2 r_\text{vir}$, where 
		$r_\text{vir}$ is the virial radius of haloes in the sample. For BOSS LOWZ galaxies, $r_\text{vir}\lesssim1\mpch$ and 
		hence we will use $r_0=2\mpch$ or greater in our analysis.
   		As stated, using $\Upsilon$ helps in reducing the
           impact of cosmic variance, which improves the $S/N$, particularly on large scales and at least partially 
           compensates for the lost $S/N$ at small scales}.

      We will use
		 $\Upsilon$ to derive most of the constraints on galaxy bias $b_g$, galaxy lensing
         amplitude, and the galaxy-matter cross-correlation coefficient $\cgm$. In principle, for
         the combination of CMB lensing and galaxy clustering, using $\Upsilon$ is unnecessary since
         the observables at a given $r_p$ are not contaminated by information from smaller $r_p$.
         However, use of a consistent estimator for all probes is valuable.
   Moreover, while use of
         $\Upsilon$ reduces the $S/N$ to some extent (this is dominated by reduced $S/N$ at small scales, as we will show explicitly in Fig.~\ref{fig:lowz_cmass_SN}), the $S/N$ is moderately improved on
         the large scales that dominate our constraints.

	\subsection{Cosmography}\label{ssec:theory_cosmography}

		\cite{Hu2007} proposed the idea of using the ratio of CMB convergence and galaxy
		lensing convergence as a way to measure the distance ratio (distance to surface of last
        scattering relative to the distance to the source galaxy sample used to estimate the galaxy lensing) and hence
        constrain the geometry, $\Omega_k$ and the equation of
		state of dark energy. The ratio is defined as
		\begin{equation}
			\mathcal R(z_l)=\frac{\kappa(z_l,z_*)}{\kappa(z_l,z_s)}=\frac{\Sigma_c(z_l,z_s)}{\Sigma_c(z_l,z_*)}
		\end{equation}
	\referee{Similar distance ratio tests have also been proposed using galaxy or galaxy cluster lensing alone, in both strong 
		lensing \citep[eg. ][]{Link1998,Golse2002} and weak lensing regimes \citep[eg. ][]{Jain2003,Bernstein2004}. 
		Several studies have already measured the distance ratios 
		\citep[e.g.][and references therein]{Taylor2012,Diego2015,Kitching2015,Caminha2016}, though
        they are afflicted by 
		several 
		systematics such uncertainties in modeling cluster profiles and cosmic variance in case of multiple strong lens 
		systems, 
		and photometric redshift uncertainties as well as imaging systematics that cause a
        redshift-dependent shear calibration 
		in the case of weak lensing. The small redshift baseline also limits the cosmological applications of these 
		measurements using optical weak lensing alone \citep[see discussion in ][]{Hu2007,Weinberg2013}. Using CMB lensing in cosmographic measurements 
		is advantageous in several ways. First, the source redshift for the CMB (redshift of surface of last scattering) is 
		well known, so one of the two redshift slices being compared has no redshift
        uncertainty. The long redshift baseline between CMB and galaxy lensing sources also improves the 
		sensitivity of $\mathcal R$ to cosmological parameters \citep{Hu2007}. However, using CMB lensing 
		with galaxy lensing makes $\mathcal R$ become more sensitive to some of the systematics in galaxy lensing 
		(for example, multiplicative bias) and $\mathcal R$ can also be used as test for presence of
        such systematics. 
		}
		
		To measure $\mathcal R$, we work with galaxy lensing shear measured using estimator defined in
		eq.~\eqref{eqn:g_gamma_estimator}, not convergence. Instead we
		convert the convergence measurement from CMB (measured using estimator defined in
		eq.~\eqref{eqn:g_kappa_estimator}) to
		the shear.
      Motivated by the estimator in
		Sec.~\ref{ssec:ups_estimator}, we define the estimator $\upsilon_{t}$ as
		\begin{align}
			\upsilon_{t}(r_p,r_0)=\gamma_t(r_p)-\left(\frac{r_0}{r_p}\right)^2\gamma_t(r_0)
		\end{align}
		Just as $\gamma_t=\Delta\Sigma/\Sigma_c$, we can write $\upsilon_{t}=\Upsilon_{gm}/\Sigma_c$. In the limit
		that $r_0=0$, $\upsilon_t$ is simply $\gamma_t$. The CMB lensing convergence $\kappa$
        averaged around lens galaxy positions can be converted to $\upsilon_{t}$
		using $\gamma_t(r_p)=\bar\kappa(<r_p)-\kappa(r_p)$ and then converting $\gamma_t$ to
		$\upsilon_t$.

		One of the primary motivations for defining $\Upsilon_{gm}$ was to remove information for small scales which
		are more difficult to model. When measuring $\mathcal R$, we do not need to model those
        small scales (any nonlinear bias, etc.\ will cancel in the ratio) and
		hence using $\upsilon_t$ is not strictly necessary. However, when computing the convergence
        signal from the CMB, the smallest scales are smoothed (see Sec.~\ref{ssec:data_planck}) and it is desirable to
        completely remove
		information from those scales. Thus we will use $\upsilon_t$ to compute $\mathcal R$ and our final
		definition of the estimator $\widehat{\mathcal R}$ is
		\begin{equation}\label{eq:R}
			\widehat{\mathcal R}(z_l)=\frac{\upsilon_t(z_l,z_*)}{\upsilon_t(z_l,z_s)}=\frac{\Sigma_c(z_l,z_s)}
			{\Sigma_c(z_l,z_*)}
		\end{equation}
		Note that our estimator is different from the one used by \cite{Miyatake2016}, who use $\gamma_t$ to compute
		$\mathcal R$ and
		\referee{exclude} the scales which are affected by smoothing. We will show our measurement using small value of 
		$r_0$
		($\ll$ smoothing scale), in which case our estimator is equivalent to one using $\gamma_t$ and 
		\referee{excluding} 
		scales smaller than smoothing scale.
		Also in the estimator of \cite{Miyatake2016}, the galaxy-galaxy
		lensing measurement is in the numerator, so their estimator is effectively $1/\mathcal
        R$. We keep the CMB lensing measurement in the numerator since
      it is noisier. Later in this section we describe how we account for the bias that comes from
      taking the expectation value of the  ratio of noisy quantities.

		To model the measurement, we begin by computing the
		galaxy position-convergence and galaxy position-shear cross-correlations
		\begin{align}
			\langle g\kappa\rangle(r_p)&=\int \mathrm{d}z_l\, p(z_l) 
			\frac{\Sigma(z_l)(r_p)}
			{\Sigma_c(z_l,z_*)}\label{eq:g_kappa}\\
			\langle g\gamma_t\rangle(r_p)&=\int \mathrm{d}z_l \, p(z_l)\int_{z_l}^\infty
			\mathrm{d}z_s p(z_s|z_\text{ph}>z_l)\nonumber\\ & 
			\frac{\Delta\Sigma(z_l)(r_p)}
			{\Sigma_c(z_l,z_s)}\frac{1}{\sigma_\gamma^2+\sigma_{SN}^2}\label{eq:g_shear}
		\end{align}
		where source weights $\frac{1}{\sigma_\gamma^2+\sigma_{SN}^2}$ are defined in Sec.~\ref{ssec:estimator_galaxy_galaxy_lensing}
		The theory computations can be converted to $\upsilon_t$ using similar method as in the data. Note that due to
		variations in the number density of source galaxies, galaxy-CMB and galaxy-source cross correlations
		measurement are at different effective lens redshifts. In principle this can be modeled, but a desirable
		feature in $\mathcal R$ is to define it as a simple ratio without requiring complicated modeling for small
		scale signals. To get both signals at the same effective redshift, we explicitly compute the
        weights from galaxy-galaxy lensing as a function of lens redshift, and use them as weights
        when computing the galaxy-CMB lensing cross-correlations. These weights decrease strongly with redshift due to
        the decrease in the number of source galaxies behind a lens for increasing lens redshift. Since CMB lensing kernel increases with redshift, these
        weights are suboptimal for galaxy-CMB lensing cross correlations and increase the noise in $g\kappa$
        measurements. Since the $g\kappa$ measurement is noisier and dominates the noise in $\mathcal R$, it is desirable
        to modify the weights to give higher weight to higher redshift galaxies. Thus we add an additional factor of
        $\Sigma_c(z_l,z_*)^{-2}$ to weights. This increases the noise in the $g\gamma_t$ measurement from galaxy-galaxy
        lensing, but this increase is more than balanced by the reduced noise in $g\kappa$ from galaxy-CMB lensing cross correlations.
         The final lens weights are defined as
		\begin{equation}
			\mathcal W_{\mathcal R}(z_l)=\int_{z_l}^\infty \mathrm{d}z_s p(z_s|z_\text{ph}>z_l)
			\frac{1}{\sigma_\gamma^2+\sigma_{SN}^2} \Sigma_c(z_l,z_*)^{-2}
		\end{equation}
		Finally $\mathcal R$ is given as
		\begin{equation}\label{eq:R_theory}
			\mathcal R=\frac{\int \mathrm{d}z_l D(z_l)^2 \mathcal W_{\mathcal R}(z_l)\Sigma_c^{-1}(z_l,z_*) }
			{\int \mathrm{d}z_l D(z_l)^2 \Sigma_c(z_l,z_*)^{-2}
			\int_{z_l}^\infty \mathrm{d}z_s \frac{p(z_s|z_\text{ph}>z_l)}{\sigma_\gamma^2+\sigma_{SN}^2}
			\Sigma_c^{-1}(z_l,z_s)}
		\end{equation}
		$D(z_l)^2$ is the matter growth function and enters because $\Sigma$ scales as $D(z_l)^2$ in linear theory.
		In principle, due to non-linear effects in $\Sigma$, the weights in $\mathcal R$ can vary with the scale. Since
		non-linear effects also evolve with redshift, the measurement of $\mathcal R$ at different scales can be at
		somewhat
		different effective lens redshifts, $z_l$, and hence $\mathcal R$ can in principle be scale-dependent. However,
		this effect is likely to be small given the narrow redshift range of LOWZ sample, and should be subdominant to 
		the noise in our measurements. \referee{When comparing with data, we show this effect by replacing $D(z_l)$ 
		with scale dependent growth function ($D(z_l,r_p)$) estimated from correlation function using linear theory with 
		halofit nonlinear evolution. We do note that halofit does not capture the full non-linear evolution at small scales and thus our 
		estimate will only be approximate.}

      To estimate the bias that comes from  taking the ratio of noisy quantities
      before taking the expectation value, we make mock realizations of the numerator and
      denominator in Eq.~\eqref{eq:R_theory}, using the signal-to-noise ratio from CMB measurements for numerator and galaxy lensing measurements for the denominator.
      \begin{equation}
         {\delta X}(r_p)=\left|X\frac{\delta\upsilon_t(r_p)}{\upsilon_t(r_p)}\right|
      \end{equation}
      where $X$ is either the numerator or denominator and $\upsilon_t$ is measured from CMB (galaxy) lensing for numerator (denominator). Using $\delta X(r_p)$, we generate random realizations,
      $\tilde{X}(r_p)$ assuming gaussian distribution with mean $X$ and standard deviation $\delta
      X$. We then recompute the $\tilde R$ taking the ratio of    $\tilde X$ and then compute the
      mean, $\langle\tilde R\rangle$, using the same scales as $\widehat{R}$. We include the bias
      estimated from this exercise
      in the theory prediction before comparing with the data.

\section{Data}\label{sec:data}
	\subsection{SDSS}
		 The SDSS \citep{2000AJ....120.1579Y} imaged roughly $\pi$ steradians
		of the sky, and the SDSS-I and II surveys followed up approximately one million of the detected
		objects spectroscopically \citep{2001AJ....122.2267E,
		  2002AJ....123.2945R,2002AJ....124.1810S}. The imaging was carried
		out by drift-scanning the sky in photometric conditions
		\citep{2001AJ....122.2129H, 2004AN....325..583I}, in five bands
		($ugriz$) \citep{1996AJ....111.1748F, 2002AJ....123.2121S} using a
		specially-designed wide-field camera
		\citep{1998AJ....116.3040G} on the SDSS Telescope \citep{Gunn2006}. These imaging
		data were used to create
		the  catalogues of shear estimates that we use in this paper.  All of
		the data were processed by completely automated pipelines that detect
		and measure photometric properties of objects, and astrometrically
		calibrate the data \citep{Lupton2001,
		  2003AJ....125.1559P,2006AN....327..821T}. The SDSS-I/II imaging
		surveys were completed with a seventh data release
		\citep{2009ApJS..182..543A}, though this work will rely as well on an
		improved data reduction pipeline that was part of the eighth data
		release, from SDSS-III \citep{2011ApJS..193...29A}; and an improved
		photometric calibration \citep[`ubercalibration',][]{2008ApJ...674.1217P}.

		\subsection{SDSS-III BOSS}\label{ssec:data_boss}
			Based on the photometric catalog,  galaxies were selected for spectroscopic
			observation
			\citep{Dawson:2013}, and the BOSS spectroscopic survey was performed
			\citep{Ahn:2012} using the BOSS spectrographs \citep{Smee:2013}. Targets
			are assigned to tiles of diameter $3^\circ$ using an adaptive tiling
			algorithm \citep{Blanton:2003}, and the data were processed by an
			automated spectral classification, redshift determination, and parameter
			measurement pipeline \citep{Bolton:2012}. In this paper we use the BOSS data release 12 galaxies \citep{Alam2015}.

\begin{figure}
             \centering
             \includegraphics[width=1\columnwidth]{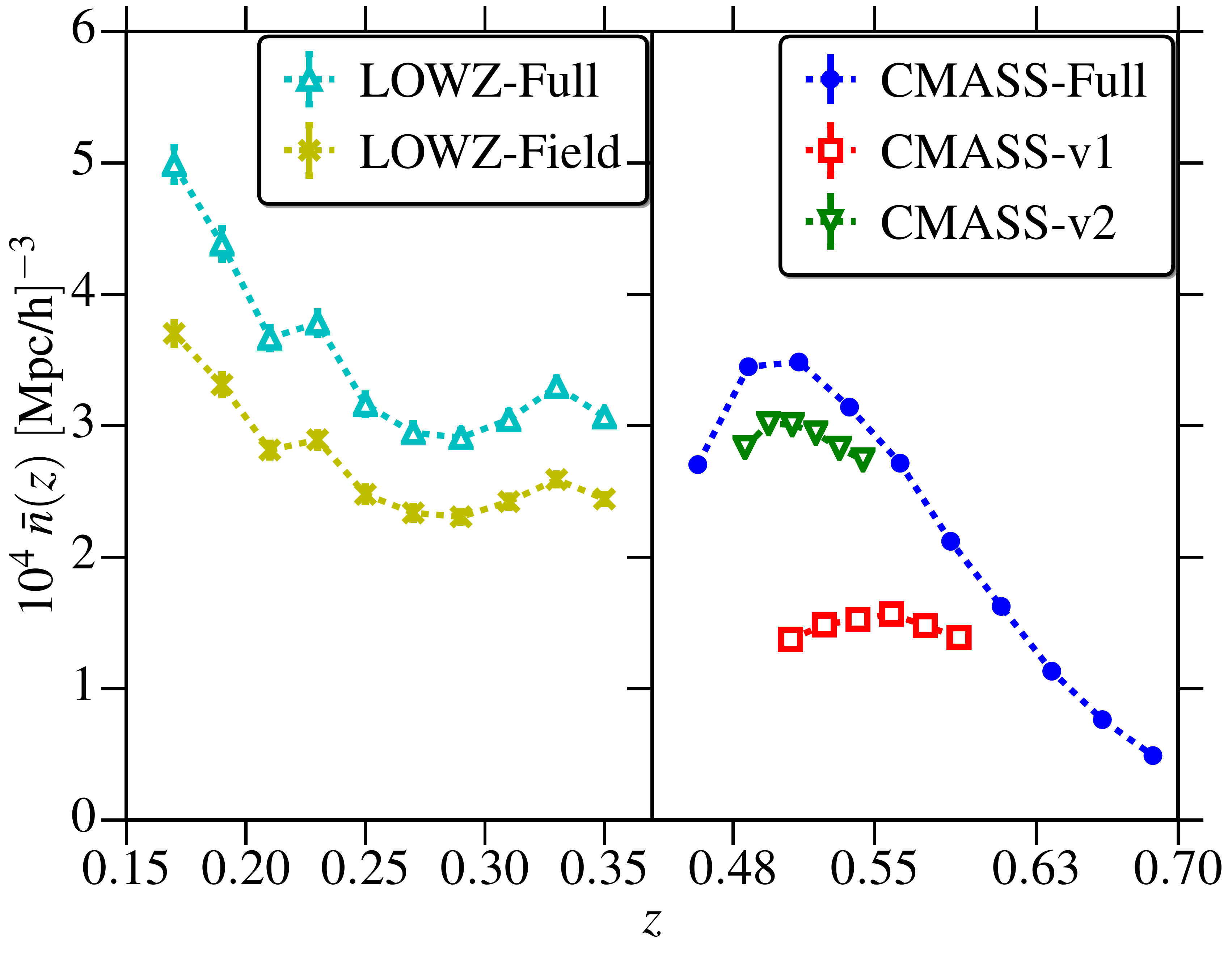}
             \caption{Number density as a function of redshift for different samples. LOWZ and field sample are defined in
             $0.16<z<0.36$, while CMASS, CMASS-v1 and CMASS-v2 are defined in $0.45<z<0.7$. No weights were used in
             computing the number densities presented here.
    			}
             \label{fig:nbar_z}
          \end{figure}

         The number densities of the BOSS samples and of various subsamples used in this work are
            shown in Fig.~\ref{fig:nbar_z}.

		\subsubsection{LOWZ}
			The LOWZ sample consists of Luminous Red Galaxies (LRGs) at $z<0.4$, selected
			from the SDSS DR8 imaging data and observed
			spectroscopically in the BOSS survey \citep{Reid2016}. The sample is approximately volume-limited in the
			redshift range $0.16<z<0.36$, with a number
			density of $\bar{n}\sim 3\times10^{-4}~h^3\text{Mpc}^{-3}$ \citep{Manera2015,Reid2016}. BOSS DR12 has
			249~938
			LOWZ galaxies within the  redshift
			range used in this work, $0.16<z<0.36$. After combining with the Planck lensing map mask and the SDSS shape
			catalog
			mask, which masks out certain regions that have higher Galactic
			extinction or poor imaging quality \citep{Reyes2012}, we are left with
			225~181 LOWZ galaxies, with redshift-dependent number density shown in Fig.~\ref{fig:nbar_z}.

			We also define a sample of field galaxies using the Counts in Cylinders methods \citep{Reid2009},
			which was used in \cite{Singh2015} to define a sample of groups from LOWZ galaxies. Here we apply the
			same technique to DR12, and select field galaxies by requiring that they are in groups
            of one galaxy and that their fiber collision and redshift failure weights are equal to 1
            (see Sec.~\ref{ssec:data_boss_weights}). These cuts ensure that
			there is no neighboring target LOWZ galaxy within $r_p<0.8\mpch$ and $|\Pi|<20\mpch$ for the field galaxies.
			Field galaxies by construction have no redshift failure weights (see Sec.~\ref{ssec:data_boss_weights}) and thus will provide a test on any possible systematics from these weights. They also tend to reside in lower mass halos, and are less likely to be satellites
            (which have lensing profiles that are more complex to interpret on small scales due to
            contamination from the host halo lensing profile).

		\subsubsection{CMASS}
			The BOSS CMASS sample consists of higher redshift galaxies ($0.4<z<0.7$) targeted using color and magnitude
			cuts intended to
			select a uniform sample of massive galaxies \citep{Reid2016}. The DR12 CMASS sample has 849~637 galaxies, of
			which we use
			682~298 after applying the masks and a redshift cut $z\in[0.45,0.7]$. The number
			density of the CMASS
			sample varies across the redshift range (see Fig.~\ref{fig:nbar_z}), which can bias the inferences of its properties
            from stacked galaxy position-lensing cross correlations. To overcome this problem, we also define two
            volume-limited samples: CMASS-v1, with $
			z
			\in[0.5,0.6]$ and $M_r\in[-23,-22]$, and CMASS-v2, with $z\in[0.48,0.55]$ and $
            M_r\in[-21.5,-22.8]$, where $M_r$ is the absolute magnitude $k+e$ corrected to $z=0$ using method described
            in
            \cite{Wake2006}. 	CMASS-v1
			has 188~586 galaxies with $\bar{n}=(1.55\pm0.10)\times10^{-4}~h^3\text{Mpc}^{-3}$, where $
			\pm0.10$ denotes the maximum variation across the redshift range. CMASS-v2 has 236~676
			galaxies with $\bar{n}=(3\pm0.2)\times10^{-4}h^3\text{Mpc}^{-3}$. Since both volume-limited
			samples
			have a narrow redshift range, we use $\Pi_\text{max}=60\mpch$ when calculating the
			clustering signal for these samples.

		\subsubsection{Weights}\label{ssec:data_boss_weights}
			In their large scale structure (LSS) samples the BOSS collaboration provides
			several weights for each
            galaxy to correct for systematics when estimating the
			galaxy clustering\citep{Reid2016}. The most important of these are the fiber collision and
            incompleteness (redshift failure)
			weights. Due to the finite size of the spectroscopic fibers, it is impossible to simultaneously take
			spectra of BOSS galaxies that are separated by less than $62\arcsec$. Many of these cases
			are resolved by revisiting the field multiple times. However, some target galaxies
			lack spectroscopic redshifts either due to fiber collisions and redshift failures. This
			introduces a bias in the clustering measurements, since the fiber-collided galaxies are
			preferentially located in overdense regions. An approach that has been shown to work on
            large scales ($\theta\gtrsim2'$) \citep{Reid2014} is to
			upweight the nearest neighbor of the fiber-collided galaxies, based on the assumption
            that they are likely to be in the same group due to their proximity on the sky.

			In addition, we also use systematics weights, which correct for the effects of varying
            target density as a function of stellar density for the CMASS sample \citep{Ross2012}
            The final weights used for CMASS are
			\begin{equation}
				w=w_\text{sys}(w_\text{no-z}+w_\text{cp}-1)
			\end{equation}
			where $w_{cp}$ corrects for fiber collisions,
            and $w_\text{sys}=1$ for LOWZ. While these weights have been shown to correct
			for biases in clustering on large scales, it is not clear whether this approach works well
			for lensing calculations \citep[see][]{More2015}. $w_\text{sys}$ is not expected to
            change the lensing measurements done by stacking procedure, as long as these weights do not alter the overall properties of the sample.
           These systematic weights do depend on the apparent surface brightness of the galaxies, as
           those with low surface brightness are more likely to be missed in regions of high stellar
           density \citep{Ross2012}. However, the dependence on surface brightness is sufficiently
           mild that their inclusion does not significantly alter the properties of the sample. We
           checked that using these weights changes the absolute magnitude and redshift distribution
           of the sample by $\lesssim0.1\%$ and thus we do not expect any significant changes in the
           lensing measurement from $w_\text{sys}$.
           Still, we do use these weights for all samples and subsamples of CMASS.
            Redshift
            failure weights do change both the lensing and clustering measurements by up-weighting
            the higher density regions. However, they do not mitigate the bias below the fiber
            collision scale, and even at slightly larger scales, the measurements will be biased
            since we are stacking on the wrong galaxy. Still, at scales $r_p\gtrsim2\mpch$, these
            weights should not lead to any significant change other than changing the effective galaxy bias $b_g$.
         In Appendix~\ref{app:sysweights}, we directly show the effect of using these weights using the CMASS sample.
         For field galaxies, these weights are all equal to one and for volume
			limited sub-samples of CMASS, it is not guaranteed whether the galaxies missed from redshift failures will pass the magnitude cuts. Hence we omit the results using redshift failure weights for these sub-samples.


		\subsection{SDSS shear catalog}\label{ssec:data_source_sample}

		For galaxy-galaxy lensing (shear measurements), we use the SDSS re-Gaussianization shape  catalog
		that was introduced in \cite{Reyes2012}. Briefly, these shapes are measured
		using the re-Gaussianization algorithm \citep{Hirata2003}. The
		algorithm is a modified version of early ones that used ``adaptive moments'' (equivalent to fitting
        the light intensity profile to an elliptical Gaussian), determining shapes of the
        PSF-convolved galaxy image based on adaptive moments and then correcting the resulting
        shapes based on adaptive moments of the PSF.   The re-Gaussianization method involves
        additional steps to correct for  non-Gaussianity of both the PSF and the galaxy surface
        brightness profiles \citep{Hirata2003}. The components of the PSF-corrected distortion are defined as
		\begin{equation}\label{eqn:distortion}
			(e_+,e_\times)=\frac{1-(b/a)^2}{1+(b/a)^2}(\cos 2\phi,\sin 2\phi),
		\end{equation}
		where $b/a$ is the galaxy minor-to-major axis ratio and $\phi$ is the position angle of the major
		axis on the sky with respect to the RA-Dec coordinate system. The ensemble average of the
        distortion is related to the shear as
		\begin{align}
			\widehat{\gamma}_+,\widehat{\gamma}_\times&=\frac{\langle e_+,e_\times\rangle}{2
			R}\label{eqn:regauss_shear}\\
			 \widehat{R}&=1-\frac{1}{2}\langle e_{+,i}^2+e_{\times,i}^2-2\sigma_i^2\rangle\label{eq:Responsivity}
		\end{align}
		where $\sigma_i$ is the per-component measurement uncertainty of the galaxy distortion, and
		$\widehat{R}\approx 0.87$ is the shear responsivity representing the response of an ensemble of
        galaxies with some intrinsic distribution of distortion values
		to a small shear \citep{Bernstein2002}.

		For this sample, we use photometric redshifts derived from the template fitting code ZEBRA
		\citep{Feldman2006}, as described and characterized in \cite{Nakajima2012}. Using photometric redshifts can also
		introduce bias in
		galaxy-galaxy lensing measurements. \cite{Nakajima2012} showed that this bias can be large, but can be
		determined to $2$ percent accuracy using {\em representative} spectroscopic calibration samples. Using the
		calibration method described in
		\cite{Nakajima2012}, we estimate the calibration bias for the galaxy-galaxy lensing by the LOWZ sample to be
		$\sim-10\%$, and thus multiply our
		lensing signal by a factor of 1.1 before plotting it or fitting models to it.

		When estimating the galaxy lensing-CMB lensing cross-correlations, we  do not
		need redshifts for individual source galaxies. To derive the theoretical predictions for
        this quantity, we directly use
        the $\mathrm{d}n/\mathrm{d}z$ obtained
        from the representative spectroscopic redshift dataset from \cite{Nakajima2012} after
        applying the same cuts that
		were applied to data during measurements.

	\subsection{Planck Lensing Maps}\label{ssec:data_planck}
		We use the Planck 2015 lensing map provided by the Planck collaboration
		\citep{Planck2015lensing}. We convert
		the provided $\kappa_{l,m}$ values to a convergence map using {\sc healpy} \citep{Gorski2005},
        with
		\nside$=1024$ (pixel size of $3.43\arcmin$), where \nside\ determines the resolution of the {\sc healpy} map
		(higher
		\nside\ means smaller pixels; $n_\text{pix}=12n^2_\text{side}$ over the full sky).
          When constructing the convergence map, we use
		$\kappa_{l,m}$
		in the  range $8<\ell<2048$, which corresponds to modes from $\sim25^\circ$ to $\sim6\arcmin$.
		\cite{Planck2015lensing} found some evidence of systematics in the high $\ell$ range, and used $40<\ell<400$
		(`conservative') for their main cosmological constraints, though using the `aggressive'
        range that is adopted here gives a very similar
		amplitude of the lensing power spectrum and constraints for cosmological parameters except for
		$\sigma_8\Omega_m^{0.25}$, which shifts by $\sim1\sigma$ between the two $\ell$
        ranges.

		While our primary results use \nside$=1024$, we test the effects of changing
		the pixel size (using \nside$=512$ and 2048, with pixel sizes of $6.9$ and $1.71\arcmin$
		respectively) and applying smoothing on the convergence maps (Gaussian beams with $\sigma=1$
        and $10\arcmin$).
		Since the pixel size with \nside$=512$ is somewhat greater than the smoothing scale in the lensing map, we
		do expect to gain some information by using  smaller pixels with \nside$=1024$. Going to
        even higher resolution with \nside$=2048$
		should not make a very significant difference except at very small scales, in case there is some
		information left in the lensing maps at those scales. Similarly, smoothing with a Gaussian
        beam with $\sigma=1'$
		should not significantly affect our results given the resolution of the Planck maps and the scales used for
        our measurements, though $\sigma=10'$ should change the signal
		on scales up to the FWHM ($\approx25'$) of the smoothing kernel. Hence we will omit the
        measurements with $\sigma=1\arcmin$ and will show results with $\sigma=10\arcmin$
        smoothing for comparison with the main results, which have no additional smoothing applied to maps.

		When calculating the cross-correlations, we apply the common galaxy and Planck mask on both
        the galaxy shear and Planck convergence maps. This     reduces the area within the BOSS mask
        by $\sim3\%$, primarily driven by the Planck point source mask which selectively masks the
        very massive clusters. This can change the effective linear bias of the galaxy samples and hence we use the same mask when computing galaxy clustering as well as galaxy-galaxy lensing.

		To perform null tests, we generate a map by shuffling the pixels in the convergence map
        (with the Planck
		mask applied) and then applying
		the galaxy mask. We also generate a realization of a noise map using the noise power spectrum provided
		with the lensing
		maps. Throughout this work, $\kappa_{sh}$ and $\kappa_N$ will be used to represent the
        shuffled map and noise map convergences, respectively.

\section{Results}\label{sec:results}
	\subsection{Lensing of the CMB by galaxies}\label{ssec:results_sigma}

		In this section we present the results from cross-correlating the Planck convergence maps
        with the lens galaxy samples described in Sec.~\ref{ssec:data_boss}.

		\begin{figure*}
		     \begin{subfigure}{\columnwidth}
		         \centering
		         \includegraphics[width=0.9\columnwidth]{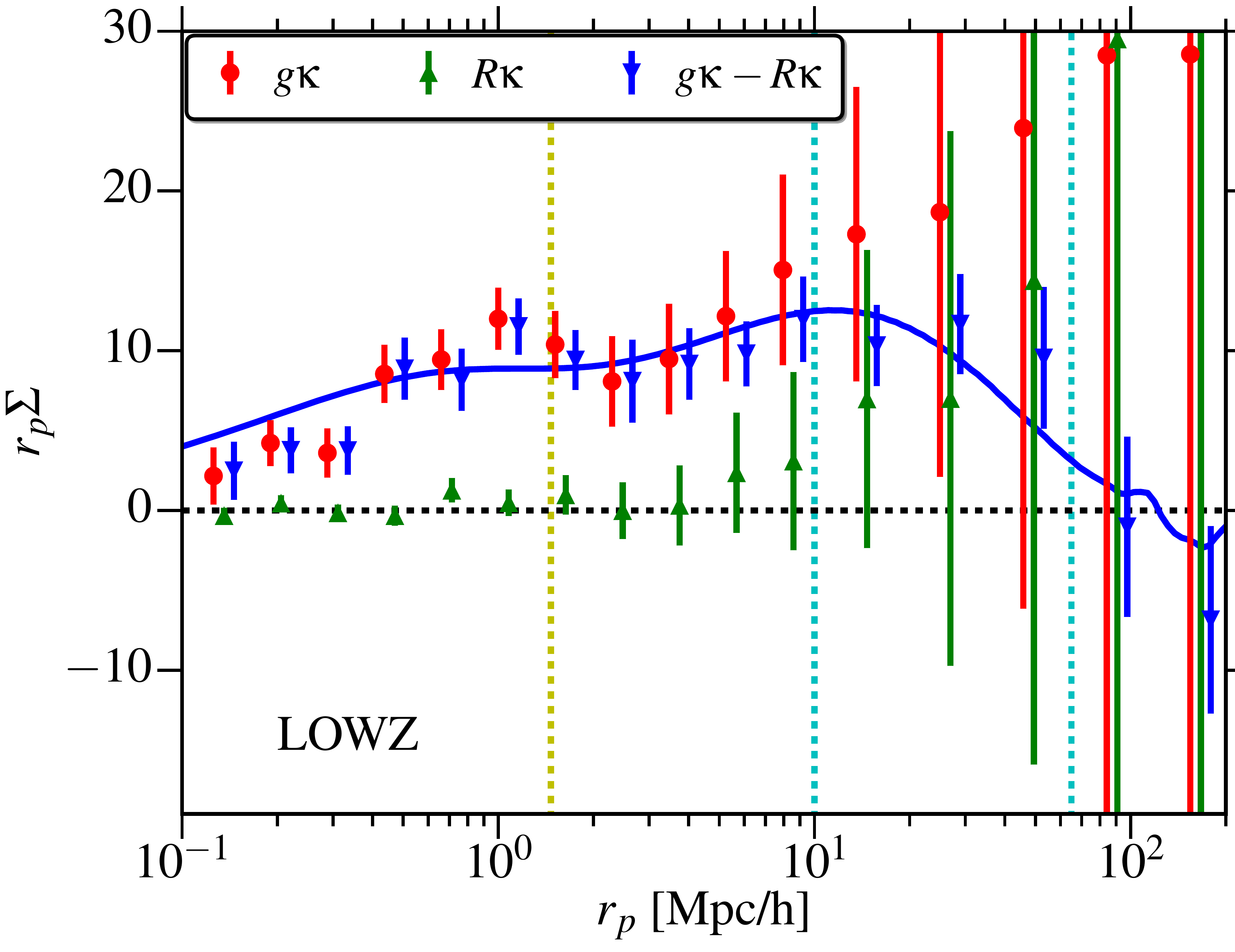}
		         \caption{}
		      \end{subfigure}
		     \begin{subfigure}{\columnwidth}
		         \centering
		         \includegraphics[width=.9\columnwidth]{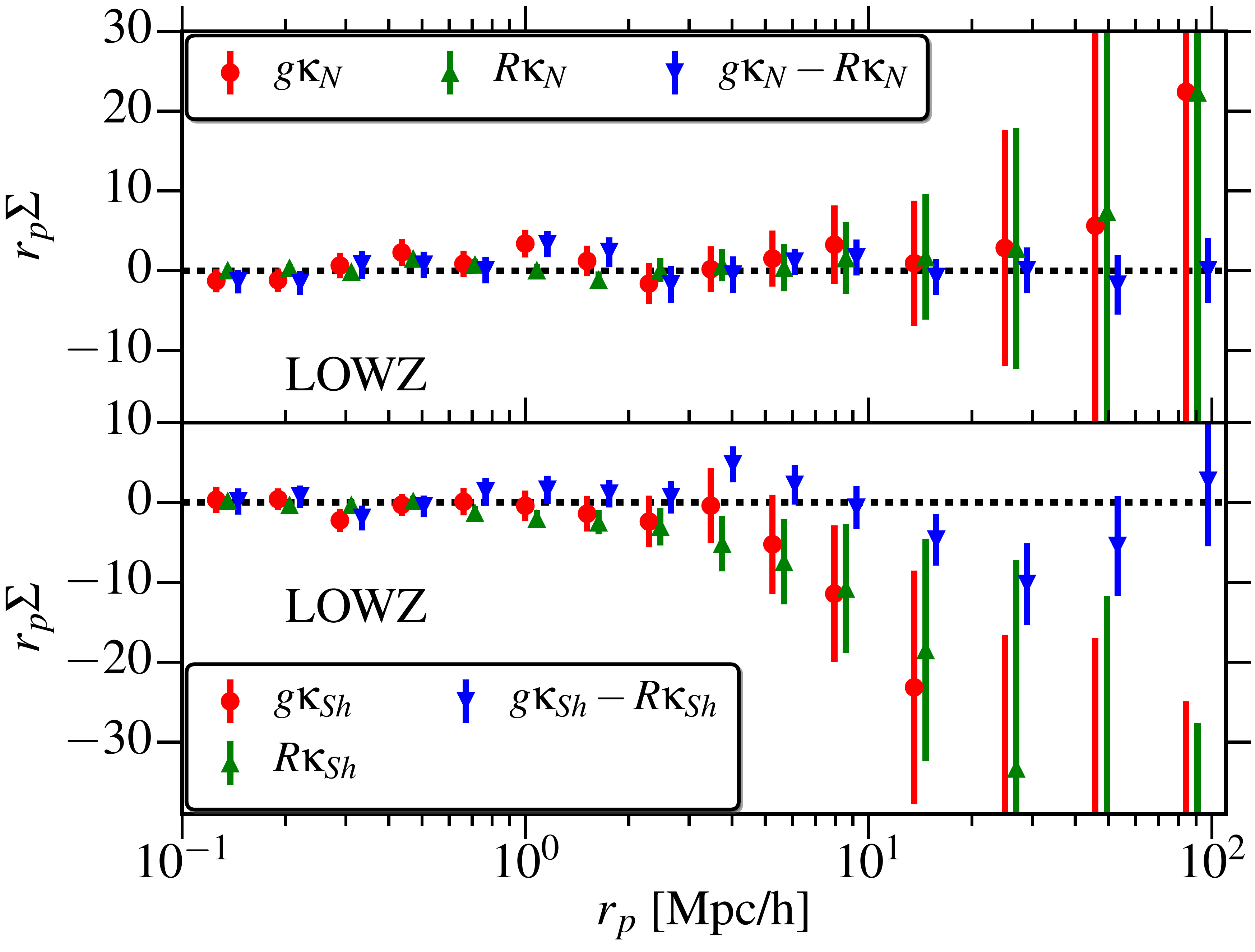}
		         \caption{}

		      \end{subfigure}
		     \begin{subfigure}{\columnwidth}
		         \centering
		         \includegraphics[width=0.9\columnwidth]{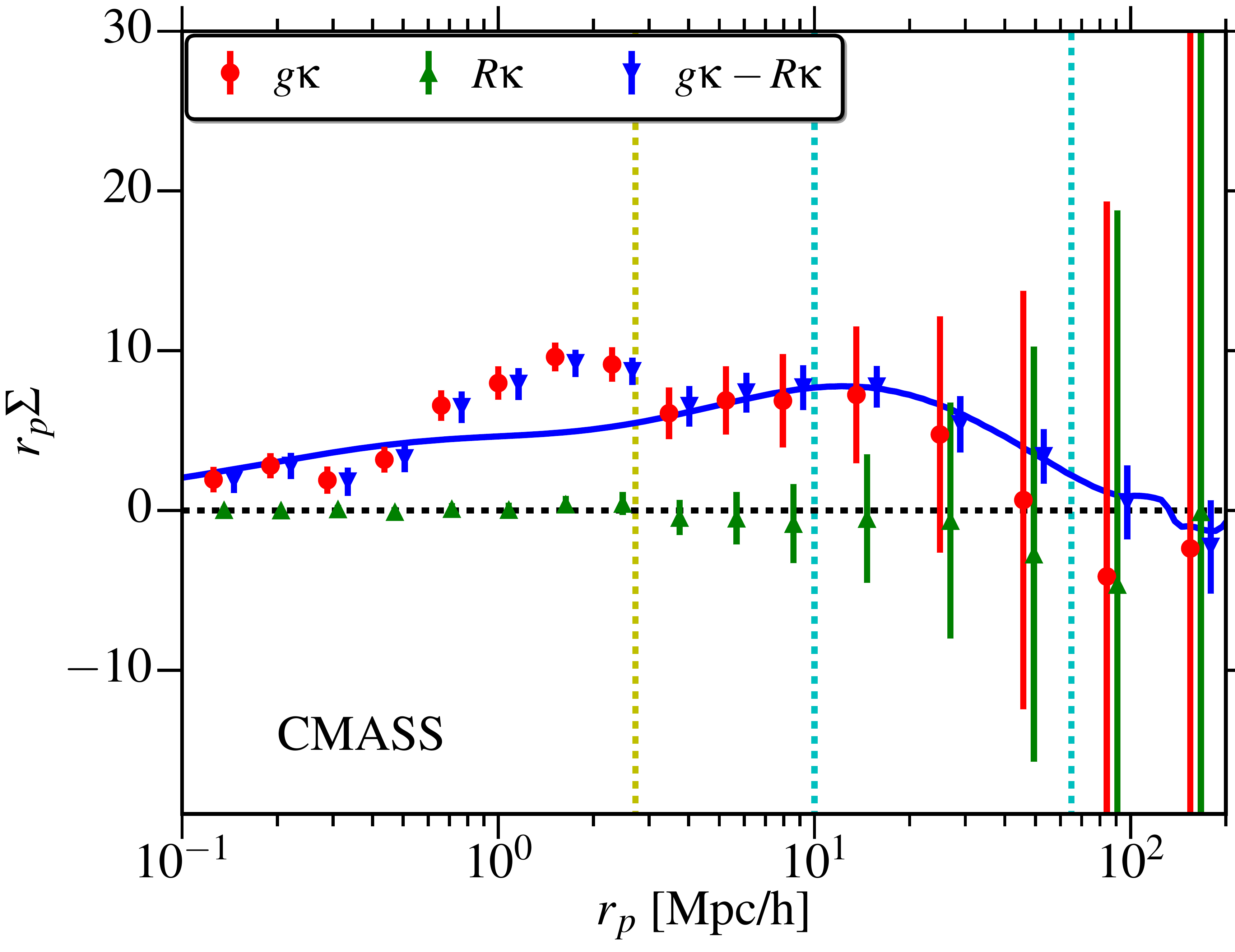}
		         \caption{}
		      \end{subfigure}
		     \begin{subfigure}{\columnwidth}
		         \centering
		         \includegraphics[width=.9\columnwidth]{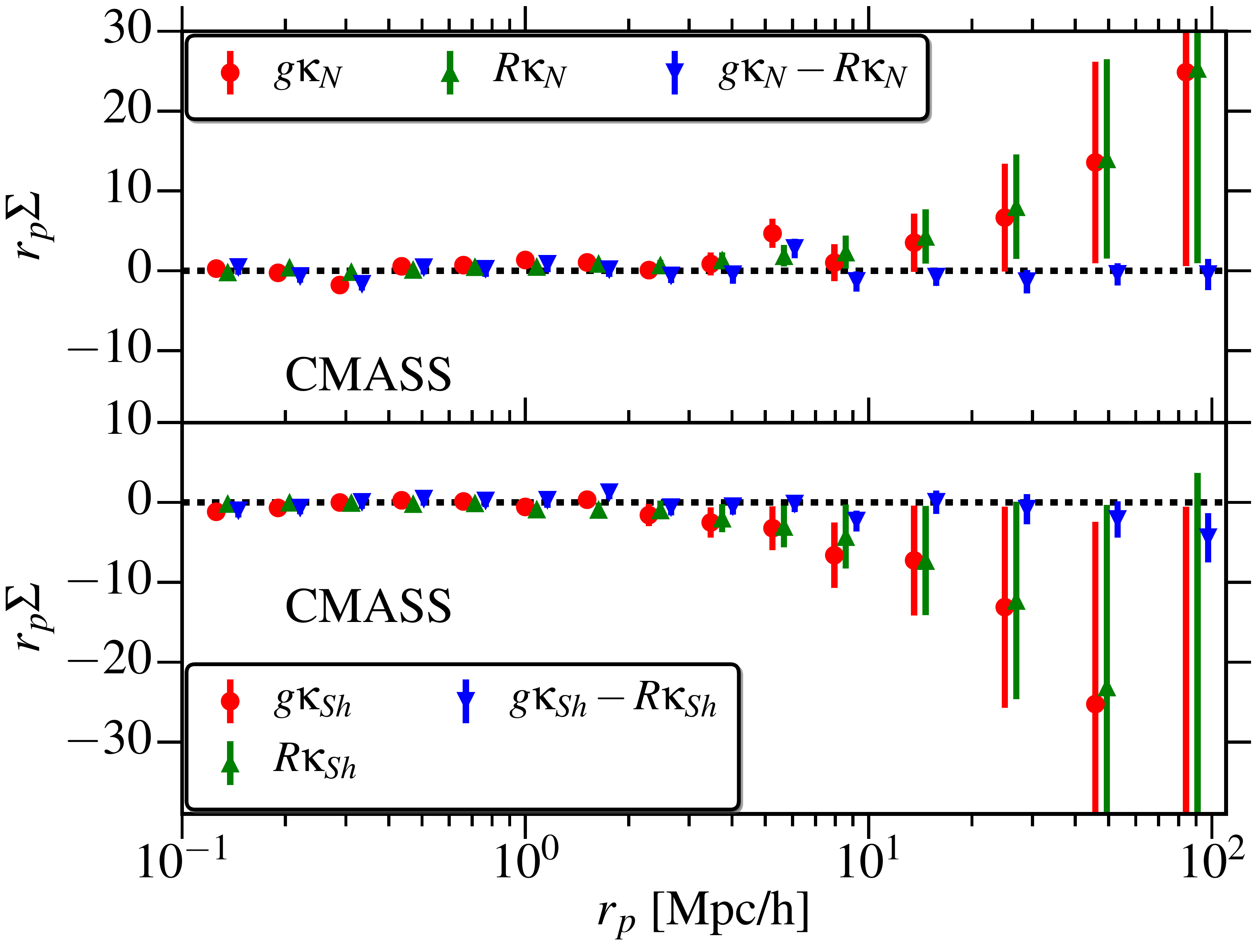}
		         \caption{}
		      \end{subfigure}
		      \caption{$r_p\Sigma$ (in units of $10^{6} M_\odot/\text{pc}$)
                    measurements for the LOWZ (top row) and CMASS (bottom row) samples,  using maps
                    with \nside$=1024$. The left column
		      		shows the measurements around galaxies (red points), random points (green points) and the
					final signal after taking the difference (blue), along with the Planck
                    $\Lambda$CDM model with best-fitting bias. Vertical yellow lines mark the scales corresponding to
                    $6\arcmin$ at the maximum redshift of the sample while vertical cyan lines mark the fitting range
                    for the model.      At
					$r_p\gtrsim30\mpch$ ($\gtrsim2^\circ$), the noise from the convergence map
					starts dominating the signal around galaxies, but is removed by subtracting out the
					signal around random points.  The right column
					shows the systematics tests, where the Planck convergence is replaced by the
                    convergence from the noise map
					$\kappa_N$ and the shuffled map $\kappa_{sh}$. The errors in $\kappa_n$ measurements are very
					similar
					to the errors in $\kappa$ measurements, though the errors in $\kappa_{sh}$ show different behavior
					since the
					noise correlations are broken by shuffling the measurements.  The points at
                    large radius are moderately correlated; see Fig.~\ref{fig:lowz_corr}.
					}
				\label{fig:gk_all}
		\end{figure*}

\begin{figure}
         \centering
         \includegraphics[width=1\columnwidth]{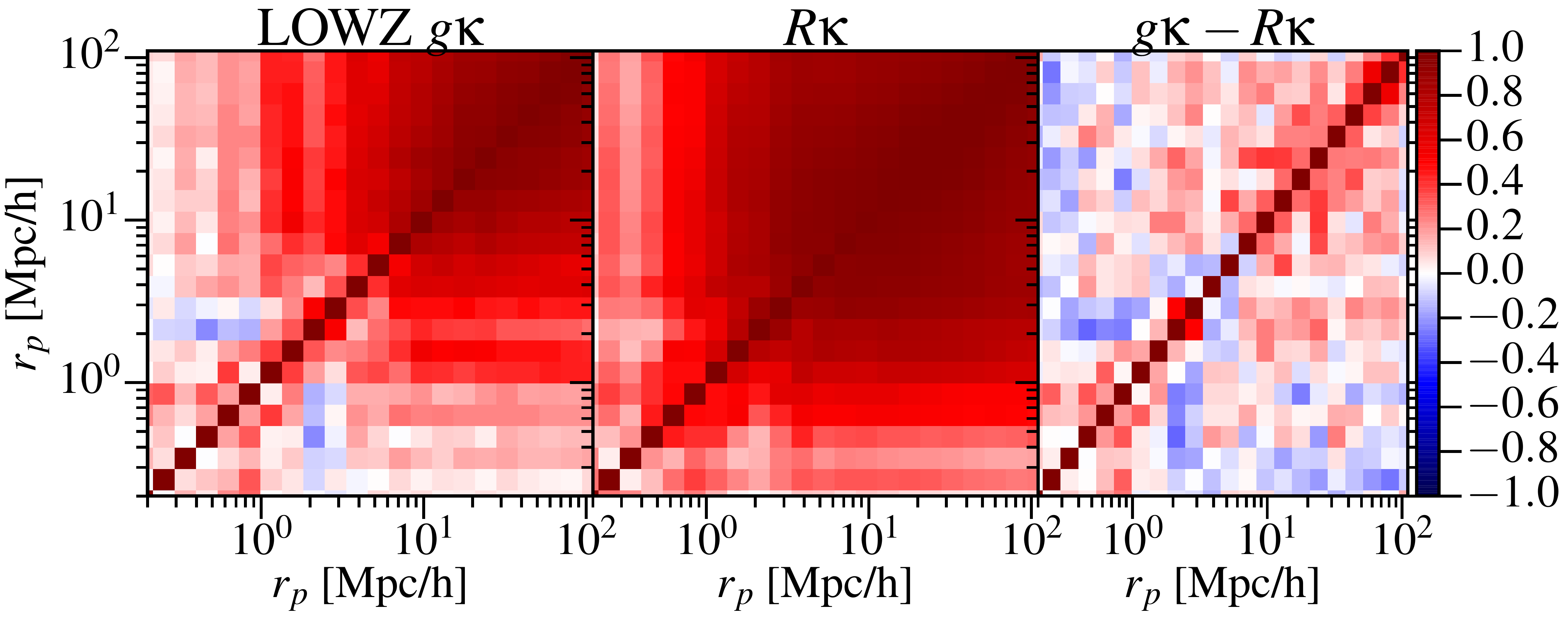}
          \includegraphics[width=1\columnwidth]{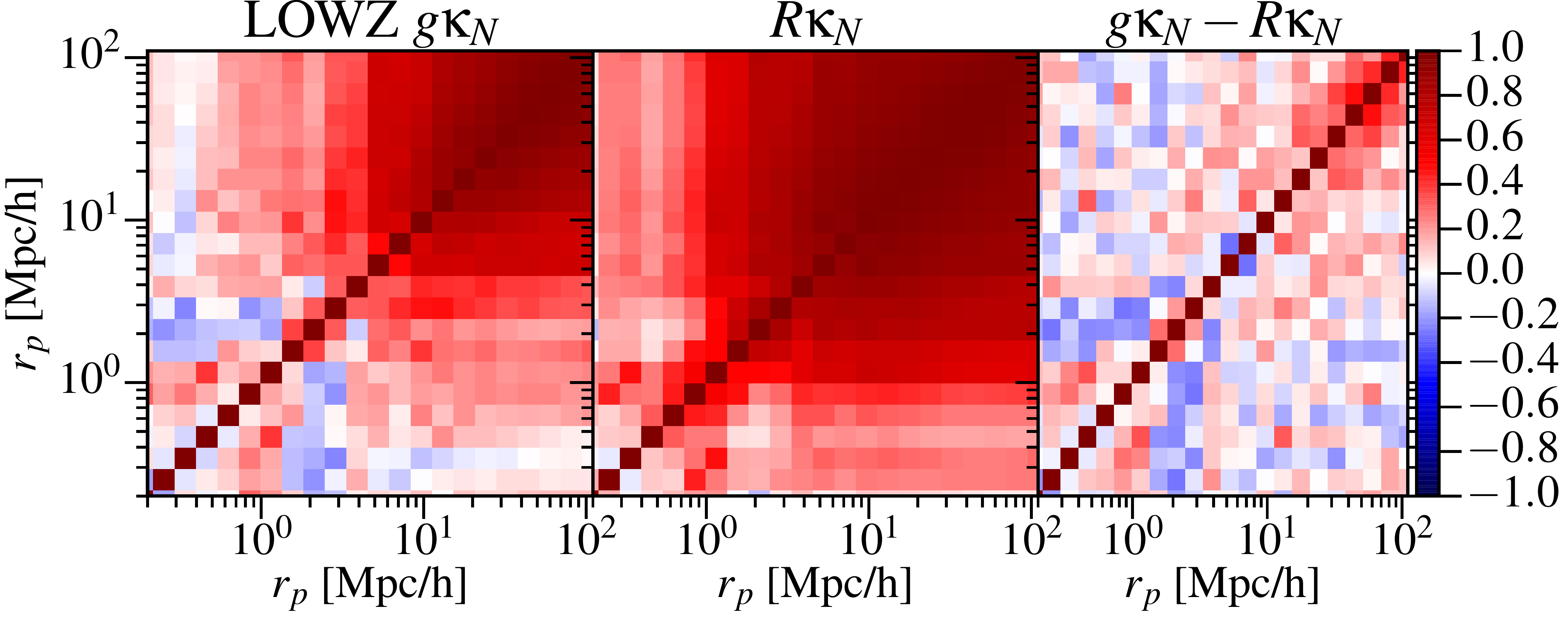}
         \caption{Correlation matrix for the galaxy position vs.\ CMB lensing cross-correlation
           measurement for the LOWZ sample, with \nside$=1024$.
         We show
         correlation matrices for the signal measured around galaxies ($g\kappa$, left column),
         around
         random points ($R\kappa$, middle column), and the difference between the two
         ($g\kappa-R\kappa$, right column) using
         both CMB ($\kappa$, top row) and noise
         convergence ($\kappa_N$, bottom row) maps. The correlation and covariance matrices obtained using both maps are
         consistent. The
         correlations in $g\kappa$ and $R\kappa$ at large scales are caused by the CMB \referee{lensing reconstruction 
         noise}, which can be removed by subtracting the signal around random points.
}
         \label{fig:lowz_corr}
      \end{figure}

		Figure~\ref{fig:gk_all} shows the cross-correlation signal for galaxy position vs.\ CMB
        lensing for both the LOWZ and CMASS samples. We
		show the signal measured around the galaxies and around random points. At small scales, there is
		significant signal
		around galaxies, and the the measurements in different $r_p$ bins are uncorrelated.
      At large scales, the lower signal and the fact that the stack in each bin includes almost all
      pixels in the map leads to the signal being dominated by the noise in the CMB convergence, and
      hence the bins are very strongly correlated (we are effectively measuring the mean and standard deviation of the map in every bin).
      This noise can be removed by subtracting out
		the signal measured around random points from the signal measured
		around the galaxies, as shown in Eq.~\eqref{eqn:sigma_cmb}. Another way to understand the
        effect of this random subtraction is that similar to the Landy-Szalay estimator in clustering, we want to correlate the CMB map with a mean zero quantity ($\langle D-R\rangle=0$), so that any additive systematics are removed to first order \citep{Mandelbaum2006}.
      After subtracting, the different $r_p$ bins exhibit less substantial
      but still noticeable correlations even at large scales, as shown in
		Fig.~\ref{fig:lowz_corr}.

		The final measurements after subtracting the signal around random points are consistent with the Planck 2015
		$\Lambda$CDM model predictions. The solid lines in Fig.~\ref{fig:gk_all} show
		the linear theory $ +$ halofit correlation functions with the best-fitting galaxy bias, $b_g$; the parameters of these fits are presented in
		Table~\ref{tab:sigma_params}.
      We only fit this model for scales $r_p>10\mpch$, as at smaller scales, the effects
		of non-linear bias and stochasticity in the galaxy-matter cross-correlation are expected to cause
		deviations between the data and the model.
          However, as shown in Fig.~\ref{fig:gk_all}, there
        is no evidence of tension between the theory and the data even down to
		$r_p\sim5\mpch$.  On the scales used for the fits, \referee{$\chi^2_\text{red}\sim0.8~(0.6)$ for LOWZ (CMASS)}. 
		Given the large minimum radius for the fits,
		 the constraints on galaxy bias are independent of the pixel size and of the smoothing
		imposed by the cutoff in the CMB $\kappa$ map at $l_\text{max}=2048$. We remind the reader
        that the minimum radius for the fits is larger than the resolution of the convergence maps,
        and hence the results from fitting to theoretical models in this section do not have any
        smoothing applied to the models. We will discuss the small scale signal, where smoothing is necessary, in Sec.~\ref{ssec:results_small_scale}.

\begin{figure}
             \centering
             \includegraphics[width=1\columnwidth]{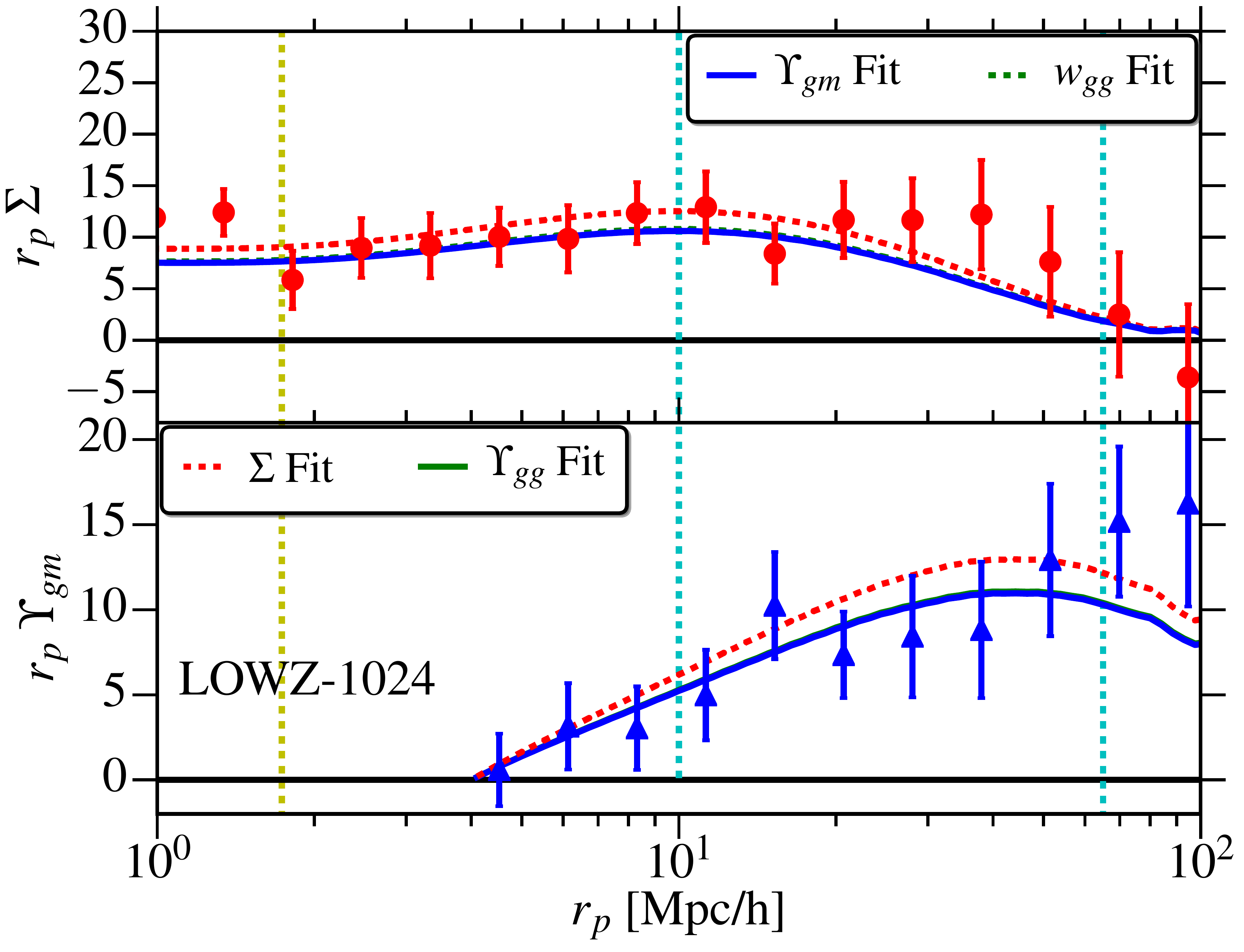}
             \caption{Comparison of $\Sigma$ and \ugm\ obtained from CMB lensing with  Planck theory
               predictions using best fitting bias from $\Sigma$ and \ugm\ as well as from galaxy
               clustering. Note that the bias from $\wgg$ and \ugg\ is consistent with the bias from
               \ugm\ and hence the models overlap on the plot. The vertical yellow line marks the
               smoothing scale $6\arcmin$ at $z=0.36$, and the vertical cyan lines show the range over which models are fitted, $10<r_p<70\mpch$.
             }
            \label{fig:lowz_sigma_ups_theory_comp}
         \end{figure}

      In Fig.~\ref{fig:lowz_sigma_ups_theory_comp} we show the comparison of
      $\Sigma$ and \ugm\ with the predictions from Planck theory model using the
      best fitting bias from $\Sigma$, \ugm, \wgg and \ugg.
      As shown in Fig.~\ref{fig:lowz_sigma_ups_theory_comp} and Table~\ref{tab:sigma_params}, there is some discrepancy between the $b_g$ obtained from $\Sigma$ and
		$\Upsilon_{gm}$. These discrepancies are not very significant ($\lesssim1.5\sigma$ after accounting for
		correlations).
        This is mostly caused by the noise in the measurements, which leads to mild tension between
		theory and data as they have slightly different scale dependences. Differences in the scale
        dependence of the theory and
		data can in general lead to different bias measurements from $\Upsilon$ and $\Sigma$, since $\Upsilon$ at any
		scale $r_p$ depends on values at scales smaller than $r_p$ (but greater than $r_0$).

		\begin{table*}
           \begin{tabular}{|c|c|c|c|c|c|}
                 \hline
Lens sample  & $\kappa$  & Pixel  & $b_g (\Upsilon)$  & $b_g (\Sigma)$ & $M_h$ $[10^{12}M_\odot/h]$\\
			&			&	size	&					&				&	\sigmaNFW \\	\hline
LOWZ & $\kappa$ & 6.9\arcmin & 1.80$\pm$0.30 & 2.20$\pm$0.50 & 13.5$\pm$6.5 \\ \hline
LOWZ & $\kappa$ & 3.4\arcmin & 1.80$\pm$0.28 & 2.12$\pm$0.46 & 13.9$\pm$3.9 \\ \hline
LOWZ & $\kappa_{\sigma=10\arcmin}$ & 3.4\arcmin & 1.87$\pm$0.27 & 2.20$\pm$0.40 & 17.1$\pm$7.4 \\ \hline
LOWZ & $\kappa$ & 1.7\arcmin & 1.75$\pm$0.28 & 2.21$\pm$0.46 & 11.8$\pm$3.6 \\ \hline
LOWZ &	$\gamma$&		&			&				&$10.1\pm0.6 (\Delta\Sigma)$		\\ \hline
Field & $\kappa$ & 3.4\arcmin & 1.61$\pm$0.26 & 1.90$\pm$0.40 & 13.5$\pm$3.9 \\ \hline
Field &	$\gamma$&		&			&				&  $8.6\pm0.6(\Delta\Sigma)$			\\ \hline
CMASS & $\kappa$ & 6.9\arcmin & 1.50$\pm$0.20 & 1.46$\pm$0.28 & 24.9$\pm$5.7 \\ \hline
CMASS & $\kappa_{\sigma=10\arcmin}$ & 3.4\arcmin & 1.75$\pm$0.15 & 1.60$\pm$0.20 & 34.3$\pm$6.4 \\ \hline
CMASS & $\kappa$ & 1.7\arcmin & 1.56$\pm$0.19 & 1.51$\pm$0.27 & 6.9$\pm$2.3 \\ \hline
CMASS-v1 & $\kappa$ & 3.4\arcmin & 1.64$\pm$0.29 & 1.80$\pm$0.40 & 4$\pm$5 \\ \hline
CMASS-v2 & $\kappa$ & 3.4\arcmin & 1.53$\pm$0.29 & 1.31$\pm$0.39 & 16.3$\pm$4.5 \\ \hline
			\end{tabular}
             \caption{Measurement of halo mass (in units of $10^{12}h^{-1}\Msun$) and linear galaxy bias $b_g$ from
             	lensing alone
             	($\cgm=1$ fixed), using
             	different estimators. When measuring $b_g$, the signals are fit using $r_p>10\mpch$, with $r_0=4\mpch$
				for
				$\Upsilon_{gm}$. We show results for different choice of pixel size and smoothing applied to maps
            and the rows called ``$\gamma$'' use the
                optical galaxy lensing shear instead of CMB lensing convergence
                maps.
				Our linear galaxy bias constraints are not significantly impacted by the choice of
                pixel size or smoothing scale.  When measuring halo masses, the NFW profiles are fit between
                $r_p<0.3\mpch$ (in case of $\Delta\Sigma$) while \sigmaNFW\ are fit for
				$r_p<5\mpch$.
				When fitting a smoothed NFW profile, we
				apply $\sigma=10'$ smoothing in cases where convergence map has also been smoothed by $10'$
				($\kappa_{\sigma=10'}$).
				}
              \label{tab:sigma_params}
         \end{table*}

		The right column of Fig.~\ref{fig:gk_all} also shows the null tests for these
		measurements. The signal around random points is consistent with zero at all scales,
		though the noise at large scales is dominated by the \referee{reconstruction noise in CMB lensing}. When using the
		shuffled CMB lensing map, the signal around random points, galaxies, and their difference is
		consistent with zero. Finally, the signal measured using the noise map is also consistent with zero. The noise
		map
		also serves as a diagnostic for our covariance estimates as shown in
        Fig.~\ref{fig:lowz_corr}, since it has the same
		correlated noise properties. The Jackknife covariance and correlation matrices obtained from
		measurements using the CMB map and noise map are consistent. In
        Appendix~\ref{app:covariance_test} we show the consistency between the jackknife covariance
        matrix and the covariance matrix obtained using 100 independent realizations of the noise maps.

       \begin{figure}
         \centering
         \includegraphics[width=1\columnwidth]{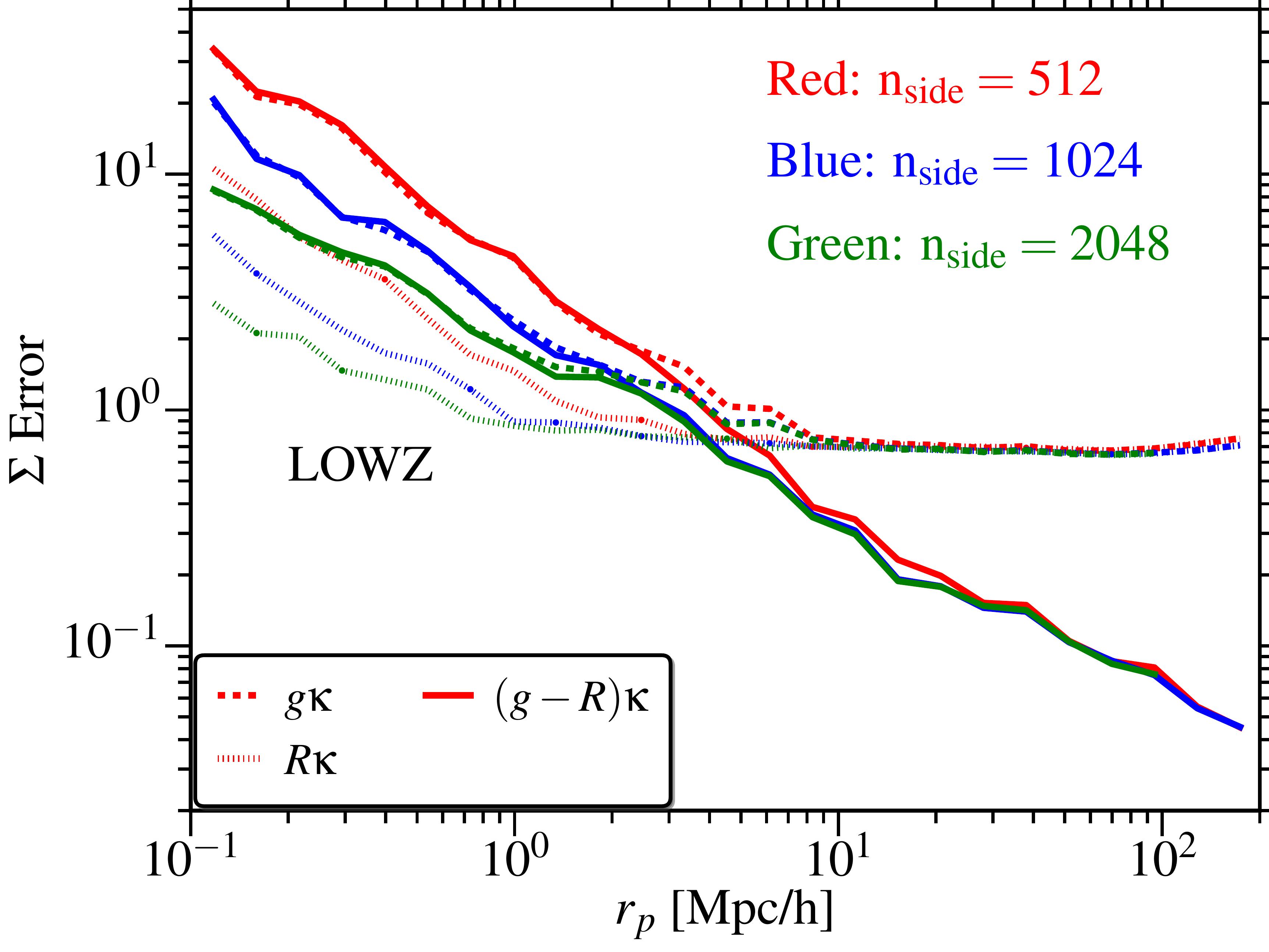}
         \caption{Comparison of error bars (square root of the diagonal part of the
           covariance matrix) in $\Sigma$ measurement for LOWZ. The color indicates the pixel size
           as indicated on the plot. Solid, dashed, and dotted lines for each \nside\ show errors in $g\kappa-R\kappa$,
           $g\kappa$, and
			 $R\kappa$ respectively. We use $N_R=10N_g$, hence errors are smaller for $R\kappa$, and as shown
			before, taking the difference between $g\kappa$ and $R\kappa$ reduces the errors at large scales.
			The errors obtained using $\kappa_N$ from noise maps (not shown) are very similar to those
            using
			$\kappa$.
			}
         \label{fig:lowz_gk_error}
      \end{figure}
		Fig.~\ref{fig:lowz_gk_error} shows the scaling of the errors with $r_p$, and a comparison of errors
		using different pixel sizes for LOWZ; CMASS results, which are not shown, are similar. Going to smaller pixel
		size improves the signal-to-noise ratio at small scales, which
		suggests that there is some information available at these scales
         In Fig.~\ref{fig:lowz_gk_error}, we show results for \nside$=512$, 1024 and 2048. Using
        \nside$=2048$ results in some
		improvement at small scales, but the results are comparable to those with \nside
		$=1024$ at
        scales $r_p\gtrsim1\mpch$, motivating our choice of
		\nside$=1024$ for our primary results. However, our galaxy bias
		constraints that are based on large scales are not significantly affected by the choice of \nside.

		Also in Fig.~\ref{fig:lowz_gk_error}, statistical uncertainties on the signals around
        galaxies and random points
		saturate for scales $r_p\gtrsim10\mpch$ ($\theta\gtrsim1^\circ$). This saturation results
        from the fact that once we have stacked many pixels,
		we are limited by the \referee{reconstruction} noise of the CMB lensing maps. This noise is also the
        reason for the strong bin-to-bin correlations in the first two columns in Fig.~\ref{fig:lowz_corr}.
		After subtracting the signal around random points, the final signal is still dominated by noise in
		the CMB convergence measurements, but there is no evidence
		of residual systematics from our null tests. Moreover, Fig.~\ref{fig:lowz_corr} shows that
        there are only mild to moderate
		correlations between the bins.
		\begin{figure*}
		     \begin{subfigure}{\columnwidth}
		         \centering
		         \includegraphics[width=0.9\columnwidth]{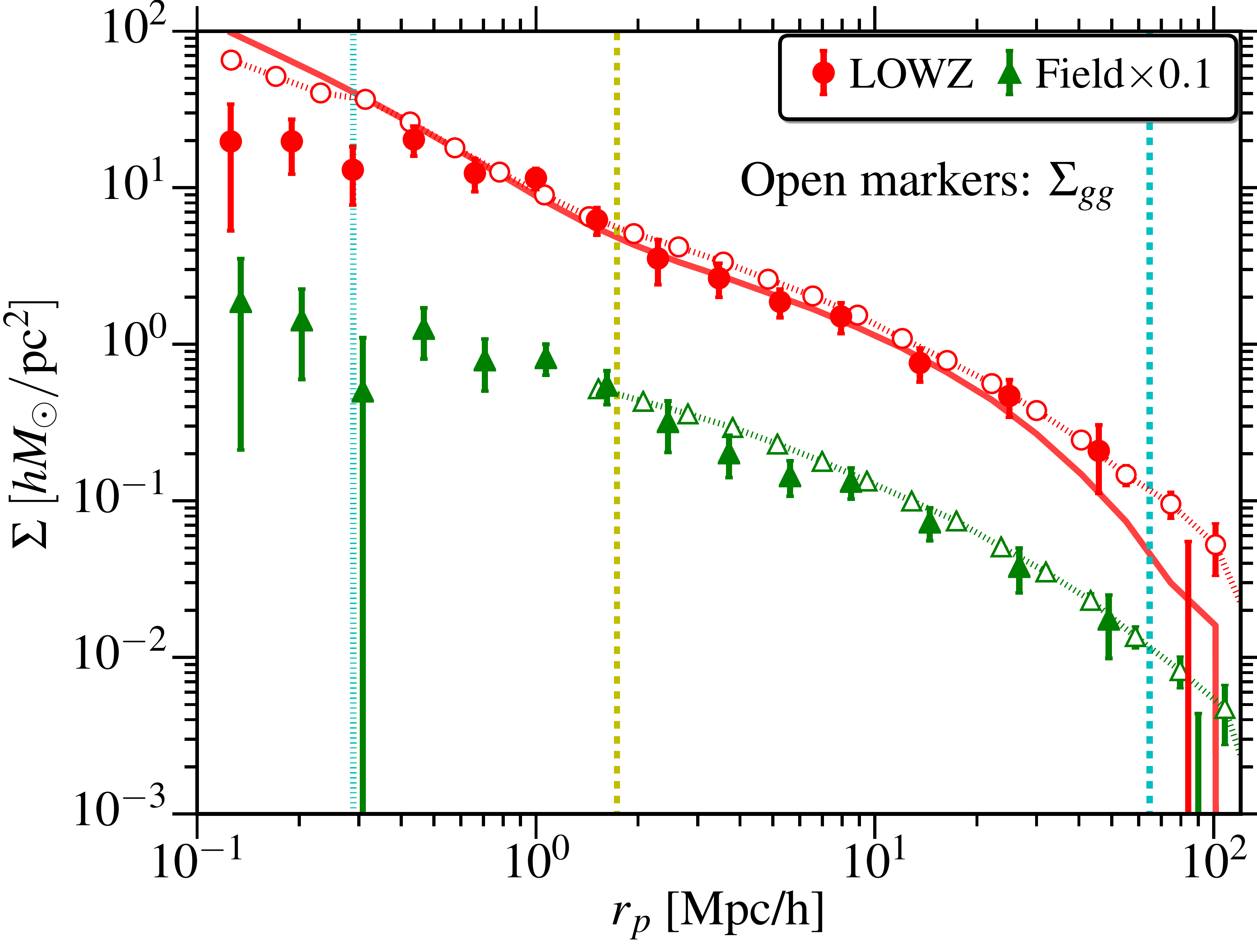}
		         \caption{}
		      \label{fig:lowz_sigma}
		      \end{subfigure}
		     \begin{subfigure}{\columnwidth}
		         \centering
		         \includegraphics[width=.9\columnwidth]{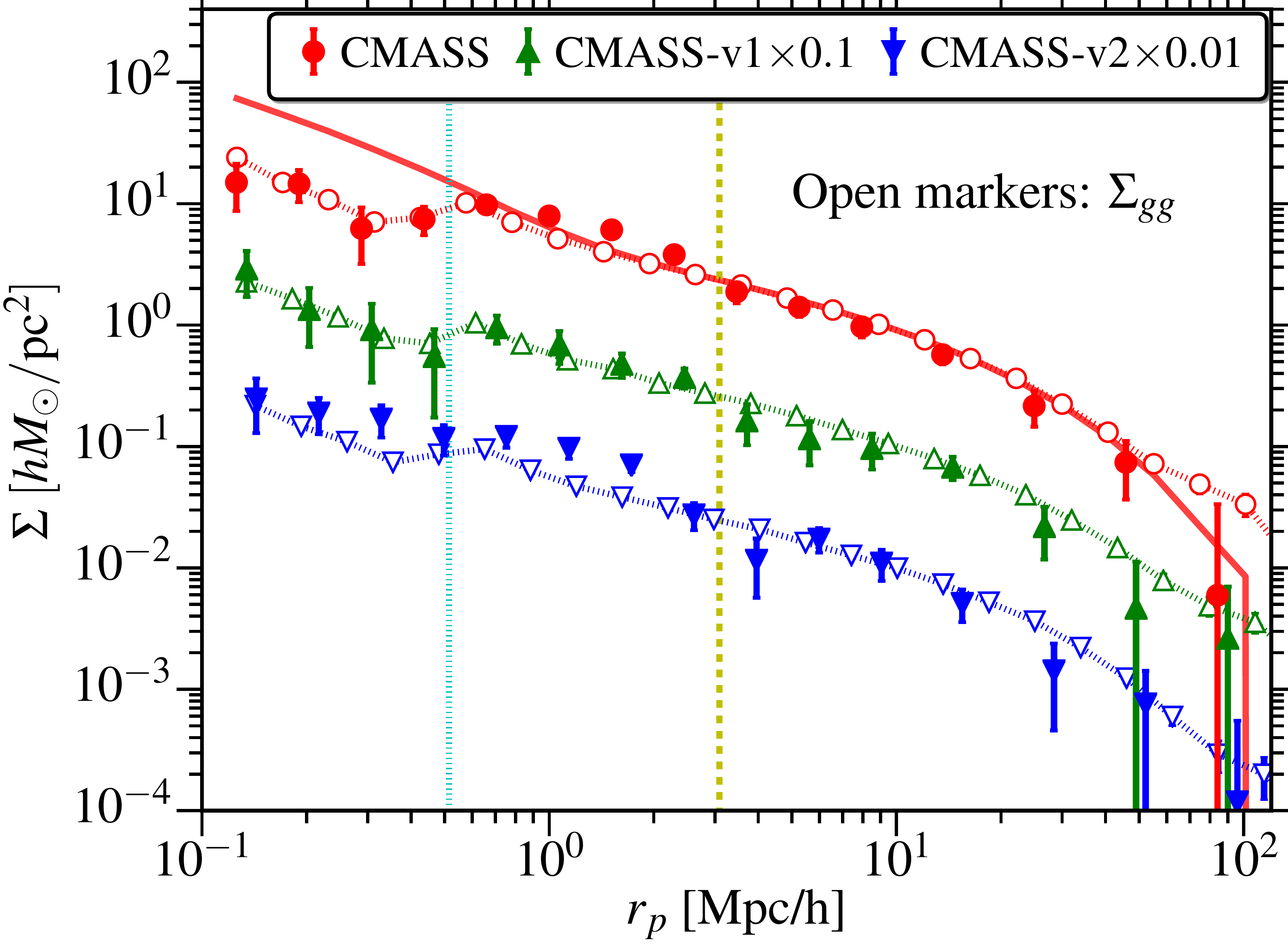}
		         \caption{}
		         \label{fig:cmass_sigma}
		      \end{subfigure}
		      \caption{ Comparison of the surface density $\Sigma$ obtained from CMB lensing (solid markers) with
                the clustering measurement,
		      	\wgg, converted
		      	into $\Sigma_{gg}=\bar{\rho}\wgg/b_g$ (open markers). Note that some samples are shifted vertically
				for easier viewing, with shift factor mentioned in the legend.
				The solid red lines for LOWZ and CMASS show
				$\Sigma_{gm}$ measured from
				simulations (no smoothing applied).
				The dashed yellow lines mark the 6$'$ scale (corresponding to $\ell_\text{max}$ cutoff) at
				$z_\text{max}=0.36~(0.7)$ for LOWZ (CMASS).
				The dashed cyan lines show the
				size of the jackknife regions at $z_\text{min}$ ($r_p\sim70\mpch)$ for LOWZ, off the
                right side of the plot for CMASS).
				The dotted cyan lines show the fiber collision scale at $z_\text{min}=0.16~(0.45)$
                for LOWZ (CMASS).
		      	}
				\label{fig:sigma_wgg}
		\end{figure*}

		In Figure~\ref{fig:sigma_wgg}, we show the $\Sigma$ measured using CMB lensing from Fig.~\ref{fig:gk_all} with
		the $\Sigma$ measured from simulations and the clustering measurement by
		converting the clustering into $\Sigma_{gg}$ using Eq.~\eqref{eqn:sigma_gg}, where we use the best fit $b_g$ to
		the clustering signal. The signals are consistent at most
		scales, though the comparison with clustering is only qualitatively valid.
		At small scales, the effects of non-linear galaxy bias and the stochasticity in the
        galaxy-matter cross-correlation can lead to differences between the two.
		Also, in the BOSS data, the clustering below $\lesssim1'$ is affected by the incompleteness due to fiber
		collisions, which biases the signal for $r_p\lesssim0.3 (0.5) \mpch$ in the case of LOWZ
        (CMASS) sample, even when weights are used (even with weights, clustering is not unbiased for $r_p\lesssim2\mpch$).
		The lensing signal is also affected since
		fiber collisions preferentially affect the higher density regions which lead them to be under-weighted in
		the lensing measurements as well. While fiber collision weights do attempt  to correct  the
        bias, the signal is still biased at small scales since we are
      stacking on the wrong galaxies.
		The CMB lensing measurement at these scales is also affected by the pixel size and smoothing of
        the Planck maps.
		At large scales, the clustering is affected by the residual RSD, while the lensing signal is not to
		first order. This effect is $\gtrsim10\%$ above $r_p\gtrsim70\mpch$. Between
        $10<r_p<50\mpch$, we do expect to find a good
		agreement ($\lesssim10\%$) between the CMB lensing and clustering measurements, as demonstrated in
		Fig.~\ref{fig:sigma_wgg}.

      In Fig.~\ref{fig:sigma_wgg}, we also show the signals measured for various LOWZ and CMASS subsamples
      defined in Sec.~\ref{ssec:data_boss}. The primary motivation for defining these samples was to test
      for the effects of redshift failure weights (using field galaxies, which should be less affected by fiber
      collisions) and variations in number densities with redshift (using volume-limited samples CMASS-v1
      and CMASS-v2). We do not find any significant tension in results using the subsamples, with the
      lensing results being largely consistent with the predictions from theory combined  with the bias
      measurements from clustering.
     \begin{figure}
         \centering
         \includegraphics[width=\columnwidth]{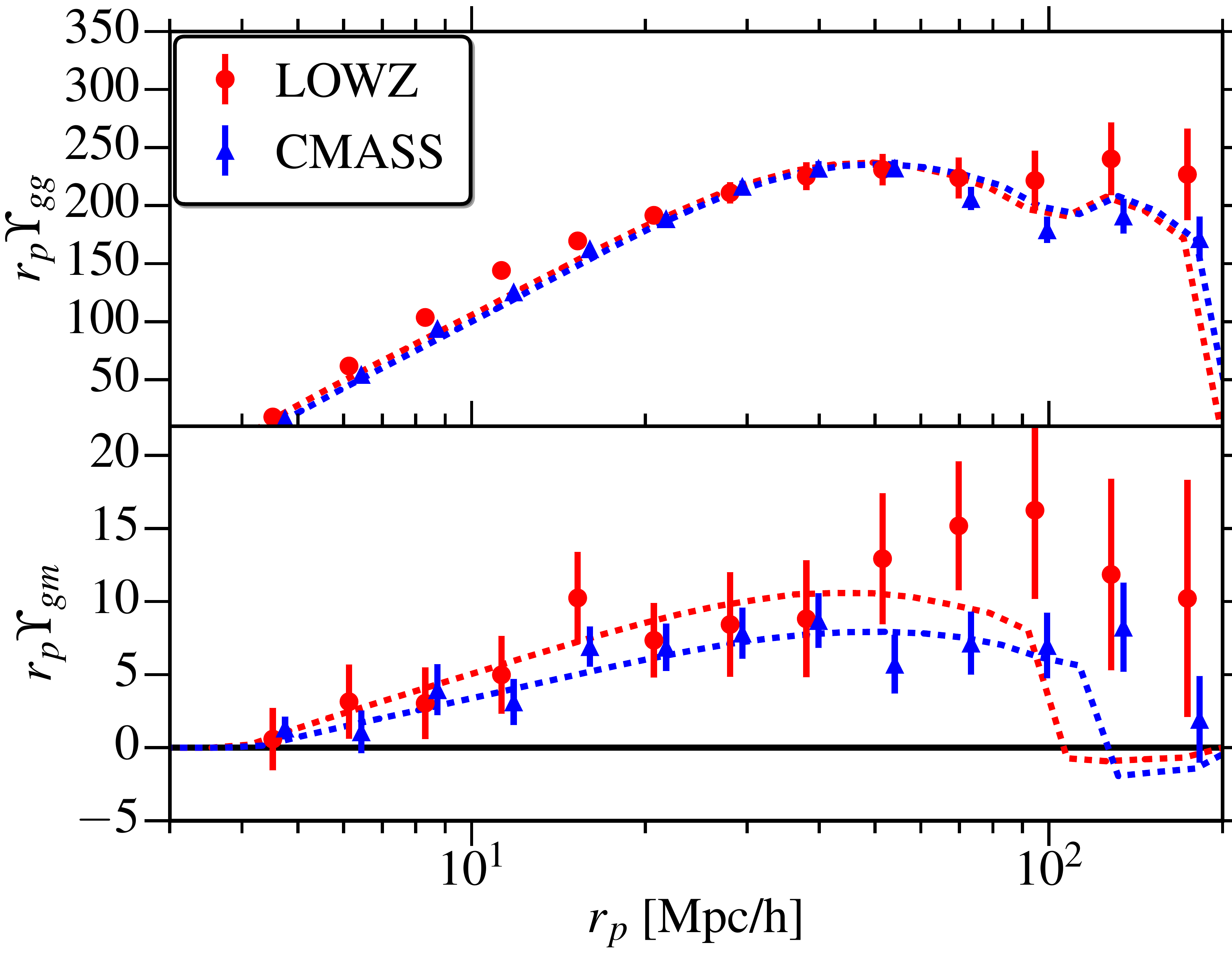}
         \caption{ $r_p\ugg$ (top; units of $(\text{Mpc}/h)^2$) and $r_p\Upsilon_{gm}$ (bottom; units of $10^{6}\Msun/\text{pc}$) measurements using CMB lensing,
         	using $r_0=4\mpch$. The dashed lines are the Planck \lcdm\ model with the best-fitting
            bias and \cgm\ from our fits.
      }
         \label{fig:lowz_cmass_upsilon}
      \end{figure}

	Fig.~\ref{fig:lowz_cmass_upsilon} shows the $\Upsilon$ measurements from the BOSS LOWZ and CMASS
    samples obtained by converting \wgg\ and $\Sigma$ measurements into $\Upsilon$ using methods described in
    Section~\ref{ssec:ups_estimator}
      This figure also shows the results from jointly fitting both clustering and lensing measurements
	with fixed Planck 2015 cosmology to get the galaxy bias
	$b_g$ and the relative lensing amplitude $\cgm$. Due to the use of weighting in lensing
    measurements, the clustering and
	lensing measurements are not at the same effective redshift.  As stated in
    Sec.~\ref{ssec:theory_galaxy_lensing}, we integrate
	the theoretical predictions over redshift using the weights, assuming redshift-independent
    linear bias and \cgm.
	The theory fits the data well within the noise, with
	$b_g=1.95\pm0.02$ and $\cgm=0.79\pm0.13$ for the CMASS sample and $b_g=1.75\pm0.03$ and $\cgm=1.0\pm0.2$ for the
	LOWZ sample (see also Table~\ref{tab:ups_joint_params}). Our \cgm\ measurements are consistent
    with the theoretical expectation of $\cgm=1$ at linear scales. The galaxy bias measurements for
    CMASS are consistent with those from \cite{Torres2016}, who measured a scale-dependent bias of
    $b_g=1.9$--$2$, using scales 10--60\mpch. For LOWZ, our bias is consistent with that measured by
    \cite{Singh2015} using the DR11 sample ($b_g=1.77\pm0.04$).

    \begin{figure}
        \centering
        \includegraphics[width=\columnwidth]{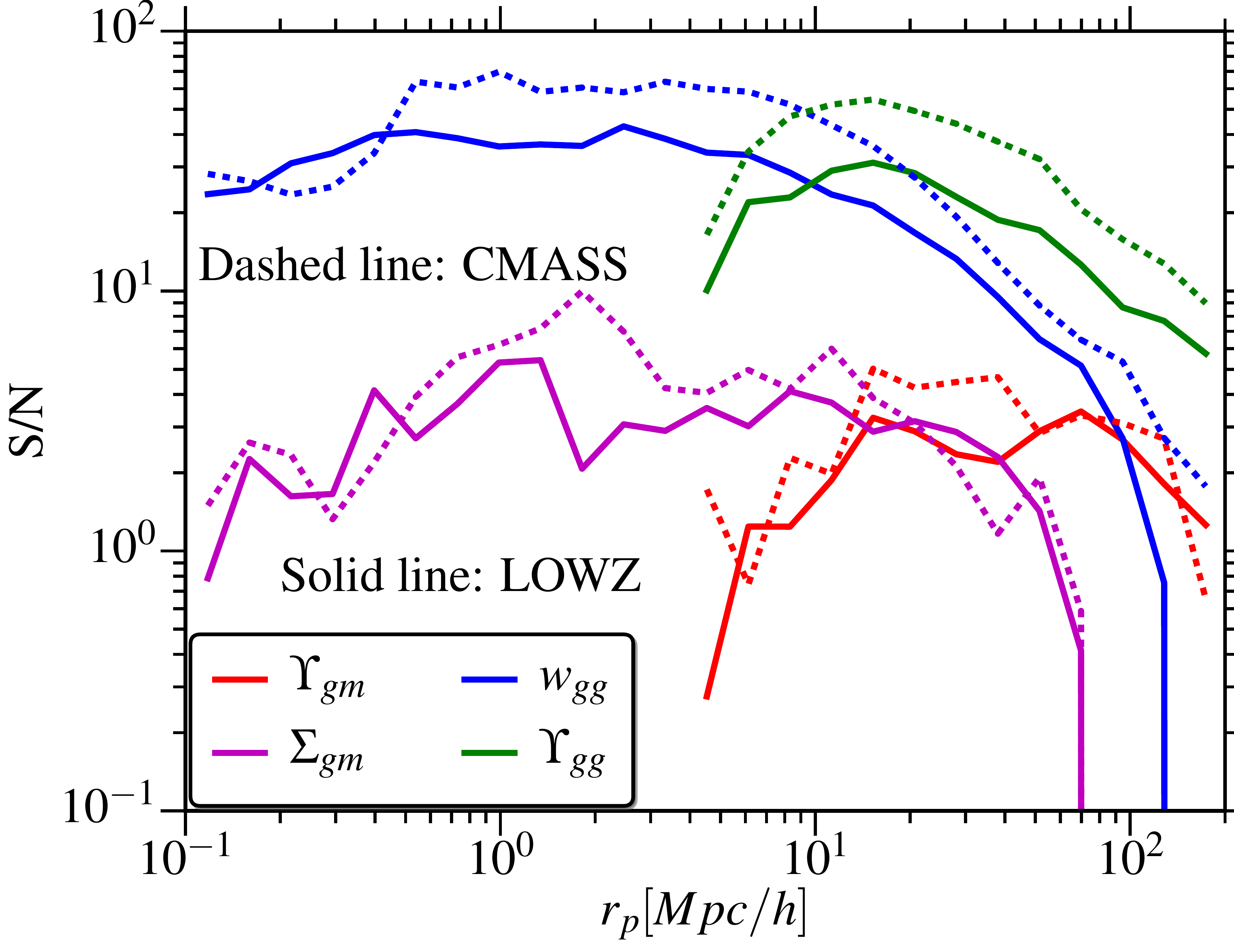}
        \caption{Comparison of the signal-to-noise ratio ($S/N$) in the clustering and CMB lensing
          measurements using different estimators for both LOWZ and CMASS samples. For $\Upsilon$,
          $r_0=4\mpch$. Using $\Upsilon$ decreases the $S/N$ for $r_p \lesssim 3r_0$ since we are
          removing some of the signal, while at large scales, the $S/N$ improves as $\Upsilon$
          reduces the impact of cosmic variance.
       }
        \label{fig:lowz_cmass_SN}
     \end{figure}

     In Fig.~\ref{fig:lowz_cmass_SN}, we compare the signal-to-noise ratio ($S/N$) for the
     clustering and CMB lensing measurements using different estimators. $\Upsilon$ has lower
     signal-to-noise for $r_p \lesssim 3r_0$  since by definition it has the signal from scales
     below $r_0$ removed. However, $\Upsilon$ also reduces the impact of cosmic variance and additive systematics in the measurement, and hence improves the $S/N$ at large scales.

		\begin{table}
           \begin{tabular}{|c|c|c|c|c|}
                 \hline
Sample-Planck  & $\kappa$  & Pix Area  & $b_g$  & $\cgm$ \\ \hline
 LOWZ & $\kappa$ & 3.4\arcmin & 1.75$\pm$0.04 & 1.0$\pm$0.2 \\ \hline
LOWZ & $\kappa_{\sigma=10'}$ & 3.4\arcmin & 1.75$\pm$0.04 & 1.1$\pm$0.2 \\ \hline
CMASS & $\kappa$ & 3.4\arcmin & 1.95$\pm$0.02 & 0.78$\pm$0.13 \\ \hline
CMASS & $\kappa_{\sigma=10'}$ & 3.4\arcmin & 1.95$\pm$0.02 & 0.8$\pm$0.1 \\ \hline
Field & $\kappa$ & 3.4\arcmin & 1.47$\pm$0.03 & 1.15$\pm$0.24 \\ \hline
CMASS-v1 & $\kappa$ & 3.4\arcmin & 2.0$\pm$0.03 & 0.9$\pm$0.2 \\ \hline
CMASS-v2 & $\kappa$ & 3.4\arcmin & 1.9$\pm$0.03 & 0.7$\pm$0.2 \\ \hline
\end{tabular}
             \caption{ Results from joint fitting of $\Upsilon_{gg}$ and $\Upsilon_{gm}$, with $r_0=10\mpch$ and
	             $r_p>20\mpch$.
             }
              \label{tab:ups_joint_params}
         \end{table}

\subsection{The small-scale signal}\label{ssec:results_small_scale}
\begin{figure}
        	 \centering
	         \includegraphics[width=\columnwidth]{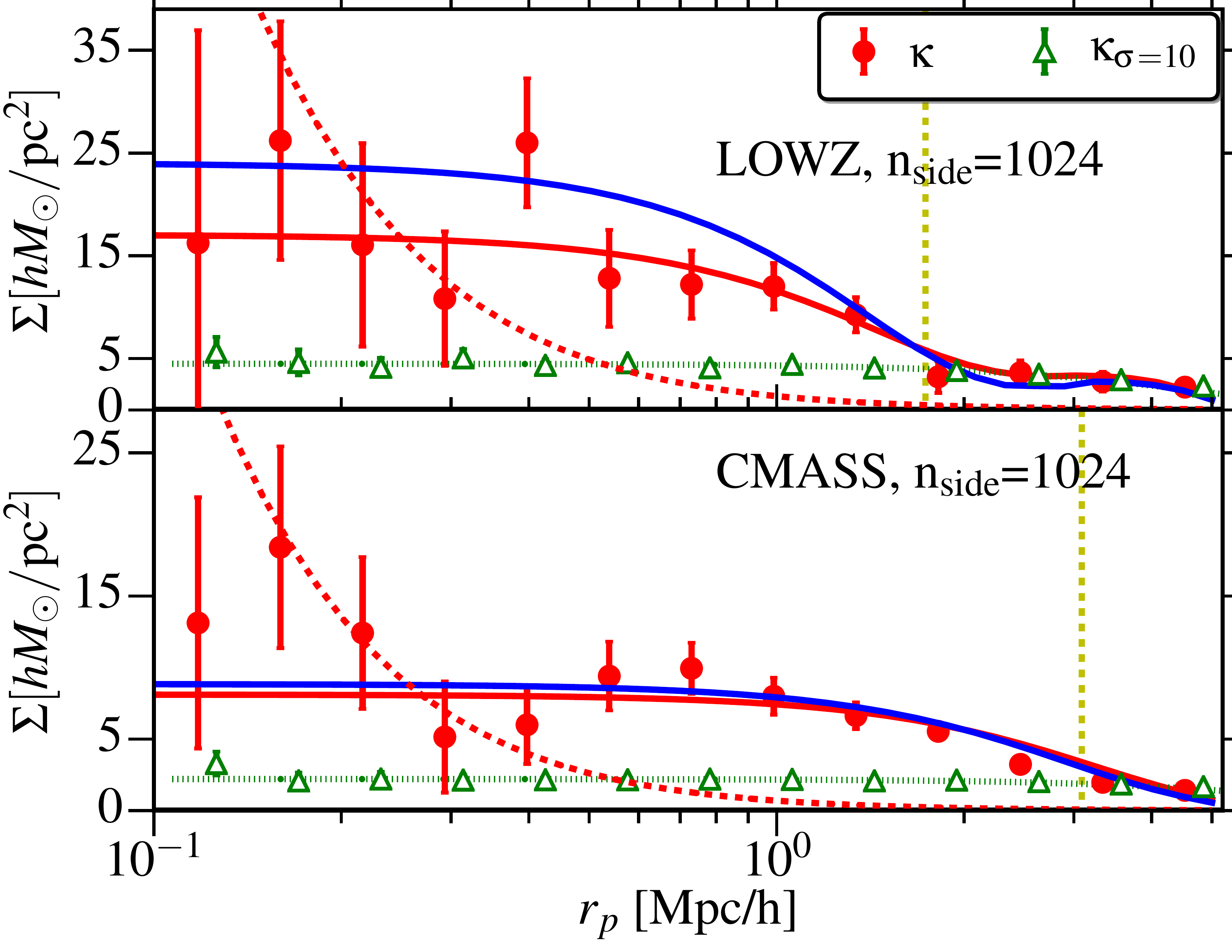}
    	     \caption{$\Sigma$ measurement for the LOWZ (top) and CMASS (bottom) samples with
               different levels of smoothing applied to the convergence
	     		map.	The solid red lines show the best-fitting
				smoothed \sigmaNFW\ profile, while dashed lines show the unsmoothed NFW model as a
            reference (incorrect model given the way the lensing map was produced).
            The dotted green line shows the \sigmaNFW\ profile smoothed with a $\sigma=10'$ gaussian
				kernel. The blue lines show the smoothed $\Sigma$ profile measured from $N$-body
                simulations with an HOD tuned to match the galaxy clustering.
				The vertical yellow line marks the $6'$ scale (corresponding to $\ell$ cutoff in Planck convergence map) at $z=0.7$ ($z=0.36$) for CMASS
				(LOWZ) sample.
}
         \label{fig:lowz_cmass_nfw}
      \end{figure}

		In Figure~\ref{fig:sigma_wgg}, we show the CMB lensing signal at small separations.
      As discussed in
		Sec.~\ref{ssec:sigma_1halo} the $\ell$ cutoff
		effectively smoothens the configuration space convergence map with a two-dimensional sinc kernel.
		In Fig.~\ref{fig:lowz_cmass_nfw} we show the measurements only at
		$r_p<5\mpch$ along with the smoothed $\Sigma_{gm}$ profile measured from the simulations (blue lines), described in Sec.~\ref{ssec:sigma_1halo}.
      We also show the best-fitting smoothed \sigmaNFW\ profile, where the NFW halo mass was set as a free parameter in the
		fitting procedure. The
		halo masses are presented in Table~\ref{tab:sigma_params}; these are consistent with the halo mass
		measured
		from galaxy-galaxy lensing in the case of the LOWZ and LOWZ-Field subsamples. The halo mass measurement using
		galaxy-galaxy lensing in this work, $M_h=(1.01\pm0.06)\times10^{13}\Msun/h$, is different
      from that of \cite{Singh2015}, $M_h=(1.5\pm0.2)\times10^{13}\Msun/h$, using BOSS DR11 sample. The difference is primarily
		driven
		by the different adopted mass-concentration relations, to which the mass estimates are sensitive when fitting scales
		$r_p<0.3\mpch$ as described in Sec.~\ref{ssec:delta_sigma_1_halo}. Using the same mass-concentration relation as
		\cite{Singh2015}, we get consistent results ($M_h=(1.68\pm0.15)\times10^{13}\Msun/h$).

		In the case of the LOWZ sample, $\Sigma_{gm}$ from simulations
      over-predicts the signal. This is because the mean halo mass from the
		simulations is
		$M_h\sim5\times10^{13}\Msun/h$ (median mass $M_h\sim2.5\times10^{13}\Msun/h$),
		which is higher than the mass preferred by data, $M_h\sim10^{13}\Msun/h$, in the case of
        both the CMB and galaxy lensing
		measurements.

		The halo mass obtained using the \sigmaNFW\ fits for the CMASS sample, $M_h\sim10^{13}\Msun/h$, is low compared to the
		values of $M_h\sim2\times10^{13}\Msun/h$ measured by
		\cite{Miyatake2015} and \cite{Madhavacheril2015} using galaxy-galaxy and galaxy-CMB lensing respectively. This discrepancy is in the expected direction since
		$\Sigma_{gg}$
		tends to over-predict the $\Sigma$ profile due to the effects of non-linear galaxy bias and hence the NFW mass will
		be suppressed. Ultimately, the proper interpretation of the lensing signal at small scales requires proper
		halo modeling, which we do not do given the noise and resolution of Planck CMB lensing maps. Instead we
		have presented a simple model to enable easy comparisons, but possible biases in this model should be kept in
		mind.   We note that the profile from simulations (after smoothing) is a reasonable
        description of the data; in the simulated CMASS sample, the mean $M_h\sim3.3\times10^{13}\Msun/h$ and median $M_h\sim1.7\times10^{13}\Msun/h$ \citep{Reid2014}.

\subsection{Lensing calibration: CMB vs.\ galaxy lensing}\label{ssec:results_lensing_calibration}

     \begin{figure}
         \centering
         \includegraphics[width=\columnwidth]{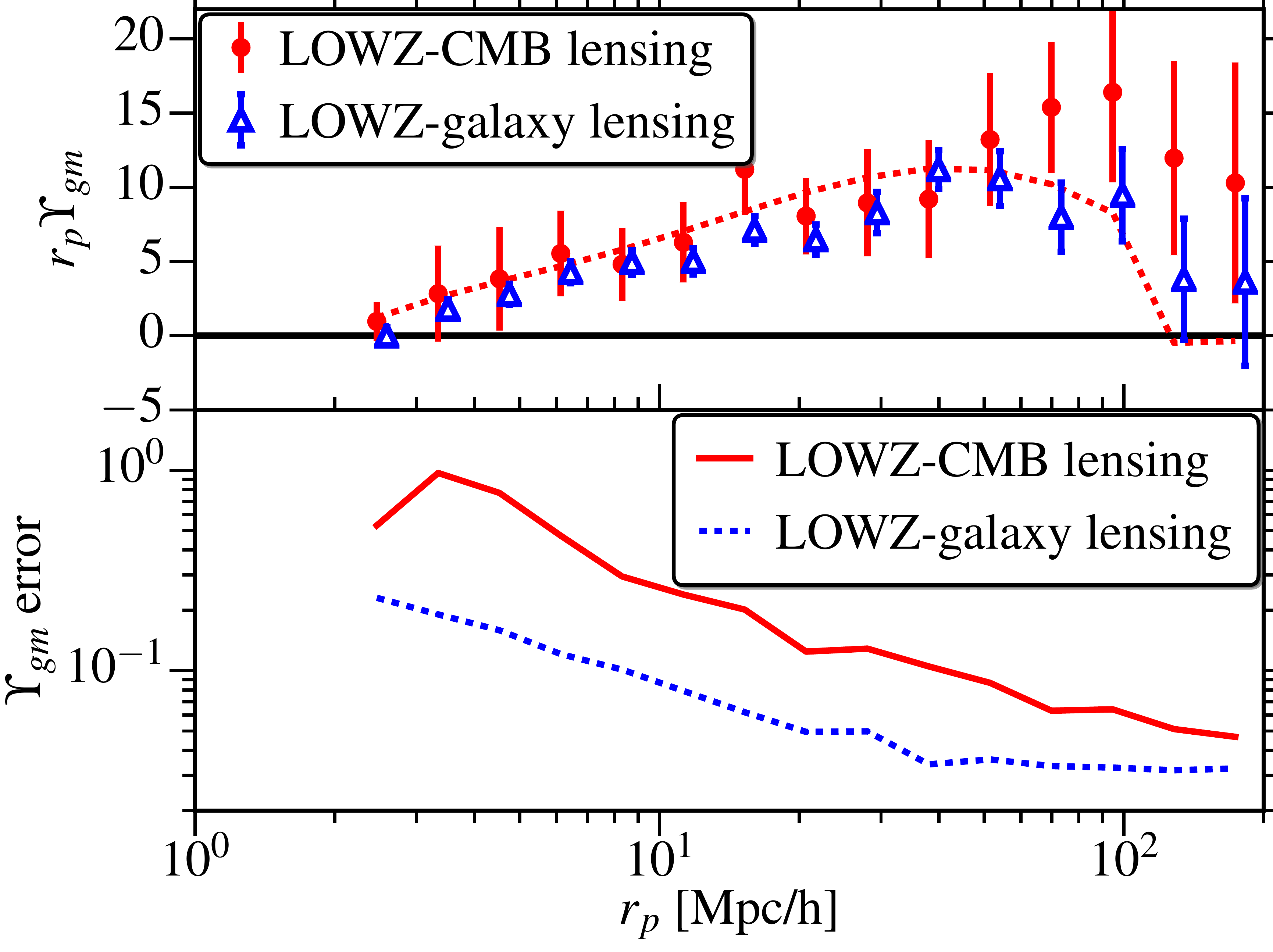}
         \caption{Comparison of $r_p\Upsilon_{gm}$ (in units of $10^{6}\Msun/\text{pc}$) and errors in $\Upsilon_{gm}$,
         obtained from galaxy-CMB lensing and galaxy-galaxy lensing. The dashed red line shows the prediction using Planck
         2015 cosmology along with best fitting bias to galaxy-CMB lensing measurement.
      }
         \label{fig:lowz_ups_gm_comparison}
      \end{figure}

	Figure~\ref{fig:lowz_ups_gm_comparison} shows the comparison of $\Upsilon_{gm}$ obtained from galaxy lensing and
	CMB lensing using LOWZ galaxies as lenses. Note that due to the different weighting used in galaxy lensing and CMB
	lensing, the
	two measurements are not at the same effective lens redshift, with $z_\text{eff}=0.24$ ($0.3$)
    for galaxy (CMB) lensing. In Fig.~\ref{fig:lowz_ups_gm_comparison}, we also compare the
	measurement uncertainties, with galaxy lensing having a higher signal-to-noise ratio by a factor
    of 2--5.

	 \begin{figure}
         \centering
         \includegraphics[width=\columnwidth]{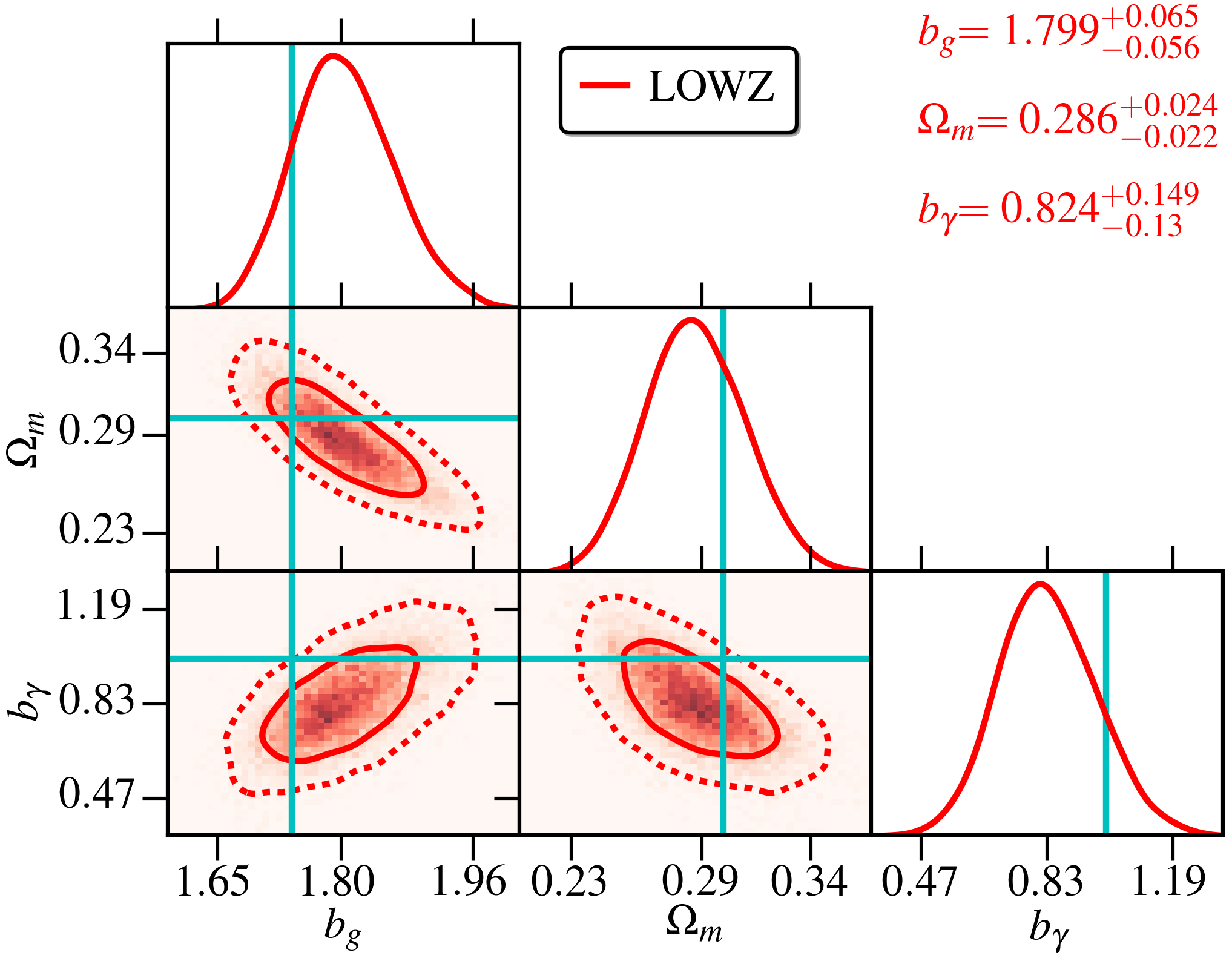}
         \caption{Results from MCMC fits to $\Upsilon_{gg},\Upsilon_{g\gamma},\Upsilon_{g\kappa}$
           for the LOWZ sample, using $r_0=10\mpch$ and $20<r_p<70\mpch$. $b_g$ is the linear
         galaxy bias and $b_\gamma$ is the relative calibration bias between galaxy lensing and CMB lensing. We use
         broad uniform priors:
         $b_g>0$, $b_\gamma>0$, and $\Omega_m>0$. Solid (dashed) contours show $1\sigma$ ($2\sigma$)
         limits. Cyan lines show the fiducial values: $b_\gamma=1$, $\Omega_m=0.309$ and
         $b_g=1.74$ ($b_g$ value is obtained with fixed cosmology jackknife best fit and no RSD correction applied).
 }
         \label{fig:mcmc}
      \end{figure}

      In Fig.~\ref{fig:mcmc}, we show the results from jointly fitting the galaxy clustering, CMB lensing and galaxy
      lensing
      signals for the LOWZ sample using the MCMC fitting method. We fit for linear galaxy bias $b_g$, $\Omega_m$ and
      $b_\gamma$, where
      $b_\gamma$ is the relative calibration bias between CMB lensing and galaxy lensing (CMB
      lensing amplitude $\propto b_g$, and galaxy lensing amplitude
      $\propto b_g b_\gamma$). Our result of $b_\gamma=0.824\pm0.15$ is consistent with $1$, which
      would imply no difference in calibration between the two lensing methods,
      at $\sim1\sigma$ level. Note that the $b_g$ value from MCMC fits, $b_g=1.80\pm0.06$ differs from the jackknife
      fits, $b_g=1.73\pm0.04$,
	  shown earlier for two main reasons: the lower value of $\Omega_m$ (fixed to $0.309$) in
      jackknife fits; and
	  we do not use the RSD corrections in the MCMC fits to speed up computation time, moving $b_g$ higher by
	  $\lesssim1\sigma$ ($b_g=1.74\pm0.04$ for jackknife fitting without RSD correction).
	  The RSD correction is $\lesssim5\%$
	  at the scales we use \citep{Baldauf2010}, which is much less than the statistical uncertainties in both
	  $\Omega_m$
	  and $b_\gamma$. A more detailed cosmological analysis using these measurements with improved modeling on small
	  scales will be presented in a forthcoming paper (Singh et al.\ {\em in prep}).

	\subsection{Cosmography}
      \begin{figure}
         \centering
         \includegraphics[width=\columnwidth]{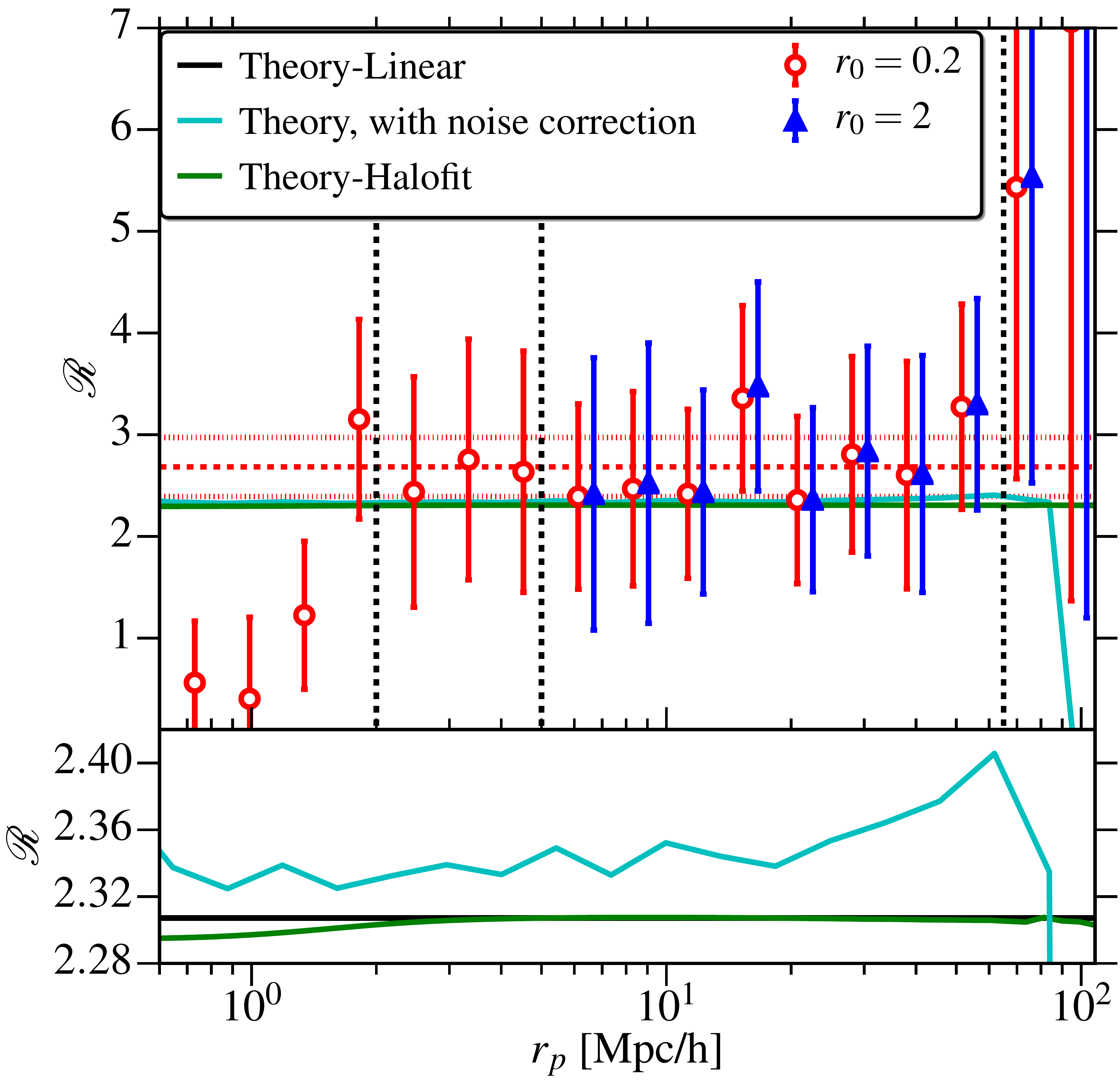}
         \caption{Distance ratio $\mathcal R$ (see Eq.~\eqref{eq:R}), measured using LOWZ galaxies. Dashed red line
         marks the mean $\mathcal R$ between $2<r_p<70\mpch$, while dotted red lines mark the $1\sigma$ limits on the
         mean. 
         \referee{Bottom panel shows the zoomed in comparison of different theory curves. 
         Solid black line is the predicted $\mathcal R$ using linear theory with Planck 2015 cosmology while solid 
         cyan line is the prediction using linear linear with the effects of observed measurement noise also included 
      	(using $r_0=0.2\mpch$ measurement). 
      	The green line is the predicted $\mathcal R$ using linear theory with halofit using Planck 2015 
         cosmology, accounting for the fact that redshift weighting will emphasize lower redshifts
         and hence suppress the predicted $\mathcal R$ on small scales.}
         }
         \label{fig:lowz_dist_ratio}
      \end{figure}
		In Fig.~\ref{fig:lowz_dist_ratio}, we present the measurement of the distance ratio $\mathcal R$ as defined in
		Eq.~\eqref{eq:R}. We present the measurement using two different values of $r_0$: $0.2$ and
		$2\mpch$. As was discussed in Sec.~\ref{ssec:theory_cosmography}, it is desirable to use $\upsilon_t$
		in estimating $\mathcal R$, to avoid the information from scales $r_p<2\mpch$ where the
        smoothing of the CMB lensing map is important.
		Using $r_0=2\mpch$ accomplishes this goal, while the $r_0=0.2\mpch$ case is equivalent to taking ratios using 
		$\gamma_t$
		when using scales $r_p>2\gg0.2\mpch$. The sudden drop in measured $\mathcal R$ below $r_p<2\mpch$
        when using $r_0=0.2\mpch$ is consistent with
		expectations from the effects of smoothing. 
		\referee{As discussed in Sec.~\ref{ssec:theory_cosmography}, the effects of non-linear growth also lead to lower 
		$\mathcal R$ on small scale by giving higher weights to lower redshifts (where $\mathcal R$
        is lower), though this is effect is estimated to be  
		much smaller than the statistical uncertainties in our measurements ($\sim1-2\%$ at $r_p<2\mpch$).}
		Using the ``aggressive'' range, $2<r_p<70\mpch$ with $r_0=0.2\mpch$,
		the mean value of $\mathcal R$ is $2.68\pm0.29$, consistent with the
		predicted value of $\mathcal R=2.31$ from the Planck 2015 cosmology.
      Including the effects of noise in our theory prediction as described in
      Sec.~\ref{ssec:theory_cosmography}, the prediction becomes $\langle\tilde R\rangle=2.35$ using $>1000$ realizations with noise and scales from $r_0=0.2\mpch$ and $2<r_p<70\mpch$.
      Using a more
      conservative range of scales, $5<r_p<70\mpch$
		with $r_0=2\mpch$, we get $\mathcal R=2.74\pm0.44$, with the prediction $\langle\tilde R\rangle=2.37$.
      However, given the strong agreement between our results with conservative and aggressive choices of $r_0$ and
      scales for the measurement, we quote $2.68\pm0.29$ as our primary result.
		We do not use $r_p>70\mpch$ as these scales are larger than the
		size of jackknife regions and hence the covariance matrix is not very reliable at these scales.

		We do not derive any cosmological
		constraints using $\mathcal R$ since it is not very sensitive to cosmological parameters given the redshift
		distribution of our lens and source redshifts \citep{Hu2007}.

   \subsection{Lensing-lensing correlations}\label{ssec:results_sk}
		In this section we present the results of cross-correlating the Planck lensing map with the galaxy shear from
		the SDSS shape sample \citep[][]{Reyes2012,Nakajima2012}, using the estimator presented in Sec.~\ref{subsec:ll-theory}.

\begin{figure}
         \centering
         \includegraphics[width=\columnwidth]{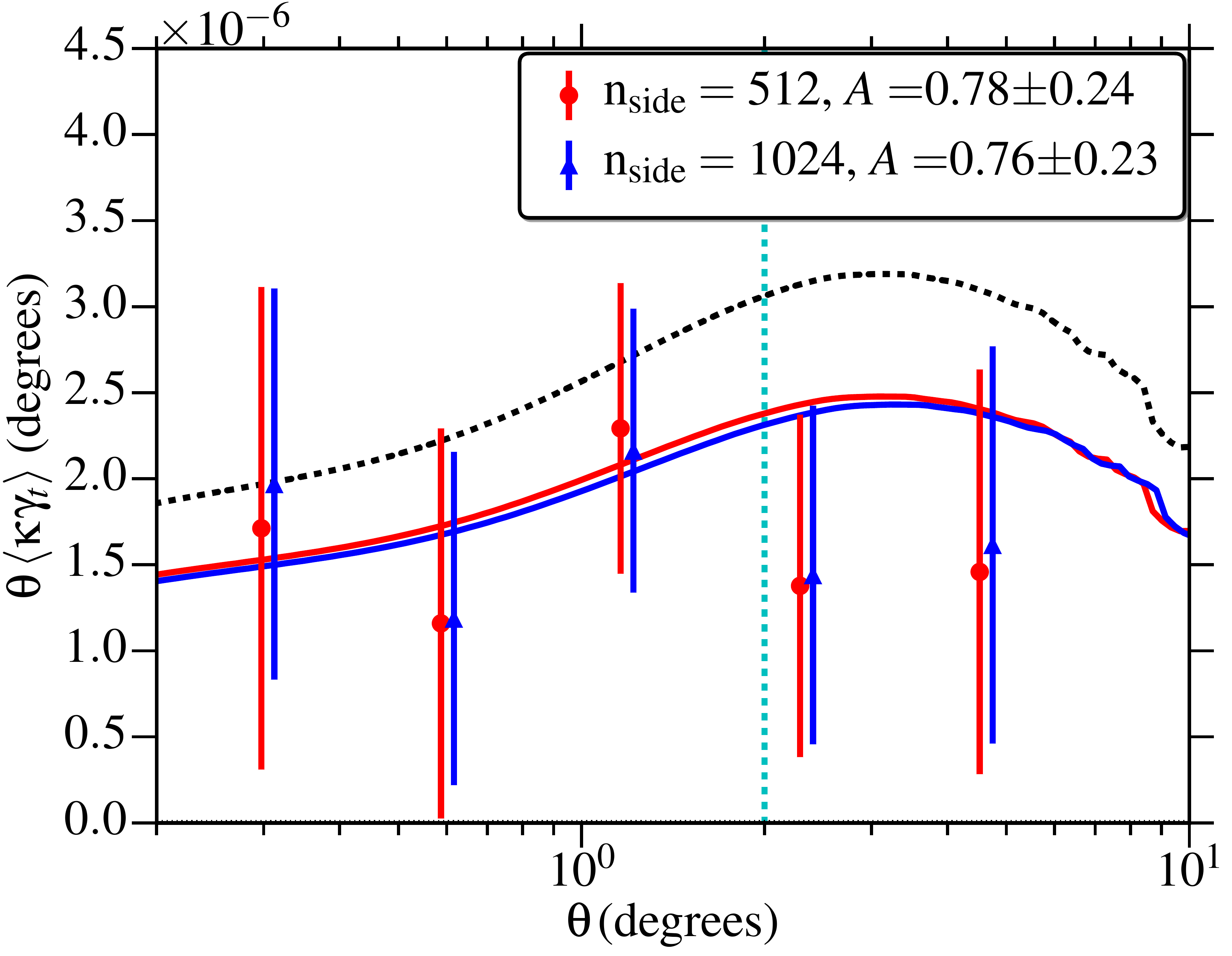}
         \caption{ The cross-correlation between the Planck CMB lensing convergence ($\kappa$) map
           and the SDSS shears. The solid
         	lines show the Planck $\Lambda$CDM model obtained using the $\mathrm{d}n/\mathrm{d}z$ from \protect\cite{Nakajima2012}, with best fit amplitude
             and fitting scale
			$\theta<2^\circ$  (marked by a vertical line). The dashed black line shows the model with amplitude $A=1$.
      		}
         \label{fig:kappa_shear}
      \end{figure}

		In Fig.~\ref{fig:kappa_shear}, we present the lensing-lensing cross-correlations with two different choices of
		pixel sizes. We fit the signal to a constant $A$ times the predictions for the Planck 2015
        cosmology, for $\theta<2^\circ$ (to avoid scales where
		noise starts dominating).  The best-fitting amplitudes, $A=0.78\pm0.24$ (\nside=512) and $A=0.76\pm0.23$ (\nside=1024), which are consistent
        with the Planck 2015 cosmological parameters ($A=1$) at
        $1\sigma$ level,
        for both pixel sizes.
        If we relax the fit limits to $\theta<5^\circ$, the amplitude decreases to $0.63\pm0.18$, which is in tension but still consistent with the Planck 2015 cosmological parameters at $2\sigma$ level.
        The shift between the fits using the more conservative and
        aggressive ranges of $\theta$ is less than $1\sigma$ after accounting for the correlations between
        the $A$ values for these two cases, and may be an effect of large-scale systematics
        at large scales. In the lensing-lensing
        cross-correlations, $\gamma_t$ is (in principle) also a quantity with zero mean and hence the estimator is
        less prone to the effects of correlated noise. However, $\gamma_t$ also has some residual
        additive systematics at large scales \citep{Mandelbaum2013}, making it a quantity with
        non-zero mean on large scales; it can therefore combine with the
        correlated noise \refereeTwo{(or systematics)} in the CMB convergence maps to give some residual systematics in 
        the cross-correlations.
 		\footnote{\refereeTwo{It is known that shear has systematics, so that measured shear, 
		 $\widehat{\gamma}=\gamma+\gamma_\text{noise}+\gamma_\text{sys}$. The measured cross-correlation with galaxies 
		 is then $\langle g\widehat\gamma_t\rangle=\langle (1+\delta_g)\widehat\gamma_t\rangle\approx\langle \delta_g
		 \gamma_t\rangle+\langle\gamma_\text{sys}\rangle$, where 
		 $\langle\gamma_\text{sys}\rangle$ is the systematics term in galaxy-galaxy lensing and is removed by 
		 subtracting measurement around randoms. Now, lets consider simple case in which CMB convergence also has 
		 some additive systematic, so that 
		 $\widehat{\kappa}=\kappa+\kappa_{noise}+\kappa_{sys}$, where we assume $\kappa_{sys}$ is constant and same in 
		 all pixels. In galaxy-CMB lensing, this systematic also gets removed when measurement around randoms is 
		 subtracted. The correlation with shear is then, $\langle \widehat\kappa\widehat\gamma_t\rangle=\langle \kappa
		 \gamma_t\rangle+\kappa_{sys}\langle\gamma_{sys}\rangle$. The $\kappa_{sys}\langle\gamma_{sys}\rangle$ can bias 
		 the shear-convergence cross correlation measurements.
 }}
        As discussed later in this section, we do not find any evidence of systematics in our null
        tests, though the uncertainties in our measurements are large. In this work we do not
        attempt to construct a better estimator given the noise in our measurements. In future
        works, with better signal-to-noise, it will be worth exploring a better estimator that removes the effects of residual additive systematics.

\begin{figure}
	         \centering
    	     \includegraphics[width=\columnwidth]{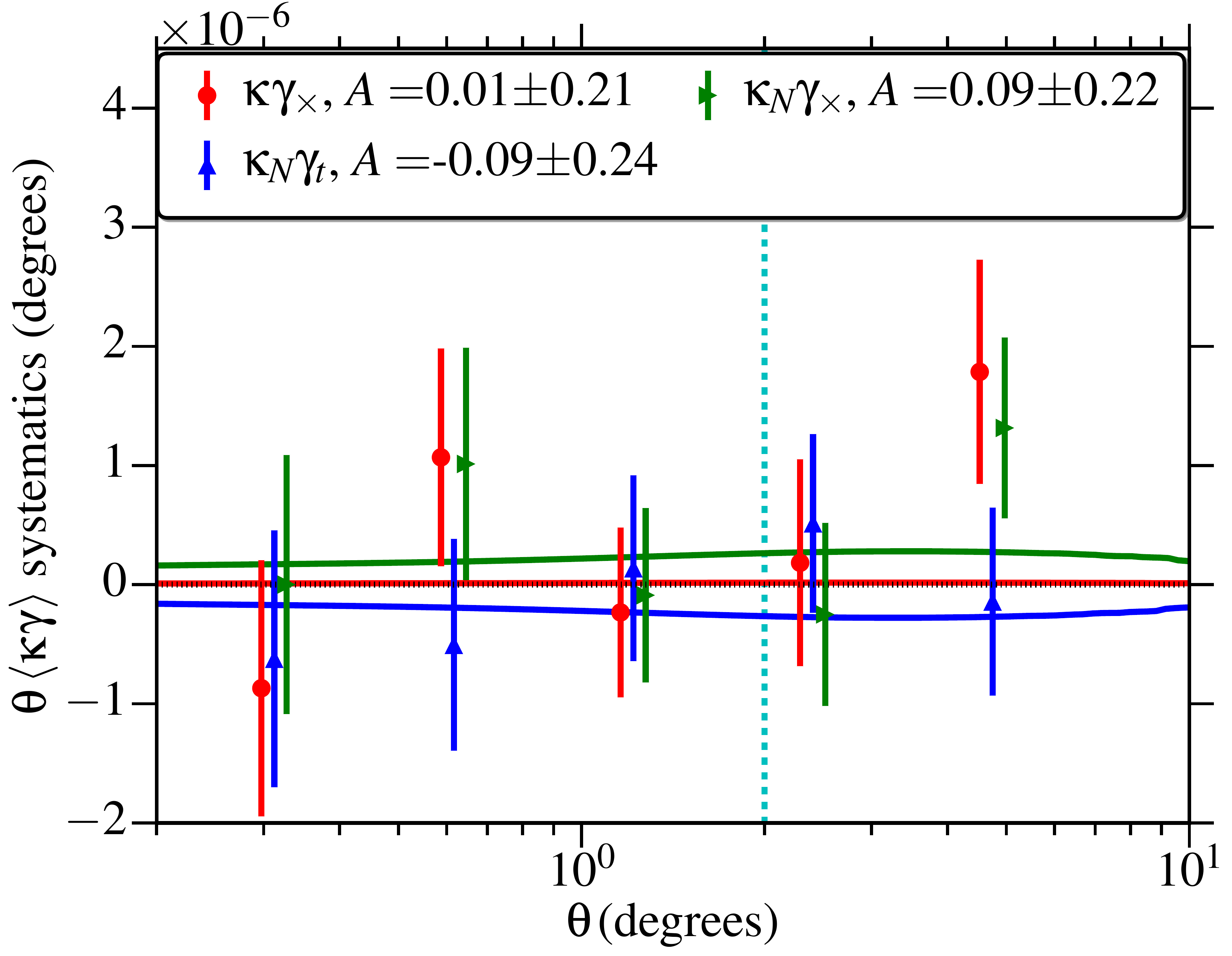}
        	 \caption{ Null tests for the lensing-lensing cross-correlations. Points show the measurements (expected to be consistent with zero) while solid lines
	 			show the $\langle\kappa\gamma_t\rangle$ model fit to these with a free overall
                amplitude $A$, for which the best-fitting value is given in the legend. The vertical cyan line marks $\theta=2^\circ$, the largest scale use in fitting the data.
            All the null
				tests we perform are consistent with zero at the $1\sigma$ level.
		      }
        	 \label{fig:kappa_shear_sys}
	     \end{figure}

      Fig.~\ref{fig:kappa_shear_sys} shows several null tests used to uncover the effects of
      systematic errors. The first is the `B-mode' signal
		$\langle\kappa\gamma_\times\rangle$, which is expected to be zero from parity conservation. As a test, we
        fit $\langle\kappa\gamma_\times\rangle$ to a model consisting of a constant $A$ times the
		prediction for $\langle\kappa\gamma_t\rangle$ from the Planck 2015 cosmology; this gives $A=0.01\pm 0.21$,
        consistent with $0$ as expected. Similarly, we repeat the measurement by replacing the
        Planck convergence with the noise map to compute $\langle\kappa\gamma_t\rangle$ and $\langle
       \kappa\gamma_\times\rangle$, and using the
		shuffled convergence map (not shown, $A=0.3\pm0.4$). All of these measurements give $A$ consistent with zero.

        While the deviations from the \lcdm\ predictions using the Planck 2015 cosmology are not
        statistically significant, it is worth noting that there are several possible systematics that could
        bias the amplitude of this cross-correlation, for example, intrinsic alignments (IA) and biases in the redshift
	     distributions.
        \cite{Chisari2015} estimated the contribution from IA contamination in the CMB vs.\ galaxy lensing
        cross-correlations to be around $\sim10\%$
        for the CFHT stripe 82 survey. Since the SDSS source sample is shallower than the stripe 82
        sample and hence at lower effective redshift, the fractional IA contamination can be
        higher. This statement assumes the validity of the linear alignment model
        \citep{Hirata2003}, which has been shown to describe the IA of red galaxies, and which
        predicts that the IA signal is approximately constant with redshift while the lensing
        cross-correlation signal with the CMB decreases at lower redshift. The IA contamination also
        depends on the galaxy luminosity (fainter galaxies have lower IA) and on the fraction of red
        galaxies in the sample (here we ignore alignments for spiral galaxies since all current
        measurements of IA for spirals are consistent with zero). Assuming a sample with $\sim20\%$
        galaxies to be LOWZ-like LRGs (effective IA amplitude $A_I=1$), we predict a contamination of
        $\sim -30\%$. This is a conservative upper limit considering that our source sample is much
        fainter than the LOWZ sample and is dominated by blue galaxies. See also \cite{Blazek2012}
        who constrained IA contamination in galaxy-galaxy lensing measurements to be less than 5\%
        using the same source sample as this work.

        Regarding photo-$z$ systematics and redshift
        uncertainties, their impact on galaxy-galaxy lensing for this shear catalogue was quantified using a complete and representative spectroscopic
        sample in \cite{Nakajima2012}, who found
         $\sim 2$ per cent uncertainties in the mean $\Sigma_c$ and therefore lensing amplitude.  To
         check whether the difference may be more significant here, we considered the difference in
         the best-fitting amplitude $A$ when we make the theoretical predictions using the redshifts
         from the calibration sample from \cite{Nakajima2012}, vs.\ when we make them from a smooth
         parametric fit to the histogram, and find $\sim 1$ per cent uncertainties.  We therefore
         conclude that redshift uncertainty is a subdominant contributor to the error budget for the
         lensing-lensing correlations.  The same may not be true in future datasets for which a
         representative spectroscopic sample is not available.

\section{Conclusions}\label{sec:conclusions}
	In this paper we have presented results from cross-correlating Planck CMB lensing maps with shear from SDSS galaxy
	lensing and galaxy positions from the SDSS-III BOSS survey using both the LOWZ and higher redshift CMASS sample.

	Cross-correlating galaxy positions with the convergence maps, we detect the CMB lensing signal around galaxies out to
	$\sim100\mpch$. The measured signals are consistent with \lcdm\ predictions using the Planck 2015 cosmology with bias measurements from clustering.
	Our null tests do not reveal any significant evidence for systematics in our
	measurements. The mild tensions between data and theory, e.g., $\cgm=0.78\pm0.13$ for CMASS
    sample, are likely due to noise fluctuations, particularly given that there is no tension
    observed for the LOWZ sample.

	We also detected the CMB lensing signal around galaxies at very small separations, well below
	the effective smoothing scale of 6$'$ in the convergence maps. Combining the clustering measurements with NFW
	profiles and then applying the smoothing kernel, we are able to constrain halo mass at the 3--4$\sigma$ level
	for different samples, though the halo masses could be biased given the simple adopted model.

	We directly compared the lensing signal around LOWZ galaxies obtained from galaxy-lensing and
    CMB-lensing.
    We find that the galaxy lensing has a better signal-to-noise ratio by a factor of 2--5, depending on the scale. Combining
	these measurements with the galaxy clustering signal, we also performed a basic cosmological
    analysis jointly fitting for $\Omega_m$, galaxy bias, and the relative calibration bias between
    galaxy and CMB lensing. We find
	$\Omega_m=0.286\pm0.024$, consistent with the Planck 2015 cosmology at the $1\sigma$ level. We
    find the relative calibration bias between galaxy lensing and CMB lensing to be
	$0.82\pm0.15$, consistent with 1 at just over $1\sigma$. In addition, we also measured the distance ratio between the lens
	and source
	galaxies and the CMB last scattering surface to within $\sim 10$\%. The ratio is consistent with the \lcdm\ prediction using Planck 2015
	cosmology; unfortunately, this ratio is not strongly sensitive to cosmology given the low lens
    redshift, and hence does not provide competitive cosmological constraints.

	For lensing-lensing cross-correlations, we detected the signal at $>3\sigma$ significance at an
    effective redshift of 0.35.
	The amplitude of the signal is consistent with \lcdm\ model predictions using Planck 2015
    cosmology.
	Given the noise in this measurement, we expect systematic errors such as
	intrinsic alignments and uncertainties in the source galaxy redshift distribution to be
    subdominant components of the error budget. We also
	performed null tests, which did not show any evidence for systematics within the errorbars.

	To conclude, our results demonstrate how CMB lensing data can be incorporated into and combined
    with the galaxy lensing
	analysis using existing lensing surveys. Even though the CMB lensing measurements are noisier
    than galaxy lensing and will perhaps remain so for the near future, existing CMB lensing measurements are already good enough to provide strong consistency checks
	on galaxy lensing measurements.  This analysis is an important proof of concept for future surveys that plan to use CMB lensing
    in conjunction with galaxy lensing, as an additional high-redshift lens plane with completely
    independent systematics.   With better resolution in the upcoming lensing results from current generation and
	Stage IV CMB surveys, CMB lensing can also develop into a unique tool to study dark matter at higher redshifts, to
	which it is most sensitive, and which will remain beyond the reach of currently planned galaxy lensing surveys.

\section*{Acknowledgments}
We thank Fran\c{c}ois Lanusse, Anthony Pullen, Alex Geringer-Sameth, S\'ebastien Fromenteau and Shirley Ho for useful discussions related to this
work. We also thank Uro\v{s} Seljak, Emmanuel Schaan, \referee{David Spergel and the anonymous referee} for helpful feedback on this work.
We thank Martin White and Beth Reid for providing us halo catalog from simulations.
{We also thank the SDSS-I/II/III and Planck collaboration for their efforts in providing the datasets used in this work. }

RM acknowledges the support of the Department of Energy Early Career Award program.
SS acknowledges support from John Peoples Jr.\ Presidential Fellowship from Carnegie Mellon University.

Some of the results in this paper have been derived using the HEALPix package \citep{Gorski2005}.

Funding for SDSS-III has been provided by the Alfred P. Sloan Foundation, the Participating Institutions, the National Science Foundation, and the U.S. Department of Energy Office of Science. The SDSS-III web site is http://www.SDSS3.org/.

SDSS-III is managed by the Astrophysical Research Consortium for the Participating Institutions of the SDSS-III Collaboration including the University of Arizona, the Brazilian Participation Group, Brookhaven National Laboratory, Carnegie Mellon University, University of Florida, the French Participation Group, the German Participation Group, Harvard University, the Instituto de Astrofisica de Canarias, the Michigan State/Notre Dame/JINA Participation Group, Johns Hopkins University, Lawrence Berkeley National Laboratory, Max Planck Institute for Astrophysics, Max Planck Institute for Extraterrestrial Physics, New Mexico State University, New York University, Ohio State University, Pennsylvania State University, University of Portsmouth, Princeton University, the Spanish Participation Group, University of Tokyo, University of Utah, Vanderbilt University, University of Virginia, University of Washington, and Yale University.

\emph{Author Contributions}: SS and RM contributed to the analysis and writing the paper. JB is an SDSS-III BOSS architect who contributed to the development of the BOSS survey.
        \bibliographystyle{mnras}
        \bibliography{sukhdeep_cmb_paper,papers}

\appendix
	\section{Effect of weights in the CMASS sample}\label{app:sysweights}
		In this section we briefly discuss the effect of using weights in measurements involving the
        CMASS sample.
		 \begin{figure}
	         \centering
    	     \includegraphics[width=\columnwidth]{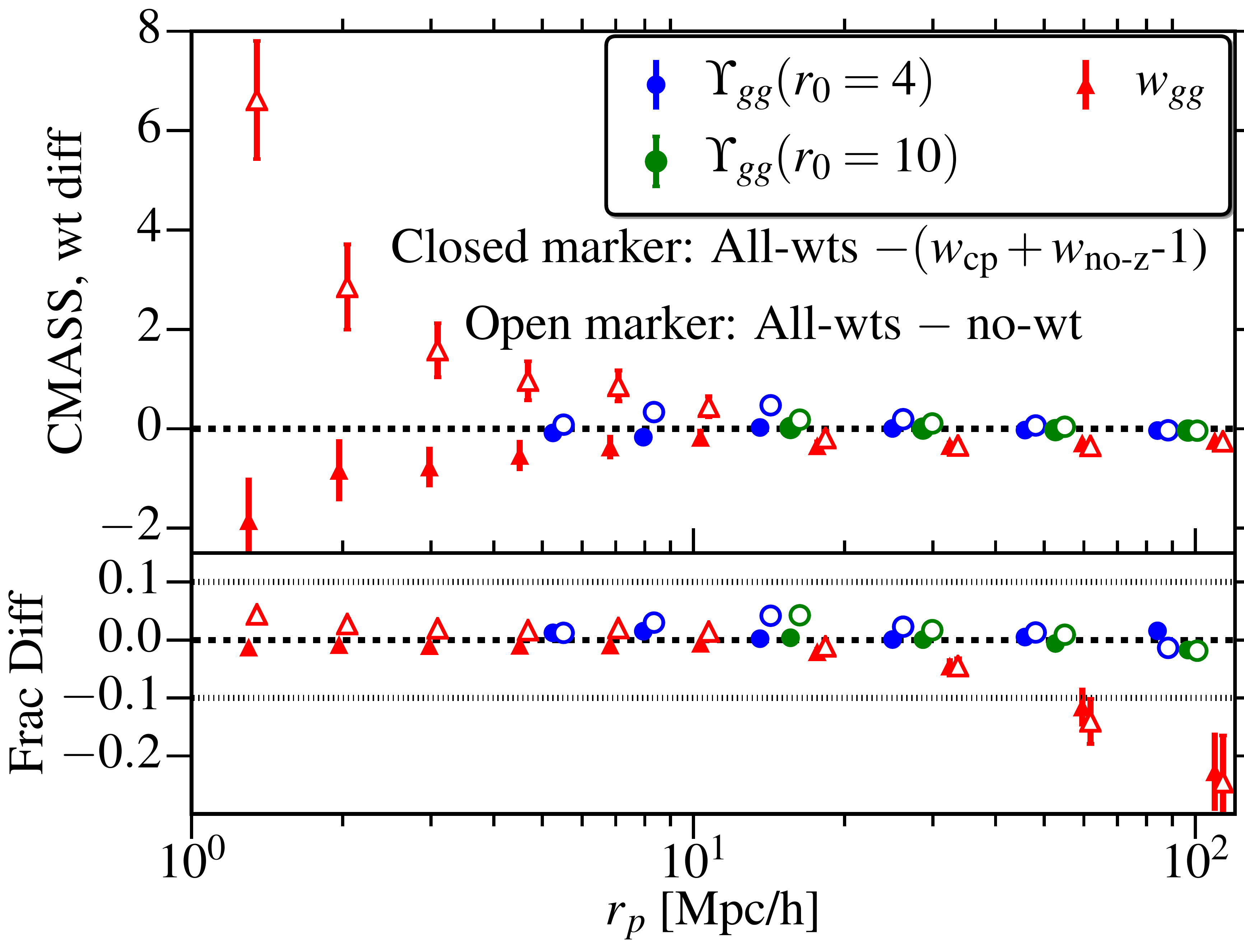}
        	 \caption{Effect of using weights in CMASS clustering measurement. In the top panel we show the difference between using all weights vs.\ using some or no weights.
            The bottom panel shows the fractional difference; the y-axis
				is $\frac{Y_\text{all-wt}-Y_\text{w}}{\langle Y_\text{all-wt} \rangle}$, where $Y_w$
                refers to calculations with no weights or with only incompleteness weights ($w_{cp}$
                and $w_{no-z}$). The systematic weights lower the clustering by an approximately
                scale independent additive factor, while the redshift incompleteness weights
                primarily change the clustering at small scales by up-weighting the higher density
                regions.
		      }
        	 \label{fig:cmass_weights}
	     \end{figure}
		Fig.~\ref{fig:cmass_weights} shows the difference in clustering measurements done with and without using
		weights. The effect of the systematics weights is to shift \wgg\ lower by an
		additive factor that has only a  weak scale dependence; this effect is the dominant impact
        of the weights on large scales. Since clustering varies very strongly with scale, the
		fractional change in clustering increases very strongly with scale. The weights for redshift
        failure up-weight the higher density regions and hence change the clustering strongly at
        small scales, while the change in the large-scale bias is small ($\lesssim1\sigma$).

		\ugg\ by definition will not be very strongly affected by additive changes in \wgg.
		Fig.~\ref{fig:cmass_weights} demonstrates the fact that the fractional change in \ugg\ from systematics weights is $<5\%$, even on scales
		where the fractional change in \wgg\ is $\ge 20\%$. At the scales that dominate our constraints in $b_g$
		($r_p\lesssim50\mpch$), the fractional change in \ugg\ without vs.\ with systematics weights
        is $<1\%$, below the statistical errors. As a result, our
		$b_g$ constraints from \ugg\ do not depend on the choice of whether or not to use
        systematics weights. The redshift failure weights change the clustering by a strongly
        scale-dependent factor, so \ugg\ is affected in nearly the same way as \wgg. However, the
        effect on scales we use for the fits for large-scale bias is small, so the change in linear galaxy bias is again $\lesssim1\sigma$.

		 \begin{figure}
	         \centering
    	     \includegraphics[width=\columnwidth]{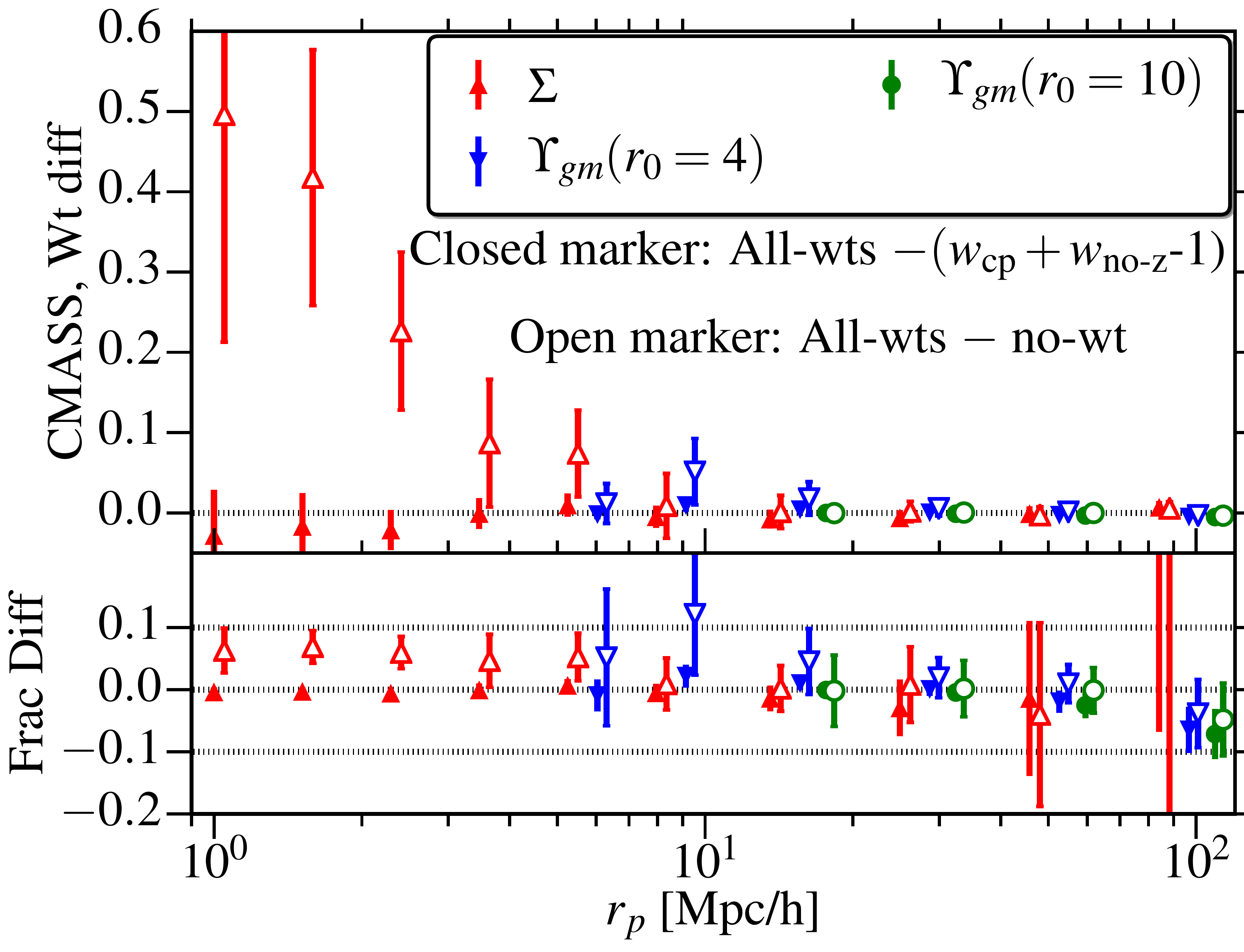}
        	 \caption{Effect of using systematic weights in CMASS lensing measurement using Planck CMB lensing map. y-
	 			axis definition is same as in fig.~\ref{fig:cmass_weights}. The systematics weights do not cause any significant change in lensing measurement, while redshift failure weights change signal primarily at small scales by up-weighting higher density regions.
		      }
        	 \label{fig:cmass_weights_lensing}
	     \end{figure}

		In Fig.~\ref{fig:cmass_weights_lensing} we show the effect of using weights on the lensing measurements with Planck
		convergence maps. Given the noise in the lensing measurements, the systematics weights do not affect the measurements very
		significantly and the fractional change in lensing is $<10\%$ at all scales, less than the statistical errors in
		the measurements. The redshift incompleteness weights change the small scale signal in a
        similar way as in the clustering measurement, by up-weighting the higher density regions,
        though the shift at large scales is small and negligible  given the larger  noise in the lensing measurements.

   \section{Comparison of different error estimates}\label{app:covariance_test}
      Here we present a brief comparison of errors in the CMB lensing measurements obtained using
      the jackknife method with the errors obtained using the scatter between 100 random realizations of the noise map
      $\kappa_N$.

      \begin{figure}
           \centering
          \includegraphics[width=\columnwidth]{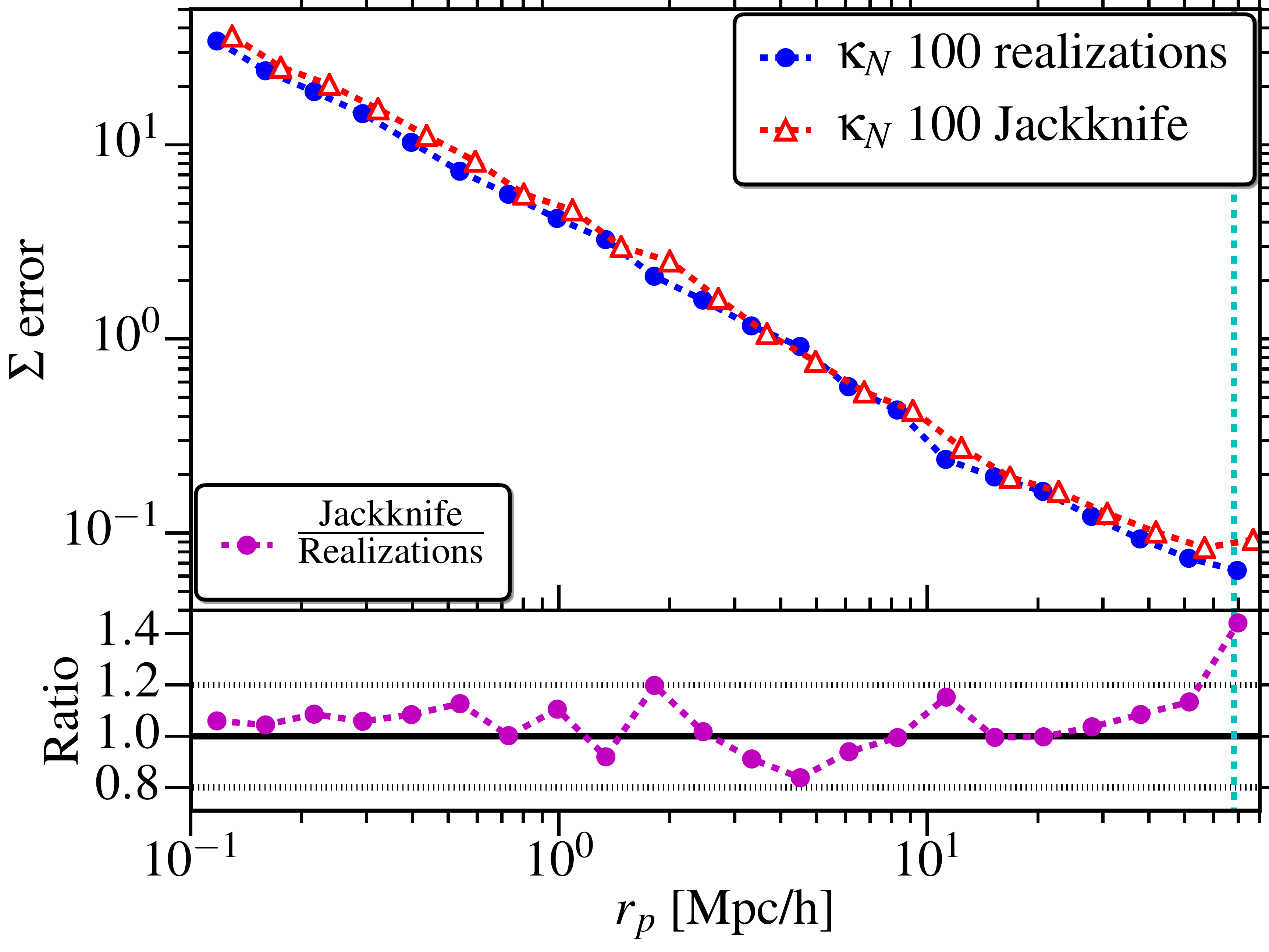}
         \caption{Comparison of errors (square root of diagonal of covariance matrix) in $\Sigma$
           measurements using simulated noise maps, estimated using 100 jackknife regions and
           separately using 100 independent realizations of the map. Up to $\sim20\%$ scatter in
           ratio is expected from the noise in estimation of errors.
           }
         \label{fig:error_comparison}
      \end{figure}
\begin{figure}
           \centering
          \includegraphics[width=\columnwidth]{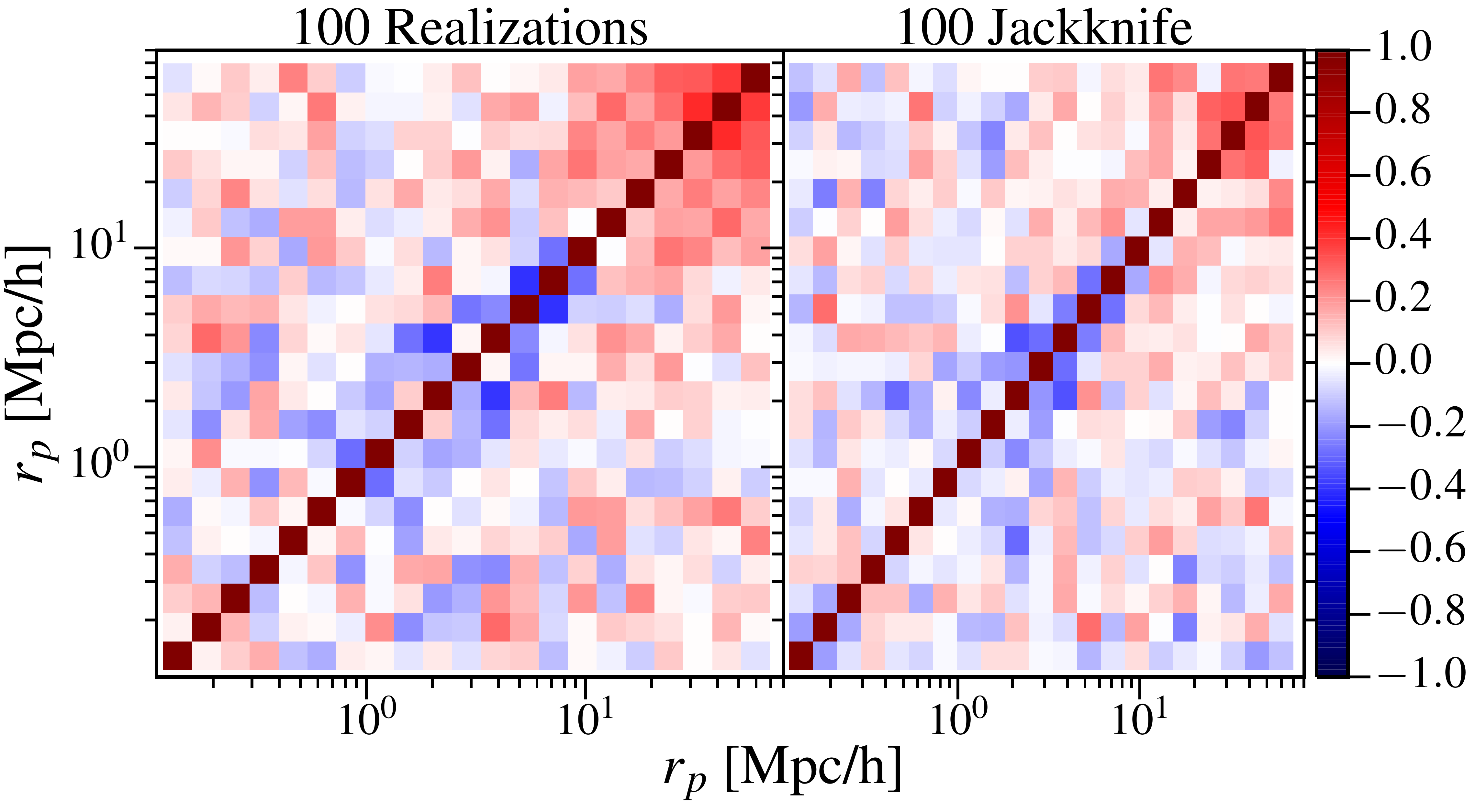}
         \caption{ Comparison of the correlation matrix estimated using the two methods.
           }
         \label{fig:corr_comparison}
      \end{figure}

      In Fig.~\ref{fig:error_comparison}, we show the comparison of the error bars (square root of
      diagonal elements of covariance matrix, $\delta\Sigma$) in
      $\Sigma$ from these two methods. The errors obtained using the two methods are
      consistent to within 20\% ($<10$\% on most scales), with the jackknife errors being larger on
      most scales. The relative uncertainty in the errors, $\delta(\delta\Sigma)/\delta\Sigma\sim
      \sqrt{2/99}\sim0.14$ \citep{Taylor2013},
      which predicts $\sim20\%$ scatter (assuming two estimates are independent) when taking the ratio of errors obtained using the two methods. Thus we can conclude that the
      errors obtained using the two methods are consistent.
      In
      Fig.~\ref{fig:corr_comparison}, we show the correlation matrix obtained from the two
      estimation methods; they are consistent within the noise.


      We also tested the covariance matrices by varying the number of random points. We find correlation
      matrices that are  consistent when using a number of random points $N_R=n\times N_G$, with
      $n=5,10,20$; the variation in the error on $\Sigma$ (square root of diagonal of covariance) is
      $\lesssim5\%$ between these cases. We use $N_R=10N_g$ for our main results.

   \end{document}

%% file: def.tex
\newcommand{\wgg}{\ensuremath{w_{gg}}}

\newcommand{\mpch}{\ensuremath{h^{-1}\text{Mpc}}}

\def\reff@jnl#1{{\rm#1\/}}

\def\aj{\reff@jnl{AJ}}                  
\def\araa{\reff@jnl{ARA\&A}}            
\def\apj{\reff@jnl{ApJ}}                        
\def\apjl{\reff@jnl{ApJ}}               
\def\apjs{\reff@jnl{ApJS}}              
\def\apss{\reff@jnl{Ap\&SS}}            
\def\aap{\reff@jnl{A\&A}}               
\def\aapr{\reff@jnl{A\&A~Rev.}}         
\def\aaps{\reff@jnl{A\&AS}}             
\def\baas{\reff@jnl{BAAS}}              
\def\jrasc{\reff@jnl{JRASC}}            
\def\memras{\reff@jnl{MmRAS}}           
\def\mnras{\reff@jnl{MNRAS}}            
\def\physrep{\reff@jnl{Phys.Rep.}}
\def\pra{\reff@jnl{Phys.Rev.A}}         
\def\prb{\reff@jnl{Phys.Rev.B}}         
\def\prc{\reff@jnl{Phys.Rev.C}}         
\def\prd{\reff@jnl{Phys.Rev.D}}         
\def\prl{\reff@jnl{Phys.Rev.Lett}}      
\def\pasp{\reff@jnl{PASP}}              
\def\pasj{\reff@jnl{PASJ}}              
\def\skytel{\reff@jnl{S\&T}}            
\def\solphys{\reff@jnl{Solar~Phys.}}    
\def\sovast{\reff@jnl{Soviet~Ast.}}     
\def\ssr{\reff@jnl{Space~Sci.Rev.}}     
\def\nat{\reff@jnl{Nature}}             

\newcommand{\Msun}{M_{\odot}}

\newcommand{\beq}{\begin{equation}}
\newcommand{\eeq}{\end{equation}}
\newcommand{\beqa}{\begin{eqnarray}}
\newcommand{\eeqa}{\end{eqnarray}}




%% file: sukhdeep_cmb_paper.bbl
\begin{thebibliography}{}
\makeatletter
\relax
\def\mn@urlcharsother{\let\do\@makeother \do\$\do\&\do\#\do\^\do\_\do\%\do\~}
\def\mn@doi{\begingroup\mn@urlcharsother \@ifnextchar [ {\mn@doi@}
  {\mn@doi@[]}}
\def\mn@doi@[#1]#2{\def\@tempa{#1}\ifx\@tempa\@empty \href
  {http://dx.doi.org/#2} {doi:#2}\else \href {http://dx.doi.org/#2} {#1}\fi
  \endgroup}
\def\mn@eprint#1#2{\mn@eprint@#1:#2::\@nil}
\def\mn@eprint@arXiv#1{\href {http://arxiv.org/abs/#1} {{\tt arXiv:#1}}}
\def\mn@eprint@dblp#1{\href {http://dblp.uni-trier.de/rec/bibtex/#1.xml}
  {dblp:#1}}
\def\mn@eprint@#1:#2:#3:#4\@nil{\def\@tempa {#1}\def\@tempb {#2}\def\@tempc
  {#3}\ifx \@tempc \@empty \let \@tempc \@tempb \let \@tempb \@tempa \fi \ifx
  \@tempb \@empty \def\@tempb {arXiv}\fi \@ifundefined
  {mn@eprint@\@tempb}{\@tempb:\@tempc}{\expandafter \expandafter \csname
  mn@eprint@\@tempb\endcsname \expandafter{\@tempc}}}

\bibitem[\protect\citeauthoryear{{Abazajian} et~al.,}{{Abazajian}
  et~al.}{2009}]{2009ApJS..182..543A}
{Abazajian} K.~N.,  et~al., 2009, \mn@doi [\apjs]
  {10.1088/0067-0049/182/2/543}, \href
  {http://adsabs.harvard.edu/abs/2009ApJS..182..543A} {182, 543}

\bibitem[\protect\citeauthoryear{{Ade} et~al.,}{{Ade} et~al.}{2014}]{Ade2014}
{Ade} P.~A.~R.,  et~al., 2014, \mn@doi [Physical Review Letters]
  {10.1103/PhysRevLett.113.021301}, \href
  {http://adsabs.harvard.edu/abs/2014PhRvL.113b1301A} {113, 021301}

\bibitem[\protect\citeauthoryear{{Ahn} et~al.,}{{Ahn} et~al.}{2012}]{Ahn:2012}
{Ahn} C.~P.,  et~al., 2012, \mn@doi [\apjs] {10.1088/0067-0049/203/2/21}, \href
  {http://adsabs.harvard.edu/abs/2012ApJS..203...21A} {203, 21}

\bibitem[\protect\citeauthoryear{{Aihara} et~al.,}{{Aihara}
  et~al.}{2011}]{2011ApJS..193...29A}
{Aihara} H.,  et~al., 2011, \mn@doi [\apjs] {10.1088/0067-0049/193/2/29}, \href
  {http://adsabs.harvard.edu/abs/2011ApJS..193...29A} {193, 29}

\bibitem[\protect\citeauthoryear{{Alam} et~al.,}{{Alam}
  et~al.}{2015}]{Alam2015}
{Alam} S.,  et~al., 2015, \mn@doi [\apjs] {10.1088/0067-0049/219/1/12}, \href
  {http://adsabs.harvard.edu/abs/2015ApJS..219...12A} {219, 12}

\bibitem[\protect\citeauthoryear{{Baldauf}, {Smith}, {Seljak}  \&
  {Mandelbaum}}{{Baldauf} et~al.}{2010}]{Baldauf2010}
{Baldauf} T.,  {Smith} R.~E.,  {Seljak} U.,   {Mandelbaum} R.,  2010, \mn@doi
  [\prd] {10.1103/PhysRevD.81.063531}, \href
  {http://adsabs.harvard.edu/abs/2010PhRvD..81f3531B} {81, 063531}

\bibitem[\protect\citeauthoryear{{Bartelmann} \& {Schneider}}{{Bartelmann} \&
  {Schneider}}{2001}]{Bartelmann2001}
{Bartelmann} M.,  {Schneider} P.,  2001, \mn@doi [\physrep]
  {10.1016/S0370-1573(00)00082-X}, \href
  {http://adsabs.harvard.edu/abs/2001PhR...340..291B} {340, 291}

\bibitem[\protect\citeauthoryear{{Baxter} et~al.,}{{Baxter}
  et~al.}{2015}]{Baxter2015}
{Baxter} E.~J.,  et~al., 2015, \mn@doi [\apj] {10.1088/0004-637X/806/2/247},
  \href {http://adsabs.harvard.edu/abs/2015ApJ...806..247B} {806, 247}

\bibitem[\protect\citeauthoryear{{Bernstein} \& {Jain}}{{Bernstein} \&
  {Jain}}{2004}]{Bernstein2004}
{Bernstein} G.,  {Jain} B.,  2004, \mn@doi [\apj] {10.1086/379768}, \href
  {http://adsabs.harvard.edu/abs/2004ApJ...600...17B} {600, 17}

\bibitem[\protect\citeauthoryear{{Bernstein} \& {Jarvis}}{{Bernstein} \&
  {Jarvis}}{2002}]{Bernstein2002}
{Bernstein} G.~M.,  {Jarvis} M.,  2002, \aj, \href
  {http://adsabs.harvard.edu/cgi-bin/nph-bib_query?bibcode=2002AJ....123..583B&db_key=AST}
  {123, 583}

\bibitem[\protect\citeauthoryear{{Bhattacharya}, {Habib}, {Heitmann}  \&
  {Vikhlinin}}{{Bhattacharya} et~al.}{2013}]{Bhattacharya2013}
{Bhattacharya} S.,  {Habib} S.,  {Heitmann} K.,   {Vikhlinin} A.,  2013,
  \mn@doi [\apj] {10.1088/0004-637X/766/1/32}, \href
  {http://adsabs.harvard.edu/abs/2013ApJ...766...32B} {766, 32}

\bibitem[\protect\citeauthoryear{{Bianchini} et~al.,}{{Bianchini}
  et~al.}{2015}]{Bianchini2015}
{Bianchini} F.,  et~al., 2015, preprint, \href
  {http://adsabs.harvard.edu/abs/2015arXiv151105116B} {} (\mn@eprint {arXiv}
  {1511.05116})

\bibitem[\protect\citeauthoryear{{Blanton}, {Lin}, {Lupton}, {Maley}, {Young},
  {Zehavi}  \& {Loveday}}{{Blanton} et~al.}{2003}]{Blanton:2003}
{Blanton} M.~R.,  {Lin} H.,  {Lupton} R.~H.,  {Maley} F.~M.,  {Young} N.,
  {Zehavi} I.,   {Loveday} J.,  2003, \mn@doi [\aj] {10.1086/344761}, \href
  {http://adsabs.harvard.edu/abs/2003AJ....125.2276B} {125, 2276}

\bibitem[\protect\citeauthoryear{{Blazek}, {Mandelbaum}, {Seljak}  \&
  {Nakajima}}{{Blazek} et~al.}{2012}]{Blazek2012}
{Blazek} J.,  {Mandelbaum} R.,  {Seljak} U.,   {Nakajima} R.,  2012, \mn@doi
  [\jcap] {10.1088/1475-7516/2012/05/041}, \href
  {http://adsabs.harvard.edu/abs/2012JCAP...05..041B} {5, 41}

\bibitem[\protect\citeauthoryear{{Bleem} et~al.,}{{Bleem}
  et~al.}{2012}]{Bleem2012}
{Bleem} L.~E.,  et~al., 2012, \mn@doi [\apjl] {10.1088/2041-8205/753/1/L9},
  \href {http://adsabs.harvard.edu/abs/2012ApJ...753L...9B} {753, L9}

\bibitem[\protect\citeauthoryear{{Bolton} et~al.,}{{Bolton}
  et~al.}{2012}]{Bolton:2012}
{Bolton} A.~S.,  et~al., 2012, \mn@doi [\aj] {10.1088/0004-6256/144/5/144},
  \href {http://adsabs.harvard.edu/abs/2012AJ....144..144B} {144, 144}

\bibitem[\protect\citeauthoryear{{Caminha} et~al.,}{{Caminha}
  et~al.}{2016}]{Caminha2016}
{Caminha} G.~B.,  et~al., 2016, \mn@doi [\aap] {10.1051/0004-6361/201527670},
  \href {http://adsabs.harvard.edu/abs/2016A%26A...587A..80C} {587, A80}

\bibitem[\protect\citeauthoryear{{Chisari}, {Dunkley}, {Miller}  \&
  {Allison}}{{Chisari} et~al.}{2015}]{Chisari2015}
{Chisari} N.~E.,  {Dunkley} J.,  {Miller} L.,   {Allison} R.,  2015, \mn@doi
  [\mnras] {10.1093/mnras/stv1655}, \href
  {http://adsabs.harvard.edu/abs/2015MNRAS.453..682C} {453, 682}

\bibitem[\protect\citeauthoryear{{Das} et~al.,}{{Das} et~al.}{2011}]{Das2011}
{Das} S.,  et~al., 2011, \mn@doi [Physical Review Letters]
  {10.1103/PhysRevLett.107.021301}, \href
  {http://adsabs.harvard.edu/abs/2011PhRvL.107b1301D} {107, 021301}

\bibitem[\protect\citeauthoryear{{Das}, {Errard}  \& {Spergel}}{{Das}
  et~al.}{2013}]{Das2013}
{Das} S.,  {Errard} J.,   {Spergel} D.,  2013, preprint, \href
  {http://adsabs.harvard.edu/abs/2013arXiv1311.2338D} {} (\mn@eprint {arXiv}
  {1311.2338})

\bibitem[\protect\citeauthoryear{{Das} et~al.,}{{Das} et~al.}{2014}]{Das2014}
{Das} S.,  et~al., 2014, \mn@doi [\jcap] {10.1088/1475-7516/2014/04/014}, \href
  {http://adsabs.harvard.edu/abs/2014JCAP...04..014D} {4, 014}

\bibitem[\protect\citeauthoryear{{Dawson} et~al.,}{{Dawson}
  et~al.}{2013}]{Dawson:2013}
{Dawson} K.~S.,  et~al., 2013, \mn@doi [\aj] {10.1088/0004-6256/145/1/10},
  \href {http://adsabs.harvard.edu/abs/2013AJ....145...10D} {145, 10}

\bibitem[\protect\citeauthoryear{{Diego} et~al.,}{{Diego}
  et~al.}{2015}]{Diego2015}
{Diego} J.~M.,  et~al., 2015, \mn@doi [\mnras] {10.1093/mnras/stu2064}, \href
  {http://adsabs.harvard.edu/abs/2015MNRAS.446..683D} {446, 683}

\bibitem[\protect\citeauthoryear{{Diemer} \& {Kravtsov}}{{Diemer} \&
  {Kravtsov}}{2015}]{Diemer2015}
{Diemer} B.,  {Kravtsov} A.~V.,  2015, \mn@doi [\apj]
  {10.1088/0004-637X/799/1/108}, \href
  {http://adsabs.harvard.edu/abs/2015ApJ...799..108D} {799, 108}

\bibitem[\protect\citeauthoryear{{Eisenstein} et~al.,}{{Eisenstein}
  et~al.}{2001}]{2001AJ....122.2267E}
{Eisenstein} D.~J.,  et~al., 2001, \aj, \href
  {http://adsabs.harvard.edu/cgi-bin/nph-bib_query?bibcode=2001AJ....122.2267E&amp;db_key=AST}
  {122, 2267}

\bibitem[\protect\citeauthoryear{{Erben} et~al.,}{{Erben}
  et~al.}{2013}]{Erben2013}
{Erben} T.,  et~al., 2013, \mn@doi [\mnras] {10.1093/mnras/stt928}, \href
  {http://adsabs.harvard.edu/abs/2013MNRAS.433.2545E} {433, 2545}

\bibitem[\protect\citeauthoryear{{Feldmann} et~al.,}{{Feldmann}
  et~al.}{2006}]{Feldman2006}
{Feldmann} R.,  et~al., 2006, \mn@doi [\mnras]
  {10.1111/j.1365-2966.2006.10930.x}, \href
  {http://adsabs.harvard.edu/abs/2006MNRAS.372..565F} {372, 565}

\bibitem[\protect\citeauthoryear{{Fukugita}, {Ichikawa}, {Gunn}, {Doi},
  {Shimasaku}  \& {Schneider}}{{Fukugita} et~al.}{1996}]{1996AJ....111.1748F}
{Fukugita} M.,  {Ichikawa} T.,  {Gunn} J.~E.,  {Doi} M.,  {Shimasaku} K.,
  {Schneider} D.~P.,  1996, \aj, \href
  {http://adsabs.harvard.edu/cgi-bin/nph-bib_query?bibcode=1996AJ....111.1748F&amp;db_key=AST}
  {111, 1748}

\bibitem[\protect\citeauthoryear{{Giannantonio} \& {Percival}}{{Giannantonio}
  \& {Percival}}{2014}]{Giannantonio2014}
{Giannantonio} T.,  {Percival} W.~J.,  2014, \mn@doi [\mnras]
  {10.1093/mnrasl/slu036}, \href
  {http://adsabs.harvard.edu/abs/2014MNRAS.441L..16G} {441, L16}

\bibitem[\protect\citeauthoryear{{Giannantonio} et~al.,}{{Giannantonio}
  et~al.}{2016}]{Giannantonio2016}
{Giannantonio} T.,  et~al., 2016, \mn@doi [\mnras] {10.1093/mnras/stv2678},
  \href {http://adsabs.harvard.edu/abs/2016MNRAS.456.3213G} {456, 3213}

\bibitem[\protect\citeauthoryear{{Golse}, {Kneib}  \& {Soucail}}{{Golse}
  et~al.}{2002}]{Golse2002}
{Golse} G.,  {Kneib} J.-P.,   {Soucail} G.,  2002, \mn@doi [\aap]
  {10.1051/0004-6361:20020448}, \href
  {http://adsabs.harvard.edu/abs/2002A%26A...387..788G} {387, 788}

\bibitem[\protect\citeauthoryear{{G{\'o}rski}, {Hivon}, {Banday}, {Wandelt},
  {Hansen}, {Reinecke}  \& {Bartelmann}}{{G{\'o}rski}
  et~al.}{2005}]{Gorski2005}
{G{\'o}rski} K.~M.,  {Hivon} E.,  {Banday} A.~J.,  {Wandelt} B.~D.,  {Hansen}
  F.~K.,  {Reinecke} M.,   {Bartelmann} M.,  2005, \mn@doi [\apj]
  {10.1086/427976}, \href {http://adsabs.harvard.edu/abs/2005ApJ...622..759G}
  {622, 759}

\bibitem[\protect\citeauthoryear{{Gunn} et~al.,}{{Gunn}
  et~al.}{1998}]{1998AJ....116.3040G}
{Gunn} J.~E.,  et~al., 1998, \aj, \href
  {http://adsabs.harvard.edu/cgi-bin/nph-bib_query?bibcode=1998AJ....116.3040G&db_key=AST}
  {116, 3040}

\bibitem[\protect\citeauthoryear{{Gunn} et~al.,}{{Gunn}
  et~al.}{2006}]{Gunn2006}
{Gunn} J.~E.,  et~al., 2006, \mn@doi [\aj] {10.1086/500975}, \href
  {http://adsabs.harvard.edu/abs/2006AJ....131.2332G} {131, 2332}

\bibitem[\protect\citeauthoryear{{Hand} et~al.,}{{Hand}
  et~al.}{2015}]{Hand2015}
{Hand} N.,  et~al., 2015, \mn@doi [\prd] {10.1103/PhysRevD.91.062001}, \href
  {http://adsabs.harvard.edu/abs/2015PhRvD..91f2001H} {91, 062001}

\bibitem[\protect\citeauthoryear{{Harnois-D{\'e}raps}
  et~al.,}{{Harnois-D{\'e}raps} et~al.}{2016}]{Harnois2016}
{Harnois-D{\'e}raps} J.,  et~al., 2016, \mn@doi [\mnras]
  {10.1093/mnras/stw947}, \href
  {http://adsabs.harvard.edu/abs/2016MNRAS.tmp..721H} {}

\bibitem[\protect\citeauthoryear{{Hirata} \& {Seljak}}{{Hirata} \&
  {Seljak}}{2003}]{Hirata2003}
{Hirata} C.,  {Seljak} U.,  2003, \mnras, \href
  {http://adsabs.harvard.edu/cgi-bin/nph-bib_query?bibcode=2003MNRAS.343..459H&db_key=AST}
  {343, 459}

\bibitem[\protect\citeauthoryear{{Hirata}, {Ho}, {Padmanabhan}, {Seljak}  \&
  {Bahcall}}{{Hirata} et~al.}{2008}]{Hirata2008}
{Hirata} C.~M.,  {Ho} S.,  {Padmanabhan} N.,  {Seljak} U.,   {Bahcall} N.~A.,
  2008, \mn@doi [\prd] {10.1103/PhysRevD.78.043520}, \href
  {http://adsabs.harvard.edu/abs/2008PhRvD..78d3520H} {78, 043520}

\bibitem[\protect\citeauthoryear{{Hogg}, {Finkbeiner}, {Schlegel}  \&
  {Gunn}}{{Hogg} et~al.}{2001}]{2001AJ....122.2129H}
{Hogg} D.~W.,  {Finkbeiner} D.~P.,  {Schlegel} D.~J.,   {Gunn} J.~E.,  2001,
  \aj, \href
  {http://adsabs.harvard.edu/cgi-bin/nph-bib_query?bibcode=2001AJ....122.2129H&amp;db_key=AST}
  {122, 2129}

\bibitem[\protect\citeauthoryear{{Hu} \& {Okamoto}}{{Hu} \&
  {Okamoto}}{2002}]{Hu2002}
{Hu} W.,  {Okamoto} T.,  2002, \mn@doi [\apj] {10.1086/341110}, \href
  {http://adsabs.harvard.edu/abs/2002ApJ...574..566H} {574, 566}

\bibitem[\protect\citeauthoryear{{Hu}, {DeDeo}  \& {Vale}}{{Hu}
  et~al.}{2007a}]{Hu2007a}
{Hu} W.,  {DeDeo} S.,   {Vale} C.,  2007a, \mn@doi [New Journal of Physics]
  {10.1088/1367-2630/9/12/441}, \href
  {http://adsabs.harvard.edu/abs/2007NJPh....9..441H} {9, 441}

\bibitem[\protect\citeauthoryear{{Hu}, {Holz}  \& {Vale}}{{Hu}
  et~al.}{2007b}]{Hu2007}
{Hu} W.,  {Holz} D.~E.,   {Vale} C.,  2007b, \mn@doi [\prd]
  {10.1103/PhysRevD.76.127301}, \href
  {http://adsabs.harvard.edu/abs/2007PhRvD..76l7301H} {76, 127301}

\bibitem[\protect\citeauthoryear{{Ivezi{\' c}} et~al.,}{{Ivezi{\' c}}
  et~al.}{2004}]{2004AN....325..583I}
{Ivezi{\' c}} {\v Z}.,  et~al., 2004, Astronomische Nachrichten, \href
  {http://adsabs.harvard.edu/cgi-bin/nph-bib_query?bibcode=2004AN....325..583I&db_key=AST}
  {325, 583}

\bibitem[\protect\citeauthoryear{{Jain} \& {Taylor}}{{Jain} \&
  {Taylor}}{2003}]{Jain2003}
{Jain} B.,  {Taylor} A.,  2003, \mn@doi [Physical Review Letters]
  {10.1103/PhysRevLett.91.141302}, \href
  {http://adsabs.harvard.edu/abs/2003PhRvL..91n1302J} {91, 141302}

\bibitem[\protect\citeauthoryear{{Kaiser}}{{Kaiser}}{1987}]{Kaiser1987}
{Kaiser} N.,  1987, \mnras, \href
  {http://adsabs.harvard.edu/abs/1987MNRAS.227....1K} {227, 1}

\bibitem[\protect\citeauthoryear{{Keck Array} et~al.,}{{Keck Array}
  et~al.}{2016}]{Keck_array2016}
{Keck Array} T.,  et~al., 2016, preprint, \href
  {http://adsabs.harvard.edu/abs/2016arXiv160601968K} {} (\mn@eprint {arXiv}
  {1606.01968})

\bibitem[\protect\citeauthoryear{{Kirk} et~al.,}{{Kirk}
  et~al.}{2016}]{Kirk2016}
{Kirk} D.,  et~al., 2016, \mn@doi [\mnras] {10.1093/mnras/stw570}, \href
  {http://adsabs.harvard.edu/abs/2016MNRAS.459...21K} {459, 21}

\bibitem[\protect\citeauthoryear{{Kitching} et~al.,}{{Kitching}
  et~al.}{2015}]{Kitching2015}
{Kitching} T.~D.,  et~al., 2015, preprint, \href
  {http://adsabs.harvard.edu/abs/2015arXiv151203627K} {} (\mn@eprint {arXiv}
  {1512.03627})

\bibitem[\protect\citeauthoryear{{Landy} \& {Szalay}}{{Landy} \&
  {Szalay}}{1993}]{landy1993}
{Landy} S.~D.,  {Szalay} A.~S.,  1993, \mn@doi [\apj] {10.1086/172900}, \href
  {http://adsabs.harvard.edu/abs/1993ApJ...412...64L} {412, 64}

\bibitem[\protect\citeauthoryear{Lewis \& Bridle}{Lewis \&
  Bridle}{2002}]{Lewis2002}
Lewis A.,  Bridle S.,  2002, Phys. Rev., D66, 103511

\bibitem[\protect\citeauthoryear{{Lewis} \& {Challinor}}{{Lewis} \&
  {Challinor}}{2006}]{Lewis2006}
{Lewis} A.,  {Challinor} A.,  2006, \mn@doi [\physrep]
  {10.1016/j.physrep.2006.03.002}, \href
  {http://adsabs.harvard.edu/abs/2006PhR...429....1L} {429, 1}

\bibitem[\protect\citeauthoryear{{Link} \& {Pierce}}{{Link} \&
  {Pierce}}{1998}]{Link1998}
{Link} R.,  {Pierce} M.~J.,  1998, \mn@doi [\apj] {10.1086/305892}, \href
  {http://adsabs.harvard.edu/abs/1998ApJ...502...63L} {502, 63}

\bibitem[\protect\citeauthoryear{{Liu} \& {Hill}}{{Liu} \&
  {Hill}}{2015}]{Liu2015}
{Liu} J.,  {Hill} J.~C.,  2015, \mn@doi [\prd] {10.1103/PhysRevD.92.063517},
  \href {http://adsabs.harvard.edu/abs/2015PhRvD..92f3517L} {92, 063517}

\bibitem[\protect\citeauthoryear{{Liu}, {Ortiz-Vazquez}  \& {Hill}}{{Liu}
  et~al.}{2016}]{Liu2016}
{Liu} J.,  {Ortiz-Vazquez} A.,   {Hill} J.~C.,  2016, preprint, \href
  {http://adsabs.harvard.edu/abs/2016arXiv160105720L} {} (\mn@eprint {arXiv}
  {1601.05720})

\bibitem[\protect\citeauthoryear{{Lupton}, {Gunn}, {Ivezi{\'c}}, {Knapp}  \&
  {Kent}}{{Lupton} et~al.}{2001}]{Lupton2001}
{Lupton} R.,  {Gunn} J.~E.,  {Ivezi{\'c}} Z.,  {Knapp} G.~R.,   {Kent} S.,
  2001, in {Harnden} Jr. F.~R.,  {Primini} F.~A.,   {Payne} H.~E.,  eds,
  Astronomical Society of the Pacific Conference Series Vol. 238, Astronomical
  Data Analysis Software and Systems X. p.~269 (\mn@eprint {}
  {astro-ph/0101420})

\bibitem[\protect\citeauthoryear{{Madhavacheril} et~al.,}{{Madhavacheril}
  et~al.}{2015}]{Madhavacheril2015}
{Madhavacheril} M.,  et~al., 2015, \mn@doi [Physical Review Letters]
  {10.1103/PhysRevLett.114.151302}, \href
  {http://adsabs.harvard.edu/abs/2015PhRvL.114o1302M} {114, 151302}

\bibitem[\protect\citeauthoryear{{Mandelbaum} et~al.,}{{Mandelbaum}
  et~al.}{2005}]{Mandelbaum2005}
{Mandelbaum} R.,  et~al., 2005, \mn@doi [\mnras]
  {10.1111/j.1365-2966.2005.09282.x}, \href
  {http://adsabs.harvard.edu/abs/2005MNRAS.361.1287M} {361, 1287}

\bibitem[\protect\citeauthoryear{{Mandelbaum}, {Hirata}, {Ishak}, {Seljak}  \&
  {Brinkmann}}{{Mandelbaum} et~al.}{2006}]{Mandelbaum2006}
{Mandelbaum} R.,  {Hirata} C.~M.,  {Ishak} M.,  {Seljak} U.,   {Brinkmann} J.,
  2006, \mn@doi [\mnras] {10.1111/j.1365-2966.2005.09946.x}, \href
  {http://adsabs.harvard.edu/abs/2006MNRAS.367..611M} {367, 611}

\bibitem[\protect\citeauthoryear{{Mandelbaum} et~al.,}{{Mandelbaum}
  et~al.}{2011}]{Mandelbaum2011}
{Mandelbaum} R.,  et~al., 2011, \mn@doi [\mnras]
  {10.1111/j.1365-2966.2010.17485.x}, \href
  {http://adsabs.harvard.edu/abs/2011MNRAS.410..844M} {410, 844}

\bibitem[\protect\citeauthoryear{{Mandelbaum}, {Slosar}, {Baldauf}, {Seljak},
  {Hirata}, {Nakajima}, {Reyes}  \& {Smith}}{{Mandelbaum}
  et~al.}{2013}]{Mandelbaum2013}
{Mandelbaum} R.,  {Slosar} A.,  {Baldauf} T.,  {Seljak} U.,  {Hirata} C.~M.,
  {Nakajima} R.,  {Reyes} R.,   {Smith} R.~E.,  2013, \mn@doi [\mnras]
  {10.1093/mnras/stt572}, \href
  {http://adsabs.harvard.edu/abs/2013MNRAS.432.1544M} {432, 1544}

\bibitem[\protect\citeauthoryear{{Manera} et~al.,}{{Manera}
  et~al.}{2015}]{Manera2015}
{Manera} M.,  et~al., 2015, \mn@doi [\mnras] {10.1093/mnras/stu2465}, \href
  {http://adsabs.harvard.edu/abs/2015MNRAS.447..437M} {447, 437}

\bibitem[\protect\citeauthoryear{{Miyatake} et~al.,}{{Miyatake}
  et~al.}{2015}]{Miyatake2015}
{Miyatake} H.,  et~al., 2015, \mn@doi [\apj] {10.1088/0004-637X/806/1/1}, \href
  {http://adsabs.harvard.edu/abs/2015ApJ...806....1M} {806, 1}

\bibitem[\protect\citeauthoryear{{Miyatake}, {Madhavacheril}, {Sehgal},
  {Slosar}, {Spergel}, {Sherwin}  \& {van Engelen}}{{Miyatake}
  et~al.}{2016}]{Miyatake2016}
{Miyatake} H.,  {Madhavacheril} M.~S.,  {Sehgal} N.,  {Slosar} A.,  {Spergel}
  D.~N.,  {Sherwin} B.,   {van Engelen} A.,  2016, preprint, \href
  {http://adsabs.harvard.edu/abs/2016arXiv160505337M} {} (\mn@eprint {arXiv}
  {1605.05337})

\bibitem[\protect\citeauthoryear{{More}, {Miyatake}, {Mandelbaum}, {Takada},
  {Spergel}, {Brownstein}  \& {Schneider}}{{More} et~al.}{2015}]{More2015}
{More} S.,  {Miyatake} H.,  {Mandelbaum} R.,  {Takada} M.,  {Spergel} D.~N.,
  {Brownstein} J.~R.,   {Schneider} D.~P.,  2015, \mn@doi [\apj]
  {10.1088/0004-637X/806/1/2}, \href
  {http://adsabs.harvard.edu/abs/2015ApJ...806....2M} {806, 2}

\bibitem[\protect\citeauthoryear{{Nakajima}, {Mandelbaum}, {Seljak}, {Cohn},
  {Reyes}  \& {Cool}}{{Nakajima} et~al.}{2012}]{Nakajima2012}
{Nakajima} R.,  {Mandelbaum} R.,  {Seljak} U.,  {Cohn} J.~D.,  {Reyes} R.,
  {Cool} R.,  2012, \mn@doi [\mnras] {10.1111/j.1365-2966.2011.20249.x}, \href
  {http://adsabs.harvard.edu/abs/2012MNRAS.420.3240N} {420, 3240}

\bibitem[\protect\citeauthoryear{{Navarro}, {Frenk}  \& {White}}{{Navarro}
  et~al.}{1996}]{Navarro1996}
{Navarro} J.~F.,  {Frenk} C.~S.,   {White} S.~D.~M.,  1996, \mn@doi [\apj]
  {10.1086/177173}, \href {http://adsabs.harvard.edu/abs/1996ApJ...462..563N}
  {462, 563}

\bibitem[\protect\citeauthoryear{{Padmanabhan} et~al.,}{{Padmanabhan}
  et~al.}{2008}]{2008ApJ...674.1217P}
{Padmanabhan} N.,  et~al., 2008, \mn@doi [\apj] {10.1086/524677}, \href
  {http://adsabs.harvard.edu/abs/2008ApJ...674.1217P} {674, 1217}

\bibitem[\protect\citeauthoryear{{Pier}, {Munn}, {Hindsley}, {Hennessy},
  {Kent}, {Lupton}  \& {Ivezi{\' c}}}{{Pier}
  et~al.}{2003}]{2003AJ....125.1559P}
{Pier} J.~R.,  {Munn} J.~A.,  {Hindsley} R.~B.,  {Hennessy} G.~S.,  {Kent}
  S.~M.,  {Lupton} R.~H.,   {Ivezi{\' c}} {\v Z}.,  2003, \aj, \href
  {http://adsabs.harvard.edu/cgi-bin/nph-bib_query?bibcode=2003AJ....125.1559P&amp;db_key=AST}
  {125, 1559}

\bibitem[\protect\citeauthoryear{{Planck Collaboration} et~al.,}{{Planck
  Collaboration} et~al.}{2014}]{Planck2013lensing}
{Planck Collaboration} et~al., 2014, \mn@doi [\aap]
  {10.1051/0004-6361/201321543}, \href
  {http://adsabs.harvard.edu/abs/2014A%26A...571A..17P} {571, A17}

\bibitem[\protect\citeauthoryear{{Planck Collaboration} et~al.,}{{Planck
  Collaboration} et~al.}{2015c}]{Planck2015cosmo}
{Planck Collaboration} et~al., 2015c, preprint, \href
  {http://adsabs.harvard.edu/abs/2015arXiv150201589P} {} (\mn@eprint {arXiv}
  {1502.01589})

\bibitem[\protect\citeauthoryear{{Planck Collaboration} et~al.,}{{Planck
  Collaboration} et~al.}{2015a}]{Planck2015lensing}
{Planck Collaboration} et~al., 2015a, preprint, \href
  {http://adsabs.harvard.edu/abs/2015arXiv150201591P} {} (\mn@eprint {arXiv}
  {1502.01591})

\bibitem[\protect\citeauthoryear{{Planck Collaboration} et~al.,}{{Planck
  Collaboration} et~al.}{2015b}]{Planck2015Clusters}
{Planck Collaboration} et~al., 2015b, preprint, \href
  {http://adsabs.harvard.edu/abs/2015arXiv150201597P} {} (\mn@eprint {arXiv}
  {1502.01597})

\bibitem[\protect\citeauthoryear{{Pullen}, {Alam}, {He}  \& {Ho}}{{Pullen}
  et~al.}{2015}]{Pullen2015}
{Pullen} A.~R.,  {Alam} S.,  {He} S.,   {Ho} S.,  2015, preprint, \href
  {http://adsabs.harvard.edu/abs/2015arXiv151104457P} {} (\mn@eprint {arXiv}
  {1511.04457})

\bibitem[\protect\citeauthoryear{{Reid} \& {Spergel}}{{Reid} \&
  {Spergel}}{2009}]{Reid2009}
{Reid} B.~A.,  {Spergel} D.~N.,  2009, \mn@doi [\apj]
  {10.1088/0004-637X/698/1/143}, \href
  {http://adsabs.harvard.edu/abs/2009ApJ...698..143R} {698, 143}

\bibitem[\protect\citeauthoryear{{Reid}, {Seo}, {Leauthaud}, {Tinker}  \&
  {White}}{{Reid} et~al.}{2014}]{Reid2014}
{Reid} B.~A.,  {Seo} H.-J.,  {Leauthaud} A.,  {Tinker} J.~L.,   {White} M.,
  2014, \mn@doi [\mnras] {10.1093/mnras/stu1391}, \href
  {http://adsabs.harvard.edu/abs/2014MNRAS.444..476R} {444, 476}

\bibitem[\protect\citeauthoryear{{Reid} et~al.,}{{Reid}
  et~al.}{2016}]{Reid2016}
{Reid} B.,  et~al., 2016, \mn@doi [\mnras] {10.1093/mnras/stv2382}, \href
  {http://adsabs.harvard.edu/abs/2016MNRAS.455.1553R} {455, 1553}

\bibitem[\protect\citeauthoryear{{Reyes}, {Mandelbaum}, {Gunn}, {Nakajima},
  {Seljak}  \& {Hirata}}{{Reyes} et~al.}{2012}]{Reyes2012}
{Reyes} R.,  {Mandelbaum} R.,  {Gunn} J.~E.,  {Nakajima} R.,  {Seljak} U.,
  {Hirata} C.~M.,  2012, \mn@doi [\mnras] {10.1111/j.1365-2966.2012.21472.x},
  \href {http://adsabs.harvard.edu/abs/2012MNRAS.425.2610R} {425, 2610}

\bibitem[\protect\citeauthoryear{{Richards} et~al.,}{{Richards}
  et~al.}{2002}]{2002AJ....123.2945R}
{Richards} G.~T.,  et~al., 2002, \mn@doi [\aj] {10.1086/340187}, \href
  {http://adsabs.harvard.edu/abs/2002AJ....123.2945R} {123, 2945}

\bibitem[\protect\citeauthoryear{{Rodr{\'{\i}}guez-Torres}
  et~al.,}{{Rodr{\'{\i}}guez-Torres} et~al.}{2016}]{Torres2016}
{Rodr{\'{\i}}guez-Torres} S.~A.,  et~al., 2016, \mn@doi [\mnras]
  {10.1093/mnras/stw1014}, \href
  {http://adsabs.harvard.edu/abs/2016MNRAS.tmp..786R} {}

\bibitem[\protect\citeauthoryear{{Ross} et~al.,}{{Ross}
  et~al.}{2012}]{Ross2012}
{Ross} A.~J.,  et~al., 2012, \mn@doi [\mnras]
  {10.1111/j.1365-2966.2012.21235.x}, \href
  {http://adsabs.harvard.edu/abs/2012MNRAS.424..564R} {424, 564}

\bibitem[\protect\citeauthoryear{{Sheldon} et~al.,}{{Sheldon}
  et~al.}{2004}]{Sheldon2004}
{Sheldon} E.~S.,  et~al., 2004, \mn@doi [\aj] {10.1086/383293}, \href
  {http://adsabs.harvard.edu/abs/2004AJ....127.2544S} {127, 2544}

\bibitem[\protect\citeauthoryear{{Sherwin} et~al.,}{{Sherwin}
  et~al.}{2012}]{Sherwin2012}
{Sherwin} B.~D.,  et~al., 2012, \mn@doi [\prd] {10.1103/PhysRevD.86.083006},
  \href {http://adsabs.harvard.edu/abs/2012PhRvD..86h3006S} {86, 083006}

\bibitem[\protect\citeauthoryear{{Singh}, {Mandelbaum}  \& {More}}{{Singh}
  et~al.}{2015}]{Singh2015}
{Singh} S.,  {Mandelbaum} R.,   {More} S.,  2015, \mn@doi [\mnras]
  {10.1093/mnras/stv778}, \href
  {http://adsabs.harvard.edu/abs/2015MNRAS.450.2195S} {450, 2195}

\bibitem[\protect\citeauthoryear{{Smee} et~al.,}{{Smee}
  et~al.}{2013}]{Smee:2013}
{Smee} S.~A.,  et~al., 2013, \mn@doi [\aj] {10.1088/0004-6256/146/2/32}, \href
  {http://adsabs.harvard.edu/abs/2013AJ....146...32S} {146, 32}

\bibitem[\protect\citeauthoryear{{Smith} et~al.,}{{Smith}
  et~al.}{2002}]{2002AJ....123.2121S}
{Smith} J.~A.,  et~al., 2002, \aj, \href
  {http://adsabs.harvard.edu/cgi-bin/nph-bib_query?bibcode=2002AJ....123.2121S&amp;db_key=AST}
  {123, 2121}

\bibitem[\protect\citeauthoryear{{Smith} et~al.,}{{Smith}
  et~al.}{2003}]{Smith2003}
{Smith} R.~E.,  et~al., 2003, \mn@doi [\mnras]
  {10.1046/j.1365-8711.2003.06503.x}, \href
  {http://adsabs.harvard.edu/abs/2003MNRAS.341.1311S} {341, 1311}

\bibitem[\protect\citeauthoryear{{Smith}, {Zahn}  \& {Dor{\'e}}}{{Smith}
  et~al.}{2007}]{Smith2007}
{Smith} K.~M.,  {Zahn} O.,   {Dor{\'e}} O.,  2007, \mn@doi [\prd]
  {10.1103/PhysRevD.76.043510}, \href
  {http://adsabs.harvard.edu/abs/2007PhRvD..76d3510S} {76, 043510}

\bibitem[\protect\citeauthoryear{{Strauss} et~al.,}{{Strauss}
  et~al.}{2002}]{2002AJ....124.1810S}
{Strauss} M.~A.,  et~al., 2002, \aj, \href
  {http://adsabs.harvard.edu/cgi-bin/nph-bib_query?bibcode=2002AJ....124.1810S&db_key=AST}
  {124, 1810}

\bibitem[\protect\citeauthoryear{{Takahashi}, {Sato}, {Nishimichi}, {Taruya}
  \& {Oguri}}{{Takahashi} et~al.}{2012}]{Takahashi2012}
{Takahashi} R.,  {Sato} M.,  {Nishimichi} T.,  {Taruya} A.,   {Oguri} M.,
  2012, \mn@doi [\apj] {10.1088/0004-637X/761/2/152}, \href
  {http://adsabs.harvard.edu/abs/2012ApJ...761..152T} {761, 152}

\bibitem[\protect\citeauthoryear{{Taylor} et~al.,}{{Taylor}
  et~al.}{2012}]{Taylor2012}
{Taylor} J.~E.,  et~al., 2012, \mn@doi [\apj] {10.1088/0004-637X/749/2/127},
  \href {http://adsabs.harvard.edu/abs/2012ApJ...749..127T} {749, 127}

\bibitem[\protect\citeauthoryear{{Taylor}, {Joachimi}  \& {Kitching}}{{Taylor}
  et~al.}{2013}]{Taylor2013}
{Taylor} A.,  {Joachimi} B.,   {Kitching} T.,  2013, \mn@doi [\mnras]
  {10.1093/mnras/stt270}, \href
  {http://adsabs.harvard.edu/abs/2013MNRAS.432.1928T} {432, 1928}

\bibitem[\protect\citeauthoryear{{Tucker} et~al.,}{{Tucker}
  et~al.}{2006}]{2006AN....327..821T}
{Tucker} D.~L.,  et~al., 2006, \mn@doi [Astronomische Nachrichten]
  {10.1002/asna.200610655}, \href
  {http://adsabs.harvard.edu/abs/2006AN....327..821T} {327, 821}

\bibitem[\protect\citeauthoryear{{Vallinotto}}{{Vallinotto}}{2012}]{Vallinotto2012}
{Vallinotto} A.,  2012, \mn@doi [\apj] {10.1088/0004-637X/759/1/32}, \href
  {http://adsabs.harvard.edu/abs/2012ApJ...759...32V} {759, 32}

\bibitem[\protect\citeauthoryear{{Wake} et~al.,}{{Wake}
  et~al.}{2006}]{Wake2006}
{Wake} D.~A.,  et~al., 2006, \mn@doi [\mnras]
  {10.1111/j.1365-2966.2006.10831.x}, \href
  {http://adsabs.harvard.edu/abs/2006MNRAS.372..537W} {372, 537}

\bibitem[\protect\citeauthoryear{{Weinberg}, {Mortonson}, {Eisenstein},
  {Hirata}, {Riess}  \& {Rozo}}{{Weinberg} et~al.}{2013}]{Weinberg2013}
{Weinberg} D.~H.,  {Mortonson} M.~J.,  {Eisenstein} D.~J.,  {Hirata} C.,
  {Riess} A.~G.,   {Rozo} E.,  2013, \mn@doi [\physrep]
  {10.1016/j.physrep.2013.05.001}, \href
  {http://adsabs.harvard.edu/abs/2013PhR...530...87W} {530, 87}

\bibitem[\protect\citeauthoryear{{York} et~al.,}{{York}
  et~al.}{2000}]{2000AJ....120.1579Y}
{York} D.~G.,  et~al., 2000, \aj, \href
  {http://adsabs.harvard.edu/cgi-bin/nph-bib_query?bibcode=2000AJ....120.1579Y&db_key=AST}
  {120, 1579}

\bibitem[\protect\citeauthoryear{{Zaldarriaga} \& {Seljak}}{{Zaldarriaga} \&
  {Seljak}}{1999}]{Zaldarriaga1999}
{Zaldarriaga} M.,  {Seljak} U.,  1999, \mn@doi [\prd]
  {10.1103/PhysRevD.59.123507}, \href
  {http://adsabs.harvard.edu/abs/1999PhRvD..59l3507Z} {59, 123507}

\bibitem[\protect\citeauthoryear{{Zheng} et~al.,}{{Zheng}
  et~al.}{2005}]{Zheng2005}
{Zheng} Z.,  et~al., 2005, \mn@doi [\apj] {10.1086/466510}, \href
  {http://adsabs.harvard.edu/abs/2005ApJ...633..791Z} {633, 791}

\bibitem[\protect\citeauthoryear{{de Putter}, {Dor{\'e}}  \& {Das}}{{de Putter}
  et~al.}{2014}]{Putter2014}
{de Putter} R.,  {Dor{\'e}} O.,   {Das} S.,  2014, \mn@doi [\apj]
  {10.1088/0004-637X/780/2/185}, \href
  {http://adsabs.harvard.edu/abs/2014ApJ...780..185D} {780, 185}

\bibitem[\protect\citeauthoryear{{van Engelen} et~al.,}{{van Engelen}
  et~al.}{2012}]{Engelen2012}
{van Engelen} A.,  et~al., 2012, \mn@doi [\apj] {10.1088/0004-637X/756/2/142},
  \href {http://adsabs.harvard.edu/abs/2012ApJ...756..142V} {756, 142}

\bibitem[\protect\citeauthoryear{{van Engelen}, {Bhattacharya}, {Sehgal},
  {Holder}, {Zahn}  \& {Nagai}}{{van Engelen} et~al.}{2014}]{Engelen2014}
{van Engelen} A.,  {Bhattacharya} S.,  {Sehgal} N.,  {Holder} G.~P.,  {Zahn}
  O.,   {Nagai} D.,  2014, \mn@doi [\apj] {10.1088/0004-637X/786/1/13}, \href
  {http://adsabs.harvard.edu/abs/2014ApJ...786...13V} {786, 13}

\bibitem[\protect\citeauthoryear{{van den Bosch}, {More}, {Cacciato}, {Mo}  \&
  {Yang}}{{van den Bosch} et~al.}{2013}]{Bosch2013}
{van den Bosch} F.~C.,  {More} S.,  {Cacciato} M.,  {Mo} H.,   {Yang} X.,
  2013, \mn@doi [\mnras] {10.1093/mnras/sts006}, \href
  {http://adsabs.harvard.edu/abs/2013MNRAS.430..725V} {430, 725}

\makeatother
\end{thebibliography}
